\documentclass[twocolumn,english,aps,prb,twocolum,superscriptaddress,bibnotes,amsmath,amssymb,floatfix]{revtex4-2}

\pdfoutput=1
\usepackage{amsfonts}
\usepackage{amsmath}
\usepackage{amssymb}
\usepackage[colorlinks=true,citecolor=blue,linkcolor=blue,breaklinks=true]{hyperref}
\usepackage{graphicx}

\usepackage{soul}
\usepackage{url}
\usepackage{dcolumn}
\usepackage{bm}
\usepackage{stmaryrd}
\usepackage[final]{changes}
\usepackage{multirow}
\usepackage{rotating}
\usepackage{titlesec}
\usepackage{tabularx}
\usepackage{textcomp}
\usepackage[flushleft]{threeparttable}
\usepackage{ragged2e}
\usepackage{caption}
\DeclareCaptionJustification{justified}{\justifying}
\usepackage{titlesec}
\renewcommand{\thesection}{\arabic{section}}

\newcommand{\figref}[1]{\mbox{Fig.~\ref{#1}}}
\newcommand{\figsref}[1]{\mbox{Figs.~\ref{#1}}}
\newcommand{\tabref}[1]{\mbox{Table~\ref{#1}}}
\newcommand{\secref}[1]{\mbox{Sec.~\ref{#1}}}

\begin{document}

\title{Quantum collective motion of macroscopic mechanical oscillators}
\author{Mahdi Chegnizadeh}\thanks{These authors contributed equally.}
\author{Marco Scigliuzzo}\thanks{These authors contributed equally.}
\author{Amir Youssefi}
\author{Shingo Kono}
\author{Evgenii Guzovskii}
\author{Tobias~J.~Kippenberg}
\email{tobias.kippenberg@epfl.ch}
\affiliation{Institute of Physics, Swiss Federal Institute of Technology Lausanne (EPFL); CH-1015 Lausanne, Switzerland.}
\affiliation{Institute of Electrical and Micro Engineering, Swiss Federal Institute of Technology Lausanne (EPFL); CH-1015 Lausanne, Switzerland.}
\affiliation{Center for Quantum Science and Engineering, EPFL, Lausanne, Switzerland.}

\maketitle

\noindent \textbf{Collective phenomena arise from interactions within complex systems, leading to behaviors absent in individual components. Observing quantum collective phenomena with macroscopic mechanical oscillators has been impeded by the stringent requirement that oscillators be identical. Here, we demonstrate the quantum regime for collective motion of $N=6$ mechanical oscillators, a hexamer, in a superconducting circuit optomechanical platform. By increasing the optomechanical couplings, the system transitions from individual to collective motion, characterized by a $\sqrt{N}$ enhancement of cavity-collective mode coupling, akin to super-radiance of atomic ensembles. Using sideband cooling, we prepare the collective mode in the quantum ground state and measure its quantum sideband asymmetry, with zero-point motion distributed across distant oscillators. This regime of optomechanics opens avenues for studying multi-partite entanglement, with potential advances in quantum metrology.}

Understanding the collective dynamics of complex multimode systems in physics~\cite{andersonMoreDifferent1972} allows to describe fundamental emergent phenomena such as phase transitions~\cite{onuki2002phase} in the classical regime and Bose-Einstein condensation~\cite{anderson1995observation} in the quantum regime. Collective phenomena can alter the light-matter interaction, leading to super-radiance of atomic ensembles~\cite{dicke1954coherence} and the collective coupling enhancement in Tavis-Cummings systems~\cite{tavis1968exact}. While atomic ensembles have historically been the natural testbed for studying collective phenomena~\cite{skribanowitz1973observation}, engineered artificial atoms, such as superconducting qubits~\cite{fink2009dressed}, offer new avenues for exploring collective behaviors due to their design flexibility. 

Mechanical oscillators, with compact size and long coherence times, are proposed for exploring quantum collective dynamics. Controlling their quantum states has become possible via optomechanical coupling to electromagnetic resonators~\cite{aspelmeyer2014cavity} or piezoelectric coupling to superconducting qubits~\cite{o2010quantum}. Optomechanical systems enable ground state cooling~\cite{chan2011laser} of mechanical oscillators, strong coupling~\cite{verhagen2012quantum}, and squeezing of light~\cite{safavi2013squeezed}, even at room temperature~\cite{huang2024room}. Superconducting circuit-based optomechanical systems~\cite{teufel2011sideband} are notable for low decoherence, maintaining sufficient optomechanical coupling, and residing in the resolved-sideband regime, allowing the demonstration of a wide range of optomechanical phenomena such as entanglement~\cite{ockeloen2018stabilized,kotler2021direct} or squeezing of mechanical oscillators~\cite{wollman2015quantum,youssefi2023squeezed}. Recent topological lattice implementation~\cite{youssefiTopological2022} has laid the foundation for studying the collective behavior of mechanical oscillators.

Observing collective phenomena in optomechanical systems requires mechanical oscillators to be degenerate, i.e., identical. Investigations using degenerate oscillators have revealed long-range interactions~\cite{xuereb2012strong}, synchronization~\cite{heinrichCollective2011}, topological phases of sound~\cite{peano2015topological}, hybridization~\cite{shkarin2014optically}, mechanical dark and bright modes~\cite{massel2012multimode,ockeloen2019sideband,kharel2022multimode}, and exceptional points~\cite{xu2016topological,patil2022measuring}. However, experimental demonstrations have largely been limited to two nearly degenerate oscillators~\cite{massel2012multimode,ockeloen2019sideband,shkarin2014optically} or the modes of a bulk-acoustic-wave resonator~\cite{kharel2022multimode}, due to fabrication limitations. Similar efforts with two levitated nanoparticles have demonstrated non-reciprocal~\cite{rieser2022tunable} and long-range interactions~\cite{vijayan2024cavity}. Nevertheless, collective phenomena scale with the number of units, underscoring the importance of large-scale arrays of identical oscillators.

\subsection*{Nearly-degenerate mechanical oscillators}
\label{sec:sec1}
\begin{figure*}[hbt!]%
	\includegraphics[width=\linewidth]{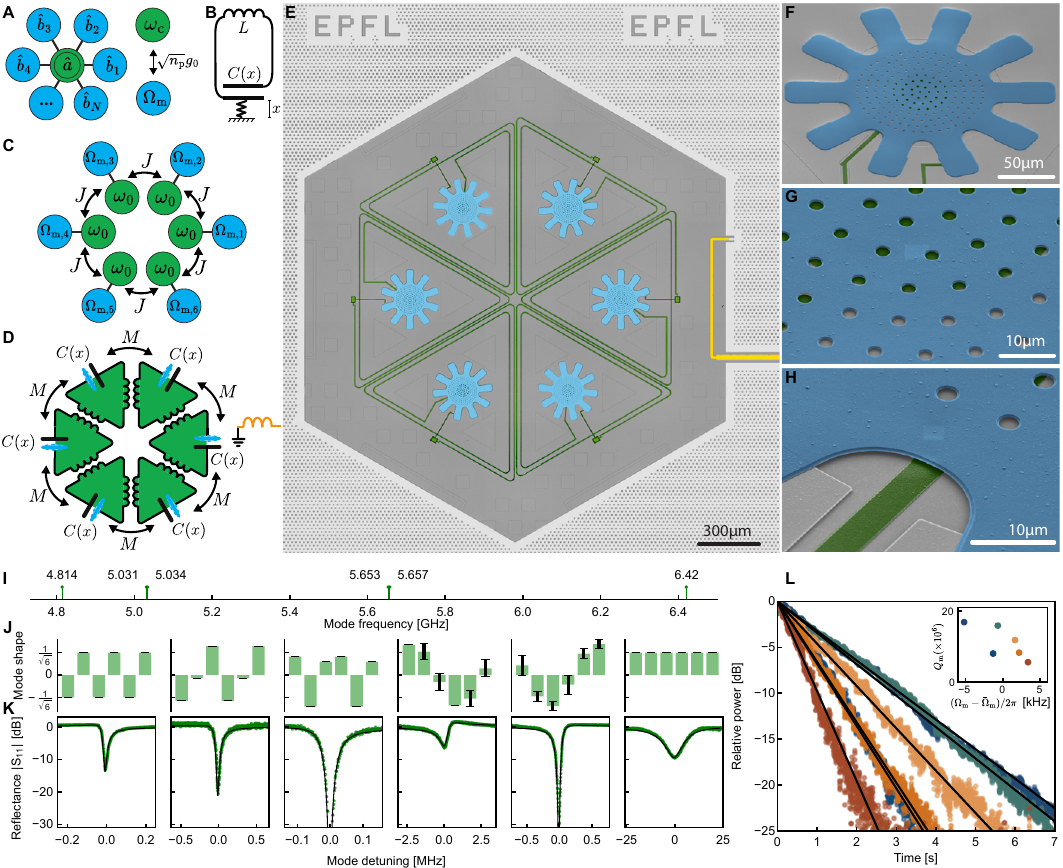}
	\caption{\textbf{A superconducting circuit optomechanical hexamer for studying collective phenomena.}
		\textbf{A}, Mode diagram for a system of $N$ mechanical oscillators (blue) optomechanically coupled to a cavity mode (green). \textbf{B}, Circuit equivalent of a single optomechanical system, where a mechanical oscillator modulates the capacitance $C(x)$ of an LC resonator. \textbf{C}, Mode diagram, and \textbf{D}, equivalent electrical circuit of the implemented system: six identical optomechanical unit cells coupled with rotational symmetry. A waveguide is inductively coupled to drive the system. \textbf{E}, False-color optical micrograph of the device, with details of the drum in \textbf{F}, \textbf{G}, and \textbf{H}. \textbf{I}, \textbf{J}, and \textbf{K}, Microwave mode frequencies, modeshapes, and reflection traces (circles: data, solid line: fit), respectively. \textbf{L}, Ringdown measurements of the mechanical oscillators (circles: data, solid line: fit) with their quality factors shown in the inset.}
	\label{fig:device}
\end{figure*}

We study the role of degeneracy for observing collective dynamics in an optomechanical system, comprising $N$ degenerate mechanical oscillators $\hat b_i$ with frequency $\Omega_{{\rm m}}$ optomechanically coupled with equal coupling rates $g$ to a common photonic mode $\hat a$ with frequency $\omega_{\rm c}$ (Fig.~1\textbf{A}). The Hamiltonian of the system in the presence of a pump with frequency $\omega_{\rm p}$ red-detuned from the cavity by $\Delta=\omega_{\rm c}-\omega_{\rm p}=\Omega_{\rm m}$ in the rotating frame of the pump is given by: $\hat H/\hbar= \Delta \hat a^\dagger \hat a +\Omega_{\rm m} \sum_{i=1}^N \hat b_i^\dagger\hat b_i + g\left(\hat a^\dagger\sum_{i=1}^N  \hat b_i+\hat a\sum_{i=1}^N  \hat b_i^\dagger\right)$. The optomechanical coupling rate $g=\sqrt{n_{\rm p}}g_0$ can be tuned by changing the intra-cavity photon number $n_{\rm p}$, where $g_0$ is the single photon optomechanical coupling rate. 

The Hamiltonian can be recast into an effective beam-splitter interaction between the microwave mode and a single mechanical mode $\hat \beta_1=\sum_{i=1}^N \hat b_i/\sqrt{N}$, that will be referred to as a bright collective mode, 
\begin{equation}
	\label{eq:interaction}
	\hat H/\hbar=\Delta \hat a^\dagger \hat a +\Omega_{\rm m} \sum_{i=1}^N \hat \beta_i^\dagger\hat \beta_i + g\sqrt{N}\left(\hat a^\dagger \hat \beta_1+\hat a \hat \beta_1^\dagger\right),
\end{equation}
while other collective modes $\hat \beta_{i\neq 1}$ are decoupled from the cavity, referred to as dark collective modes. Notably, such decomposition is only possible due to the degeneracy of the mechanical oscillators~\cite{SI}. The bright collective mode, $\beta_1$, is central to our investigation. Its modeshape is characterized by equal participation of each individual mechanical oscillator, with zero relative phase between them.

In contrast to the case of individual mechanical oscillators, the optomechanical interaction part in the Hamiltonian~\eqref{eq:interaction} includes an additional factor of $\sqrt{N}$. This collective enhancement of coupling arises from the identical oscillators being equally coupled to a common cavity. A similar phenomenon is observed when $N$ identical emitters are equally coupled to a common cavity, known as Tavis-Cummings systems~\cite{tavis1968exact,fink2009dressed}. Consequently, the optomechanical damping rate~\cite{aspelmeyer2014cavity} of the bright collective mode is given by
\begin{equation}
	\Gamma_{\rm opt}=N\frac{4 g^2}{\kappa},
\end{equation} 
where $\kappa$ is the linewidth of the cavity, illustrating an $N$-fold enhancement compared to a single-mode optomechanical system. A red-detuned tone therefore cools down the bright collective mode $\hat\beta_1$ while the other dark collective modes $\hat \beta_{i\neq 1}$ retain their high phonon occupation. Remarkably, in the limit $N\gg 1$, this results in individual mechanical oscillators in equilibrium with the thermal bath while their collective mode is in ground state~\cite{SI}. 

Our implementation of the model described by the Hamiltonian~\eqref{eq:interaction} is based on a circuit optomechanical platform~\cite{teufel2011sideband,youssefi2023squeezed}, where the frequency $\omega_{\rm 0}$ of a lumped element superconducting microwave LC resonator is modulated by a mechanically compliant electrode that forms the parallel-plate vacuum-gap capacitor of the resonator (Fig.~1\textbf{B}). The circuit is mounted on the mixing chamber of a dilution refrigerator with temperature $T\approx 12$~mK. In particular, we couple six microwave resonators of frequency $\omega_{\rm 0}$, each optomechanically coupled to an individual mechanical oscillator with frequency $\Omega_{{\rm m},i}$ (Figs.~1\textbf{C}, \textbf{D}).
The coupling strength $J$ between the microwave resonators is implemented through mutual inductance $M$, resulting in six hybridized microwave modes. Importantly, the rotational symmetry of the implemented hexamer assures the presence of a symmetric (highest frequency) and an anti-symmetric (lowest frequency) microwave modes, which have equal participation among the resonator sites, both effectively implementing our target Hamiltonian~\eqref{eq:interaction}~\cite{SI}. Their linewidths, however, differ appreciably: the (anti-)symmetric mode has a (narrow) broad linewidth, emanating from the (destructive) constructive interference of each microwave resonator's emission within the common bath it is coupled to, i.e., the microwave feedline~\cite{SI}. This aspect is pivotal in our implementation, as it provides at the same time a primary cavity (the anti-symmetric mode, with frequency $\omega_{\rm c}$ and linewidth $\kappa$) in the resolved-sideband regime $\Omega_{{\rm m},i}\gg \kappa$, instrumental for characterizing and cooling the bright collective mode, as well as an auxiliary cavity (the symmetric mode, with frequency $\omega_{\rm aux}$ and linewidth $\kappa_{\rm aux}$) in the bad-cavity limit $\Omega_{{\rm m},i}\ll \kappa_{\rm aux}$, necessary for probing its occupation using quantum sideband asymmetry experiment~\cite{weinstein2014observation}.
A false-colour micrograph of the system is shown in Figs.~1\textbf{E}-\textbf{H}.  

The measured frequencies and reflections of the microwave modes are shown in Figs.~1\textbf{I} and \textbf{K}, respectively. The primary and auxiliary cavities have frequencies $\omega_{\rm c}/2\pi=4.8$~GHz and $\omega_{\rm aux}/2\pi = 6.4$~GHz, with corresponding linewidths $\kappa/2\pi=28$~kHz and $\kappa_{\rm aux}/2\pi= 14.7$~MHz, respectively.
By measuring the external coupling rates to the feedline for all the microwave modes together with their frequencies, we infer the disorder among individual microwave frequencies $\omega_0$ and reconstruct the modeshape of microwave modes (Fig.~1\textbf{J}, \cite{SI}). These measurements confirm that our circuit realizes the underlying Hamiltonian~\eqref{eq:interaction}.

In Fig.~1\textbf{L}, we present the ringdown traces of mechanical oscillators and their quality factors $Q_{{\rm m},i}$, averaging 11 million. Remarkably, their frequencies fall within a narrow band of 10~kHz. Considering their average frequency $\bar{\Omega}_{{\rm m} }/2\pi= 2$~MHz and frequency standard deviation $\sigma_{\Omega_{\rm m}}/2\pi= 2.8$~kHz, the disorder is $\sigma_{\Omega_{\rm m}}/\bar{\Omega}_{{\rm m} }= 0.1\%$, enabling us to observe the collective modes of mechanical oscillators. 

\subsection*{Emergence of collective modes}
\label{sec:sec2}

\begin{figure*}[hbt!]%
		\includegraphics[width=\linewidth]{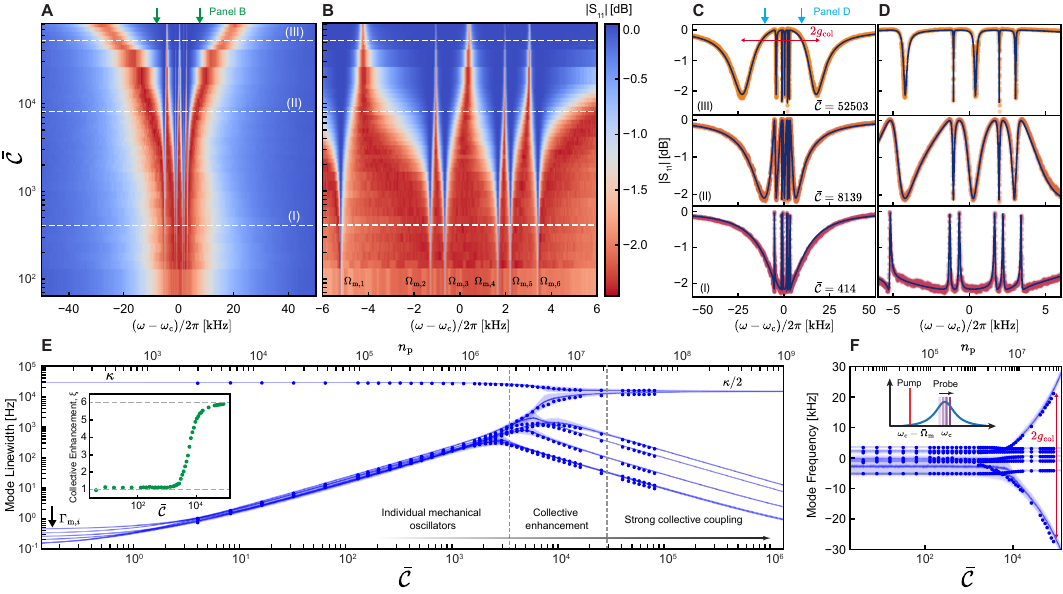}
	\caption{\textbf{Emergence of dark and bright collective modes in a circuit optomechanical hexamer}. \textbf{A}, Wide span and \textbf{B}, zoomed-in optomechanically induced transparency (OMIT) response. The bare mechanical frequencies are labeled in low cooperativity regime. Three cross sections (white lines) are shown in panels \textbf{C} and \textbf{D}. \textbf{C}, Three OMIT traces (circles) and fits (solid line) using the theoretical model. \textbf{D}, The zoomed-in OMIT traces. In \textbf{A}-\textbf{D}, we have removed the Fano interference for better visibility of the microwave response. \textbf{E}, Linewidths of the modes. Dots are extracted from the experimental data and the blue line is the theoretical expectation. The blue shaded region reflects 90\% certainty range. The inset shows the collective enhancement factor $\xi$. The first dashed line determines the region with $\xi \geq 2$. The second dashed line determines the region with $>90\%$ hybridization. \textbf{F}, Extracted mode frequencies and the theory prediction (subtracted from $\bar \Omega_{\rm m}$). Schematic of the multi-mode OMIT measurement is shown in inset.}
	\label{fig:fig2}
\end{figure*}

To investigate bright and dark collective modes, we conduct an optomechanically induced transparency (OMIT) experiment~\cite{weis2010optomechanically}, in which we send a coherent tone (pump) at the red sideband of the primary cavity, detuned by $\bar \Omega_{\rm{m}}$. We measure the reflection $S_{11}(\omega)$ around the cavity frequency using a weak probe (Fig.~2\textbf{F} inset). A wide span response and a zoomed-in one are plotted for varying pump powers (Figs.~2\textbf{A}, \textbf{B}). The theoretical model can accurately predict experimental data (Figs.~2\textbf{C}, \textbf{D}). Using this data, we extract the linewidths and frequencies of the system modes (the system contains six mechanical oscillators and one microwave mode) versus average of cooperativities $\overline {\cal C}  = 4\overline {g_i^2/\kappa {\Gamma _{{\rm{m}},i}}}$, where $\Gamma_{{\rm m},i}$ is the damping rate and $g_i$ is the optomechanical coupling rate of the $i$th mechanical oscillator (Figs.~2\textbf{E}, \textbf{F}). As the pump power increases, $g_i$ scales with $n_{\rm p}$ according to $g_i=g_{0,i}\sqrt{n_{\rm p}}$, where $g_{0,i}$ is the single-photon optomechanical coupling rate of the $i$th mechanical oscillator. In our device, $\overline{g_{0,i}}/2\pi=1.3$~Hz, which is six times smaller than that of a similar optomechanical system but with a single mechanical oscillator, as the optomechanical coupling rates inherit the microwave mode distribution of the primary cavity~\cite{SI,youssefiTopological2022}. 

At low powers, $\mathcal{\bar{C}} \approx \mathcal{O}(10^1-10^3)$, six transparency windows appear in the reflection spectrum, corresponding to the OMIT features of individual mechanical oscillators. As pump power increases, the linewidth of each oscillator grows due to the optomechanical damping rate $\Gamma_{{\rm opt},i} = 4g_{i}^2/\kappa$. While $|\Omega_{{\rm{m}},i} - \Omega_{{\rm{m}},j}| \gg \Gamma_{{\rm{opt}},i}, \Gamma_{{\rm{opt}},j}$, the modes remain individual oscillators, evidenced by a linear increase in their linewidths with pump power (Fig.~2\textbf{E}). As the pump power increases further, the frequency difference between oscillators becomes comparable to their linewidths $\Gamma_{{\rm opt},i}$. When this criterion is met for all oscillators, $\mathcal{\bar{C}} \approx \mathcal{O}(10^3-10^4)$, they become indistinguishable, and a transition occurs from independent oscillators to collective modes. In this regime, mechanical sidebands scattered from individual oscillators interfere in the feedline, causing either suppression or enhancement of optomechanical damping rates. This interference is similar to the non-reciprocal optomechanical circuits, where pathway interference creates a transparency window~\cite{bernier2017nonreciprocal}. The transition is marked by the appearance of dark and bright collective modes: dark modes show decreasing linewidths with increasing pump power, while the bright mode is distinguished by an increase in its linewidth compared to a single optomechanical system. To quantify such collective enhancement, we define the collective optomechanical coupling rate $g_{\rm col} = \left| {\left\langle {{{\psi}}} \mathrel{\left | {\vphantom {{\psi} {{g_i}}}} \right. \kern-\nulldelimiterspace} {{{g_i}}} \right\rangle } \right|$, where $\left| {{\psi}} \right\rangle $ is the vector of the bright collective mode obtained from the OMIT measurement and $\left| {{g_i}} \right\rangle  = {\left( {{g_1},...,{g_N}} \right)^t} $. It quantifies the coupling of the bright collective mode to the primary cavity and simplifies to $\sqrt{N}g$ in the ideal case. In Fig.~2\textbf{E} inset, we plot the collective enhancement $\xi = {{{g_{{\rm{col}}}^2}/\overline {g_i^2} }}$, which starts at 1, indicating no collective enhancement, and approaches $N=6$, the maximum enhancement. At higher powers, $\mathcal{\bar C}\approx\mathcal{O}(10^4-10^5)$, the optomechanical damping rate of the bright collective mode becomes comparable to the linewidth of the primary cavity, and they enter the strong coupling regime, akin to the single mode case~\cite{teufel2011circuit}. In this regime, the linewidths of the hybridized modes become equal at $\kappa/2$, and their frequencies split by $2g_{\rm col}$ (Fig.~2\textbf{F}). At the highest power ($\mathcal{\bar C}=81784$) $\xi=5.91$, the frequency splitting is $\Delta \omega /2\pi  = 48.56$~kHz, and $\Delta \omega /2{g_{{\rm{col}}}} = 0.98$.

\subsection*{Measuring collective mechanical modeshapes}

\begin{figure*}[hbt!]%
		\includegraphics[width=\linewidth]{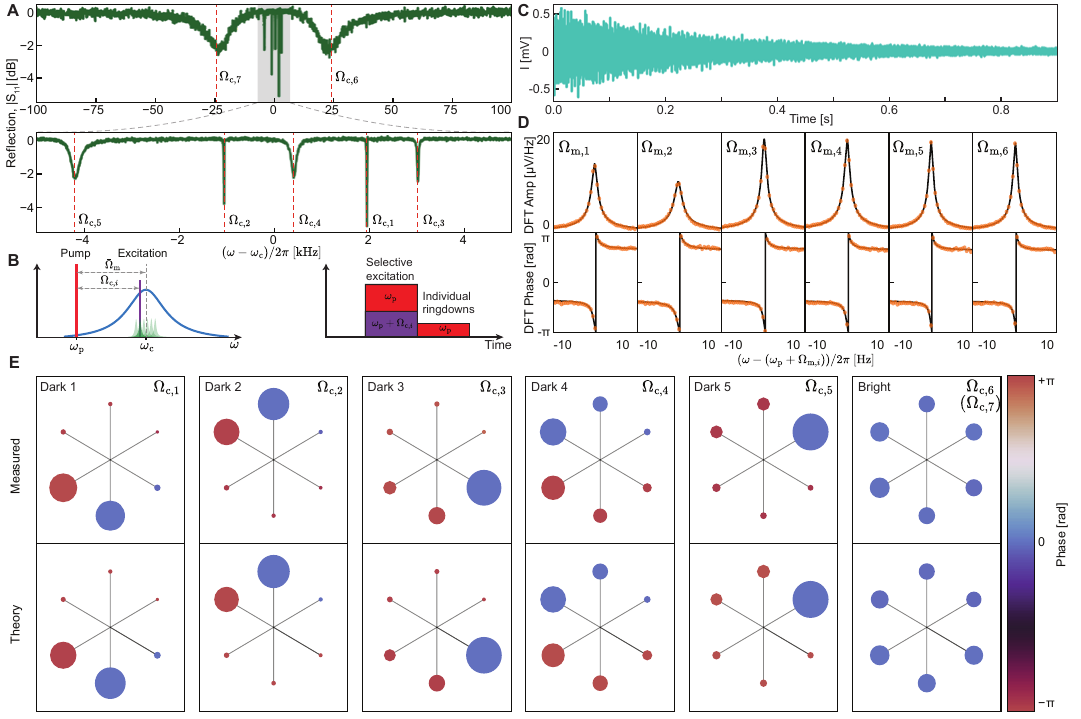} \footnotesize \linespread{1} 
	\caption{\textbf{Modeshapes of dark and bright collective modes}. \textbf{A}, OMIT response for $\mathcal{\bar C}=82685$: Broad dips indicate hybridization of the bright mode with the cavity, while narrow dips represent dark modes. Mode frequencies are shown near each resonance, sorted by linewidth. \textbf{B}, Frequency and pulse scheme: Each collective mode is selectively excited by an excitation signal that is blue-detuned from the pump by the frequency of the respective mode. The modeshape is then measured using a weak red-detuned pump. \textbf{C}, In-phase signal $I(t)$ after exciting the bright mode. \textbf{D}, Discrete Fourier Transform (DFT) of the complex signal $z(t)=I(t)+iQ(t)$ around each mechanical frequency $\Omega_{{\rm m},i}$. \textbf{E}, Reconstructed modeshapes for frequencies $\Omega_{{\rm c},i}$ in the phase space of individual mechanical oscillators. Circle radius reflects oscillator participation ratio; color shows relative phase to the oscillator with the maximum participation ratio. Top row: measured modeshapes, bottom: theory. Modes $\Omega_{{\rm c},6}$ and $\Omega_{{\rm c},7}$ have the same mechanical modeshape, so only $\Omega_{{\rm c},6}$ is shown.}
	\label{fig:fig3}
\end{figure*}

We then directly measure both the amplitudes and relative phases of the bright and dark collective modes by leveraging the finite non-degeneracy of mechanical frequencies ($0.1\%$). By switching from collective to individual dynamics, we initiate the ringdown of individual mechanical oscillators at their bare frequencies. The amplitude and relative phase of the ringdowns yield those of the collective modeshapes. For a given pump power, red detuned from the primary cavity by $\bar \Omega_{\rm m}$, we determine the bright and dark collective mode frequencies ($\Omega_{{\rm c},i}$) via an OMIT experiment, as outlined previously (Fig.~3\textbf{A}). Subsequently, we use a time-domain protocol, where a pulsed pump is sent along with a pulsed excitation signal blue-detuned from the pump by the frequency of the target collective mode $\Omega_{{\rm c},i}$ to selectively excite it (Fig.~3\textbf{B}). The beating between the pump and the excitation signal excites the mechanical oscillators according to their participation in the collective mode being excited. For example, by selectively exciting the bright collective mode, we expect to excite all the mechanical oscillators with the same amplitude and relative phase. After waiting for \(\tau \gg 1/\Gamma_{{\rm opt},i}\), we abruptly reduce the optomechanical coupling rates by turning off the excitation signal and reducing the pump power, which initiates the ringdown of the mechanical oscillators. At this power, the optomechanical damping rates satisfy \(\Gamma _{{\rm{opt}},i,j} \approx 10\Gamma _{{\rm{m}},i,j} \ll |\Omega _{{\rm{m}},i} - \Omega _{{\rm{m}},j}|\), allowing us to characterize them individually at their bare frequencies while measuring them faster than their internal decay rates. The in-phase and out-of-phase quadrature ringdown signals, $I(t)$ and $Q(t)$ respectively, around the cavity frequency are then recorded (Fig.~3\textbf{C}). With both quadratures, the complex signal in the rotating frame of the pump can be expressed as: \[	z(t\ge 0) = I(t) + iQ(t) = \sum\limits_{j = 1}^N {{A_j}{e^{-i({\Omega _{{\rm{m}},j}}t - {\varphi _j}) - {(\Gamma _{{\rm{m}},j} + \Gamma _{{\rm{opt}},j})}t/2}}},\] where $\varphi_j$ and ${A_j}$ denote the relative phase and amplitude of each mechanical oscillator within the selectively excited collective mode, respectively. We perform a Fourier transform on the complex signal to extract both the amplitude and the relative phase of the individual mechanical ringdowns~(Fig.~3\textbf{D}, \cite{SI}). The reconstructed modeshapes align closely with theoretical values (Fig.~3\textbf{E}). Notably, in the bright collective mode, all mechanical oscillators exhibit nearly identical amplitude and relative phase, while the dark collective modes are localized on one or more oscillators. 

\subsection*{Ground-state cooling of the bright collective mode}
\label{sec:cooling}
The high-quality factors of individual mechanical oscillators (averaging $11\times10^6$) and their thermal decoherence rates $\overline {{\Gamma_{{\rm{th}},i}}} \left( { \equiv \overline {{\Gamma_{{\rm{m}},i}}n_{{\rm{m}},i}^{{\rm{th}}}} } \right)/2\pi = 54.5 \pm 0.5$~Hz, where ${n_{{\rm{m}},i}^{{\rm{th}}}}$ is the occupation of the $i$th mechanical bath, enable preparation of the bright collective mode in its quantum ground state. Sideband cooling~\cite{teufel2011sideband} is employed by applying a pump red-detuned from the primary cavity by $\bar \Omega_{\rm m}$ (Fig.~4\textbf{A}), while filter cavities at room temperature suppress pump phase noise near the cavity frequency~\cite{SI}. The thermomechanical sideband around the cavity frequency is detected and fitted to a model based on quantum Langevin equations and input-output relations~\cite{SI}. The model closely matches the experimental data across different pump power ranges (Fig.~4\textbf{D}). Using system parameters obtained through fitting, we construct the covariance matrix of the mechanical quadratures, whose eigenvalues correspond to the occupations of the dark and bright collective modes~\cite{SI}. The occupation of the bright collective mode remains below one quanta over a wide range of powers, reaching a minimum of $n_{\rm m}=0.40\pm0.04$, limited by heating of the primary cavity~\cite{teufel2011sideband} (Fig.~4\textbf{B}). The quantum efficiency of the measurement chain, used to derive the occupations shown in Fig.~4\textbf{B}, is determined through a temperature sweep of the dilution fridge~\cite{SI}. To evaluate the overlap between the measured bright mode and the ideal one, we plot the modeshape fidelity based on the covariance matrix eigenvectors (Fig.4~\textbf{C})\cite{SI}. This fidelity starts from 1/6, indicating localization in one oscillator, and approaches 1, signifying equal amplitude distribution across all sites with zero relative phase.

To verify the occupation measurement, we perform an optomechanical quantum sideband asymmetry experiment~\cite{safavi2012observation,weinstein2014observation,qiu2020laser,youssefi2023squeezed}, using the asymmetry between the Stokes and anti-Stokes thermomechanical sidebands to extract the mechanical occupation. The experiment is carried out in two regimes: (i) at low pump power ($\mathcal{\bar C} \approx \mathcal{O}(10^2-10^3)$), where the mechanical oscillators remain in their individual basis, and (ii) at high pump power ($\mathcal{\bar C} \approx \mathcal{O}(10^4-10^5)$), where the bright collective mode emerges and enters the strong coupling regime with the primary cavity. To measure the sidebands, we use the auxiliary cavity and apply a probe tuned to its center frequency (Fig.~4\textbf{E}). Filter cavities at the upper and lower motional sidebands of the probe suppress source phase noise, allowing quantum backaction to dominate. The Stokes and anti-Stokes sidebands are proportional to $n_{\rm m}+1+2n_{\rm aux}$ and $n_{\rm m}-2n_{\rm aux}$, respectively, where $n_{\rm aux}$ accounts for the heating of the auxiliary cavity. In the first regime, we focus on the lowest-frequency mechanical oscillator (Fig.~4\textbf{G}). In the second regime, the sidebands reflect the strong coupling between the primary cavity and the bright collective mode~\cite{teufel2011sideband}, as evidenced by the doublet in the observed sidebands (Fig.~4\textbf{I}). Using the difference between the sideband areas and accounting for the heating of the auxiliary cavity, $n_{\rm aux}$\cite{SI}, we extract the mechanical occupations. The results from sideband asymmetry closely match those derived from the temperature-sweep-calibrated quantum efficiency (Figs.~4\textbf{F}, \textbf{H}), confirming the validity of the measurements in Fig.~4\textbf{B}. Notably, the mechanical occupation in Fig.~4\textbf{F} is higher than in Fig.~4\textbf{B}, mainly due to the non-negligible quantum backaction (compared to decoherence rate) from the resonant probe on the auxiliary cavity~\cite{SI,qiu2020laser,youssefi2023squeezed}.

\begin{figure*}[hbt!]%
	\includegraphics[width=\linewidth]{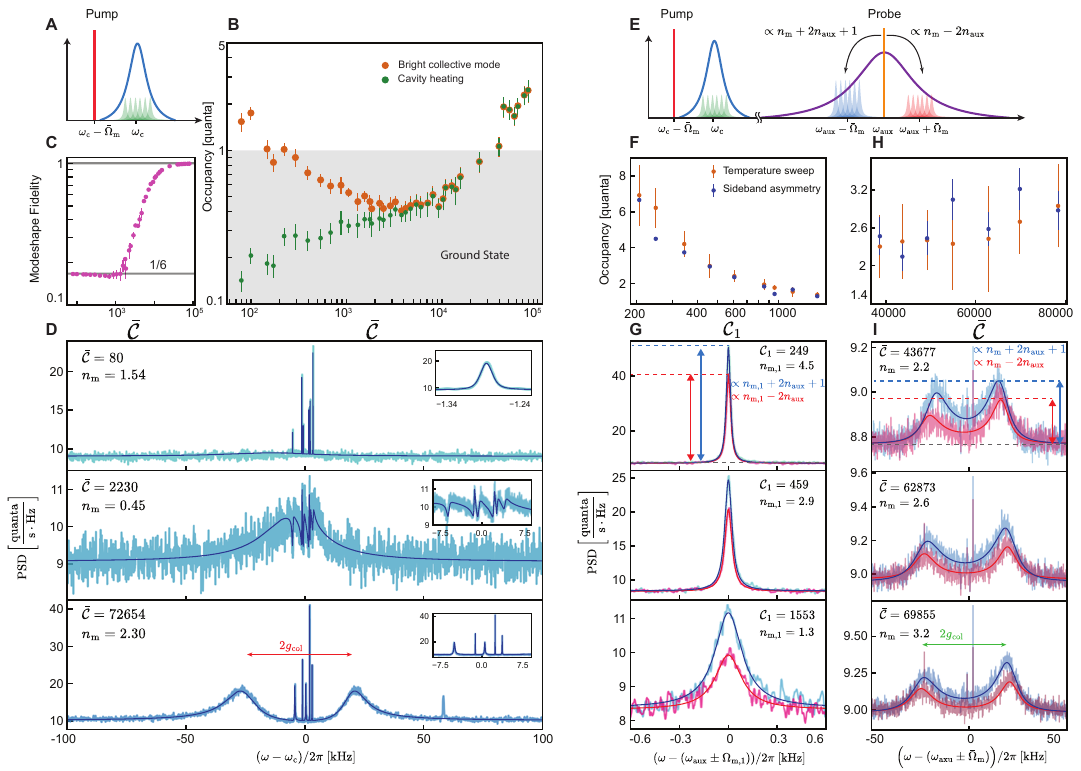}
	\caption{\textbf{Ground-state cooling and quantum sideband asymmetry of the bright collective mode}. \textbf{A}, Measurement scheme for sideband cooling of the bright collective mode, with no pump at the auxiliary cavity. \textbf{B}, Occupation of the bright collective mode (orange circles) and heating of the primary cavity (green circles), with the ground state indicated by the gray shaded region. \textbf{C}, Fidelity between the ideal and measured bright collective mode. \textbf{D}, Output power spectral density (PSD) from the cavity with fits (blue line) for three cooperativities. Insets show zoomed-in spectra. \textbf{E}, Pumping scheme for cooling (primary cavity) and sideband asymmetry (auxiliary cavity) experiments. \textbf{F}, Occupation of the lowest frequency mechanical oscillator ($n_{{\rm m},1}$) at low cooling powers via sideband asymmetry (blue circles) and temperature-sweep-calibrated quantum efficiency (orange circles). \textbf{G}, Stokes and anti-Stokes sidebands of the lowest frequency oscillator at low powers versus its cooperativity ($\mathcal{C}_1$). \textbf{H}, Occupation of the bright collective mode in the strong coupling regime using sideband asymmetry and calibrated quantum efficiency. \textbf{I}, Stokes and anti-Stokes sidebands in the strong coupling regime. The unequal peak heights result from the finite detuning of the cooling pump from the optimal frequency. The anti-Stokes sideband is mirrored relative to the center frequency.}
	\label{fig:fig4}
\end{figure*}

\subsection*{Conclusion}
Our observation of quantum collective phenomena in an optomechanical system composed of a mechanical hexamer lays the groundwork for realizing theoretical proposals for achieving multi-partite phonon-phonon~\cite{vitali2007stationary,hartmann2008steady} and photon-phonon~\cite{lai2022noise} entanglement. Additionally, it features higher-order exceptional points that could be harnessed for energy transfer~\cite{xu2016topological} or to study non-Hermitian dynamics, potentially offering advantages in system control. The spatial extension of the collective mechanical mode may provide resilience to local sources of noise or even function as a distributed sensor. Looking ahead, integrating superconducting qubits with collective mechanical modes opens avenues for simulating more advanced systems, such as spin-boson models~\cite{leggett1987dynamics} using mechanical oscillators.

\subsection*{Acknowledgement}
The authors acknowledge helpful discussion in the fabrication development with Niccol\'o Piacentini. We thank N. Akbari, N. J. Engelsen, A. Noguchi, Y. Ashida, F. Minganti, and F. Ferrari for fruitful discussion. All devices were fabricated in the Center of MicroNanoTechnology (CMi) at EPFL. This work was supported by funding from Swiss National Science Foundation (SNSF) under grant No. 204927, European Research Council (ERC) under grant No. 835329 (ExCOM-cCEO), and Quantum Science and Engineering center at EPFL. 

\subsection*{Author contributions}
M.C. designed the devices. M.C. and M.S. performed the numerical analysis and simulations. M.C. developed the theory with assistance from S.K. M.C. and M.S. developed DUV process and fabricated the devices with assistance from A.Y. and E.G. M.C. conceived the modeshape measurement technique and performed the experiments. M.C. and M.S. analyzed the data. The manuscript is written by M.C., M.S., T.J.K., and all other authors. T.J.K. supervised the project.

\subsection*{Competing interests:}
There are no competing interests to declare.

\subsection*{Data and materials availability}
The datasets and codes used to produce the results of this paper are available at a Zenodo open-access repository~\cite{dataset}.

%
\let\oldaddcontentsline\addcontentsline
\renewcommand{\addcontentsline}[3]{}

\let\addcontentsline\oldaddcontentsline

\clearpage
\onecolumngrid

\begin{center}
	\large{\textbf{Supplementary Information}}
\end{center}

\setcounter{equation}{0}
\renewcommand{\thefigure}{S\arabic{figure}}
\renewcommand{\theHfigure}{S\arabic{figure}}
\setcounter{figure}{0}
\setcounter{table}{0}

\setcounter{subsection}{0}
\setcounter{section}{0}

\tableofcontents

\pagebreak
\titleformat{\section}
{\normalfont\large\bfseries}{Supplementary Note~\thesection.}{1em}{}
\section{System parameters and variables}
\label{sec:params}
The systems parameters are given in \tabref{tab:glossary}; quality factor, single-photon optomechanical coupling rates, and frequencies of individual oscillators are given in \tabref{tab:mechanics}; (approximate) properties of the drum are give in \tabref{tab:mechanics_approx}; and all variables used throughout the paper are in given in \tabref{tab:variables}. 

{\renewcommand{\arraystretch}{1.2}%

	\begin{table}[h!]
		\caption{System parameters}
		\label{tab:glossary}
		\begin{center}
			\begin{tabular}{| l  | c | l |} \hline
				\multicolumn{1}{|c|}{\textbf{Parameter}}  & 
				\textbf{Symbol} & 
				\multicolumn{1}{|c|}{\textbf{Value}} \\ \hline\hline
				
				Microwave cavity frequency (individual sites) &
				$\omega_\mathrm{0}$ & $2\pi \; \times$ 5.427 GHz \\ \hline
				
				Microwave inter-cavity coupling &
				$J$ & $2\pi \; \times$ 367 MHz \\ \hline
				
				Primary cavity frequency &
				$\omega_\mathrm{c}$ & $2\pi \; \times$ 4.814 GHz \\ \hline
				
				Primary cavity linewidth:  $\kappa = \kappa_\mathrm{ex}+\kappa_\mathrm{0} $ &
				$\kappa$ & $2\pi \; \times$ 32 kHz \\ \hline			
				
				Primary cavity external coupling rate&
				$\kappa_\mathrm{ex}$ & $  2\pi \; \times$ 25 kHz \\ \hline			
				
				Primary cavity internal loss rate&
				$\kappa_\mathrm{0}$ & $  2\pi \; \times$ 7 kHz \\ \hline			
				
				Auxiliary cavity frequency &
				$\omega_\mathrm{aux}$ & $2\pi \; \times$ 6.420 GHz \\ \hline
				
				Auxiliary cavity linewidth: $\kappa_{\rm aux} = \kappa_\mathrm{ex}^{\rm aux}+\kappa_\mathrm{0}^{\rm aux}$ &
				$\kappa_{\rm aux}$ & $2\pi \; \times$ 14.7 MHz \\ \hline
				
				Auxiliary cavity external coupling rate&
				$\kappa_\mathrm{ex}^{\rm aux}$ & $\ 2\pi \; \times$ 11.94 MHz \\ \hline
				
				Auxiliary cavity internal loss rate &
				$\kappa_\mathrm{0}^{\rm aux}$ & $\ 2\pi \; \times$ 2.75 MHz \\ \hline
				
				Average of mechanical frequencies &
				$\bar \Omega_\mathrm{m}$ & $2\pi  \; \times$ 1.991 MHz \\ \hline
				
				Standard deviation of mechanical frequencies &
				$\sigma_{\Omega_\mathrm{m}}$ & $2\pi  \; \times$ 2.83 kHz \\ \hline
				
				Average of mechanical bare damping rates&
				$\bar \Gamma_\mathrm{m}$ & $2\pi  \; \times$ 212 mHz \\ \hline
				
				Average of mechanical quality factors&
				$\bar Q_\mathrm{m}$ & $11\times 10^6$ \\ \hline
				
				Average of single-photon optomechanical coupling rates&
				$\bar g_0$ & $2\pi  \; \times$ 1.3 Hz \\ \hline			
				
			\end{tabular}
		\end{center}
	\end{table}

	\begin{table}[h!]
		\caption{Individual mechanical oscillator frequencies $\Omega_{{\rm m},i}$, single photon optomechanical coupling rates $g_{0,i}$, and quality factors $Q_{{\rm m},i}$.}
		\label{tab:mechanics}
		\begin{center}
			\begin{tabular}{| c  || c | c | c |} \hline
				\textbf{Number $i$}  & 
				$(\Omega_{{\rm m},i}-{\bar \Omega_{\rm m}})/2\pi$ &
				$g_{0,i}(\equiv g_{0,i}'/6)/2\pi$ &
				$Q_{{\rm m},i}(\times 10^6)$
				\\ 
				
				\hline\hline
				
				1 &
				-5.19~kHz &
				$1.3\pm 0.1$~Hz &
				16.84\\ \hline					
				
				2 &
				-1.3~kHz &
				$1.3\pm 0.1$~Hz&
				7.84\\ \hline					
				
				3 &
				-0.68~kHz&
				$1.4\pm 0.1$~Hz&
				15.87 \\ \hline					
				
				4 &
				+1.63~kHz&
				$1.2\pm 0.1$~Hz&
				11.80 \\ \hline					
				
				5 &
				+2.18~kHz&
				$1.2\pm 0.1$~Hz&
				8.26 \\ \hline					
				
				6 &
				+3.37~kHz&
				$1.2\pm 0.1$~Hz&
				5.56 \\ \hline					
				
			\end{tabular}
		\end{center}
	\end{table}
	
	\begin{table}[h!]
		\caption{Radius of the mechanical oscillator $(R)$, radius of the counter electrode that forms the capacitor $(r)$, approximate gap size $(d)$, approximate gap size fluctuations $\sigma_d$, approximate tension of the top plate, approximate zero-point motion $(x_{\rm zpf})$, approximate mass of the mechanical oscillator $m$ }
		\label{tab:mechanics_approx}
		\begin{center}
			\begin{tabular}{| c || c | c | c |c |c |c|c|} \hline
				&$R$~[$\mu$m] 
				&$r$~[$\mu$m]
				&$d$~[nm]
				&$\sigma_d^{(*)}$~[nm]
				&Tensile stress$^{(\dagger)}$~[MPa]
				&$x_{\rm zpf}^{(\ddagger)}$~[fm]
				&$m^{(\S)}$~[ng]\\

				\hline\hline
				Value
				&70
				&23
				&$\approx 200$
				&$<1$
				&$\approx 350$
				&$\approx 1$
				&$\approx 5$\\
				\hline					
				
			\end{tabular}
		\end{center}
		\caption*{\raggedright $*$ Inferred from the frequency fluctuation of the individual microwave resonators, which in our case is $\approx 5$~MHz (0.1\%, see \secref{sec:MWimperfections})\newline $\dagger$ see \cite{youssefi2023squeezed}\newline $\ddagger$ Inferred from $g_0=\xi_{\rm cap}\frac{\omega_{\rm c}}{2d}x_{\rm zpf}$, where $\xi_{\rm cap}$ is participation ration of the vacuum-gap capacitor to the total capacitance (including stray capacitance), which in our case is $\xi_{\rm cap}=0.68$ (see \cite{youssefi2023squeezed} and \secref{sec:simulation})\newline$\S$ Inferred from $x_{\rm zpf}=\sqrt{\frac{\hbar}{2m\Omega_{\rm m}}}$}
	\end{table}

	{\renewcommand{\arraystretch}{1.2}%
		\begin{table}[h]
			\caption{Variables}
			\label{tab:variables}
			\begin{center}
				\begin{tabular}{| l  | c |} \hline
					\multicolumn{1}{|c|}{\textbf{Variables}}  & 
					\textbf{Symbol} \\ \hline\hline
					
					Annihilation operator for primary cavity &
					$\hat a$  \\ \hline					
					
					Annihilation operator for the $i$th mechanical oscillator&
					$\hat b_i$ \\ \hline					
					
					Noise operator for primary cavity intrinsic bath &
					$\hat a_0^{\rm{in}} $ \\ \hline									
					
					Noise operator for primary cavity external bath&
					$\hat a_{\rm{ex}}^{\rm{in}} $  \\ \hline									
					
					Noise operator for the $i$th mechanical intrinsic bath&
					$\hat b_{0,i}^{\rm{in}} $  \\ \hline			
					
					Primary cavity thermal bath occupation  &
					$n_\mathrm{c}^\mathrm{th}$\\ \hline	
					
					Auxiliary cavity thermal bath occupation  &
					$n_\mathrm{aux}^\mathrm{th}$\\ \hline	
					
					Primary cavity thermal occupation  &
					$n_\mathrm{c}$ \\ \hline	
					
					Auxiliary cavity thermal occupation  &
					$n_\mathrm{aux}$ \\ \hline	
					
					Mechanical thermal bath occupation ($i$th oscillator)  &
					$n_{\mathrm{m},i}^\mathrm{th}$ \\ \hline			
					
					Intra-cavity photon number induced by cooling pump&
					$n_\mathrm{p}$ \\ \hline			
					
					
					
					Single photon Optomechanical coupling rate (coupled to individual resonators)&
					$g_{0,i}'$\\ \hline			
					
					Single photon Optomechanical coupling rate (coupled to the primary cavity): $g_{0,i} \equiv  g_{0,i}'/6$&
					$g_{0,i}$\\ \hline			
					
					Optomechanical coupling rate induced by pump (coupled to the primary cavity): $g_i \equiv  g_{0,i}\sqrt{n_{\rm p}}$&
					$g_i$\\ \hline			
					
					Average optomechanical cooperativity: $\mathcal{\bar C}=\overline {\frac{{4g_i^2}}{{\kappa {\Gamma _i}}}} $  &
					$\mathcal{\bar C}$ \\ \hline
					
					
					Total added noise of microwave measurement chain &
					$n_{\rm{add}}$ \\ \hline	
					
					HEMT effective added noise &
					$n_{\rm{add}}^{\rm{H}}$  \\ \hline	
					
					Occupation of the bright collective mode &
					$n_{\rm m}$  \\ \hline	
					
					Occupation of the $i$th mechanical oscillator &
					$n_{{\rm m},i}$  \\ \hline

				\end{tabular}
			\end{center}
		\end{table}
		
		\newpage
		\section{Theory}
		In this section we provide the theoretical framework in which we describe the system and the experimental measurement protocols. 
		\subsection{Full microwave circuit model}
		\label{sec:full_mw}
		A complete electrical equivalent circuit for the device reported in Fig.~1 of the main text must incorporate inductive coupling between nearest neighbors (\figref{fig:stray_capacitance}) as well as stray capacitance to ground. It is important to note that we disregard the stray capacitive coupling of two adjacent sites, as our experimental and numerical analysis (see \secref{sec:Numerical}) indicates its negligible contribution.
		\begin{figure*}[hbt!]%
			\centering
			\includegraphics[width=0.5\linewidth]{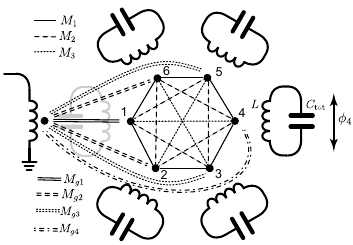}
			\caption{\textbf{Full equivalent circuit model.} The single lines correspond to coupling between different resonators, where the solid line represent nearest neighbor coupling, the dashed line is the second-nearest-neighbor coupling, and the dotted line shows the third-nearest-neighbor coupling. In the same fashion, the double line signals the coupling of individual resonators with the feedline. All coupling are dominated by the inductive part, denoted with the symbol $M_{\rm x}$. The numbers label the individual resonators.
			}
			\label{fig:fullCircuit}
		\end{figure*}
		The capacitor $\bf{C}$ and the inductor $\bf{L}$ matrices are given by
		\begin{equation}
			\begin{aligned}
				&\bf C=\left(
				\begin{array}{cccccc}
					C_{\rm d}+C_{\rm s} & 0 & 0 & 0 & 0 & 0 \\
					0 &  C_{\rm d}+C_{\rm s} & 0 & 0 & 0 & 0 \\
					0 & 0 &  C_{\rm d}+C_{\rm s} & 0 & 0 & 0 \\
					0 & 0 & 0 &  C_{\rm d}+C_{\rm s} & 0 & 0 \\
					0 & 0 & 0 & 0 &  C_{\rm d}+C_{\rm s} & 0 \\
					0 & 0 & 0 & 0 & 0 &  C_{\rm d}+C_{\rm s} \\
				\end{array}
				\right),\\ 
				&\bf L=\left(
				\begin{array}{cccccc}
					L & -M_1 & -M_2 & -M_3 & -M_2 & -M_1 \\
					-M_1 & L & -M_1 & -M_2 &  -M_3  & -M_2\\
					-M_2 & -M_1 & L & -M_1 & -M_2 & -M_3  \\
					-M_3  & -M_2 &-M_1 & L & -M_1 & -M_2 \\
					-M_2 &  -M_3  & -M_2 & -M_1 & L & -M_1 \\
					-M_1& -M_2 & -M_3  & -M_2 & -M_1 & L \\
				\end{array}
				\right),
			\end{aligned}
		\end{equation} 
		where $ C_{\rm d}$ is the drum capacitance and $C_{\rm s}$ is the stray capacitance of the electrodes to the ground. The same sign for all the mutual inductance terms (the off-diagonal terms in $\bf L$) is a direct consequence of the identical orientation of the spiral inductors in our design. The Lagrangian $\mathcal{L}$ of the circuit can be written as a function of node flux (voltage drop across each capacitor) $\bm\phi $ $=(\phi_1, \phi_2,\phi_3, \phi_4,\phi_5, \phi_6 )$: 
		\begin{equation}
			\mathcal{L}=\frac{1}{2}\bm{\dot\phi}^T\mathbf C\bm{\dot\phi} -\frac{1}{2}\bm{\phi}^T\mathbf L^{-1}\bm{\phi},
		\end{equation}
		and the normal modes are obtained by solving the equations of motion with the ansatz $\bm \phi_0 e^{i\omega t}$: $\left(\omega^2\mathbf C-\mathbf L^{-1} \right)\bm \phi_0=0$.  
		The microwave mode frequencies are given by
		\begin{equation}
			\bm\omega=\rm{Diag}(\sqrt{\mathbf {L}^{-1} \mathbf{ C}^{-1}}),
		\end{equation}
		that can be explicitly written as
		\begin{equation}
			\bm\omega=\left(
			\begin{array}{c}
				1/\sqrt{C_{\rm tot}(L-2M_1-2M_2-M_3)} \\[3pt]
				1/\sqrt{C_{\rm tot}(L-M_1+M_2+M_3)}\\[3pt]
				1/\sqrt{C_{\rm tot}(L-M_1+M_2+M_3)}  \\[3pt]
				1/\sqrt{C_{\rm tot}(L+M_1+M_2-M_3)}  \\[3pt]
				1/\sqrt{C_{\rm tot}(L+M_1+M_2-M_3)}  \\[3pt]
				1/\sqrt{C_{\rm tot}(L+2M_1-2M_2+M_3)} \\[3pt]
			\end{array}
			\right), 
		\end{equation}
		where $C_{\rm tot }=C_{\rm d}+C_{\rm s}$ and the frequencies are ordered from the largest to the smallest. 
		
		From the experimentally measured mode frequencies and the numerical simulation to estimate the ratio between the stray and vacuum gap capacitance (see \secref{sec:simulation} for more details), we can extract the circuits parameters, which are reported in table \ref{tab:circuits}.
		\begin{table}[h]
			\caption{Circuit parameters.}
			\centering
			\begin{tabular}{cc}
				\hline
				Parameter     & Value  \\ \hline \hline
				$C_{\rm d}$ & 75.2\,fF  \\
				$C_{\rm s}$ & 37.3\,fF  \\
				$L$ & 7.859\,nH \\ 
				$M_1$ & 1.010\,nH \\
				$M_2$ & 0.125\,nH \\
				$M_3$ & 0.086\,nH \\ \hline
			\end{tabular}
			\label{tab:circuits}
		\end{table}
		
		It is worth mentioning that the circuit symmetry gives rise to the degeneracy of two pairs of microwave modes. Small fluctuation in circuits parameters, experimentally accounted to capacitance variation due to gap-size fluctuations, breaks such symmetry and results in non degenerate modes. However, as we see in \secref{sec:MWimperfections}, the modeshapes of interest (the primary and auxiliary cavities) are very robust to fabrication imperfections.
		
		Following the derivation in \cite{devoret1995quantum,blais2021circuit}, the Hamiltonian is obtained from the Lagrangian using a Legendre transformation $\mathcal H=\mathbf q^T \bm{\dot \phi}-\mathcal{L}$, where $\mathbf q=(q_1, q_2, q_3, q_4, q_5, q_6)$ is the vector of conjugate variables $q_i=\partial\mathcal L/\partial\dot \phi_i$. In this case we obtain
		\begin{equation}
			\mathcal H=\frac{1}{2}\bm{q}^T\mathbf C^{-1}\bm{q} +\frac{1}{2}\bm{\phi}^T\mathbf L^{-1}\bm{\phi}.
		\end{equation}
		By introducing the frequency of the each individual resonators approximated in the first orders in $\frac{M_i}{L}$,  
		\begin{equation}
			\omega_{0}=\frac{1}{\sqrt{LC_{\rm tot}}}\left(1+30\frac{M_1}{2L}\frac{M_2}{2L}\frac{M_3}{2 L}\right),
		\end{equation}
		and by promoting $\bm q$ and $ \bm{\phi}$ to operators, such that their commutation relation reads $\left[  \hat \phi_m, \hat q_n \right]=i\delta_{mn}\hbar$, we introduce the creation and annihilation operators for the single microwave resonator
		\begin{equation}
			\begin{aligned}
				\hat \phi_m &=\phi_{\rm zpf} (\hat a_m^{\prime\dagger} +\hat a_m')= \sqrt{\frac{\hbar }{2}\frac{1}{\omega_0 C_{\rm tot}}}(\hat a_m^{\prime\dagger}+\hat a_m'), \\
				\hat q_n &=iq_{\rm zpf}(\hat a_n^{\prime\dagger} -\hat a_n') =i\sqrt{\frac{\hbar }{2}\omega_0 C_{\rm tot}}(\hat a_n^{\prime\dagger}-\hat a_n').
			\end{aligned}
		\end{equation}
		From these expressions, we can introduce the nearest-neighbor coupling $J_1$, the second-nearest-neighbor coupling $J_2$, and the third-nearest-neighbor coupling $J_3$, as
		\begin{equation}
			\begin{aligned}
				J_1&=\omega_0\left(\frac{M_1}{2L}+3\frac{M_1}{2L}\frac{M_2}{2L}+3\frac{M_2}{2L}\frac{M_3}{2L}\right), \\
				J_2&= \omega_0\left(\frac{M_2}{2L}+3\frac{M_1}{2L}\frac{M_3}{2L}+15\frac{M_1}{2L}\frac{M_2}{2L}\frac{M_3}{2L}\right),\\
				J_3&=\omega_0  \left(\frac{M_3}{2L}+6\frac{M_1}{2L}\frac{M_2}{2L}\right),
			\end{aligned}
		\end{equation}
		which are approximated in the first order in $M_i/L$. With these elements, the final Hamiltonian for the microwave part reads
		\begin{equation}
			\hat H_0/\hbar=\sum_{n=1}^N \omega_0 \hat a^{\prime\dagger}_n \hat a_n' +J_1\left(\hat a_n^{\prime \dagger} \hat a_{n+1}' +\hat a_n' \hat a_{n+1}^{\prime\dagger}\right)+J_2 \left(\hat a_n^{\prime\dagger} \hat a_{n+2}' +\hat a_n' \hat a_{n+2}^{\prime\dagger}\right) +\frac{1}{2}J_3 \left(\hat a_n^{\prime\dagger} \hat a_{n+3}' +\hat a_n' \hat a_{n+3}^{\prime\dagger}\right) 
			\label{eq:fullMW_Hamiltonian}
		\end{equation}
		with the convention $\hat a_{N+1}'=\hat a_1'$,  $\hat a_{N+2}'=\hat a_2'$ and $\hat a_{N+3}'=\hat a_3'$. Hamiltonian \eqref{eq:fullMW_Hamiltonian} can be written in the matrix form as 
		\begin{equation}
			\hat H_0=\hbar\bm{ \hat a}^{\prime\dagger}\bm{ H_0}\bm{\hat a}'=\hbar
			\left(
			\begin{array}{cccccc}
				\hat a_1^{\prime\dagger} & \hat a_2^{\prime\dagger} & \hat a_3^{\prime\dagger} & \hat a_4^{\prime\dagger} & \hat a_5^{\prime\dagger} & \hat a_6^{\prime\dagger}\\
			\end{array}
			\right) \left(
			\begin{array}{cccccc}
				\omega _0 & J_1 & J_2 & J_3 & J_2 & J_1 \\
				J_1 & \omega _0 & J_1 & J_2 & J_3 & J_2 \\
				J_2 & J_1 & \omega _0 & J_1 & J_2 & J_3 \\
				J_3 & J_2 & J_1 & \omega _0 & J_1 & J_2 \\
				J_2 & J_3 & J_2 & J_1 & \omega _0 & J_1 \\
				J_1 & J_2 & J_3 & J_2 & J_1 & \omega _0 \\
			\end{array}
			\right)
			\left(
			\begin{array}{c}
				\hat a_1' \\ \hat a_2' \\ \hat a_3'\\ \hat a_4' \\ \hat a_5' \\ \hat a_6'\\
			\end{array}
			\right).
			\label{eq:ham_unper}
		\end{equation}
		Introducing the base $\left| n\right\rangle=(\delta_{1n}, \delta_{2n}, \delta_{3n}, \delta_{4n}, \delta_{5n}, \delta_{6n})^{\rm t}$, where $\delta_{ij}$ is the Kronecker delta and $\delta_{ij}=1$ if $i=j$,
		the Hamiltonian matrix $\bm{ H_0}$ is diagonalized by the matrix $U$, where its elements are given by
		\begin{equation}
			U_{nm}=\frac{1}{\sqrt{6}}e^{in\frac{ 2\pi m}{N}},\qquad {\rm for}  \quad n\in\{0, \pm1,\pm2,3\},\quad m\in\{1,2, 3, 4, 5, 6\},
		\end{equation}
		resulting in the modes frequencies 
		\begin{equation}
			\bm\omega=\left(
			\begin{array}{c}
				\omega_0+2J_1+2J_2+J_3 \\[3pt]
				\omega_0+J_1-J_2-J_3\\[3pt]
				\omega_0+J_1-J_2-J_3 \\[3pt]
				\omega_0-J_1-J_2+J_3  \\[3pt]
				\omega_0-J_1-J_2+J_3    \\[3pt]
				\omega_0-2J_1+2J_2-J_3   \\[3pt]
			\end{array}
			\right),
			\label{eq:ws}
		\end{equation}
		and modes shape
		\begin{equation}
			\left|u_n\right\rangle=U_{nm} \left| m\right\rangle \qquad {\rm for}  \quad n\in\{0,\overbrace{ \pm1,\pm2}^\text{degenerate pairs},3\},\quad m\in\{1,2, 3, 4, 5, 6\}.
		\end{equation}
		As shown in the main text, we experimentally measure the frequencies and the linewidths of such modes. Importantly, due to small frequency fluctuation of the microwave resonators, the degeneracy is lifted, but the frequency average of such split modes is statistically resonant with the ideal degenerate case (see \secref{sec:MWimperfections} for more information). Therefore, by solving the linear system of four measured frequencies and 4 unknown parameters we extract $\omega_0/2\pi=5.436$\,GHz, $J_1/2\pi=372.6$\,MHz, $J_2/2\pi=92.2$\,MHz, and $J_3/2\pi=61.2$\,MHz. 
		
		Importantly, for our experimental implementation, the ideal mode shape for the auxiliary and primary cavities are 
		\begin{equation}
			\begin{split}
				&\left|\text{auxiliary cavity}\right\rangle=U_{0m} \left| m\right\rangle=\frac{1}{\sqrt{6}}\left(1,1, 1, 1, 1, 1\right)^t \\ 
				&\left|\text{primary cavity}\right\rangle=U_{3m} \left| m\right\rangle=\frac{1}{\sqrt{6}}\left(-1, 1, -1, 1,- 1, 1\right)^t.
			\end{split}
		\end{equation}
		
		To include the coupling to the feedline, we need to take into account that all individual microwave resonators are coupled to the same microwave bath (see \figref{fig:fullCircuit}). From the quantum Langevin equations for the microwave modes $\hat a_n'$ we have
		\begin{equation}
			\begin{aligned}
				\dot{\hat a}_n'= &\left(i\omega_0-\frac{\kappa_n}{2}\right) {\hat a}_n' +iJ_1 \left(\hat a_{n-1}'+\hat a_{n+1}'\right)+iJ_2 \left(\hat a_{n-2}'+\hat a_{n+2}'\right)+iJ_3 \left(\hat a_{n+3}'\right)+\sqrt{\kappa_n}\hat a_{\rm in}\\
				&+\sum_{m\neq n}^N\frac{1}{2}\sqrt{\kappa_m\kappa_n}\hat a_{m}',
			\end{aligned}
			\label{eq:LangevinMW}
		\end{equation}
		where $\kappa_i$ represents the individual decay rates of the resonators into the waveguide. It is crucial to note, that the last term in Eq.~\eqref{eq:LangevinMW} produces an effective \emph{dissipative coupling} between the resonators, and it arises due to the emission of the $m$th resonator to the feedline followed by the excitation of the $n$th one, similar to~\cite{toth2017dissipative} for two coupled resonators. Moreover, by a symmetry argument (confirmed by microwave simulations), resonators equally close to the feedline have the same decay rates, hence $\kappa_2=\kappa_5$ and $\kappa_5=\kappa_6$. We therefore introduce the non-Hermitian contribution to the Hamiltonian
		\begin{equation}
			\hat H_{\rm nH}/\hbar=\bm{ \hat a}^{\prime \dagger}\bm{ H_{\rm nH}}\bm{\hat a}',
		\end{equation}
		where
		\begin{equation}
			\bm{ H_{\rm nH}}=
			-\frac{i}{2}\left(
			\begin{array}{cccccc}
				\kappa _1 & \sqrt{\kappa _1 \kappa _2} & \sqrt{\kappa _1 \kappa _3} & \sqrt{\kappa _1 \kappa _4} & \sqrt{\kappa _1 \kappa _3} & \sqrt{\kappa _1 \kappa _2} \\
				\sqrt{\kappa _1 \kappa _2} & \kappa _2 & \sqrt{\kappa _2 \kappa _3} & \sqrt{\kappa _2 \kappa _4} & \sqrt{\kappa _2 \kappa _3} & \sqrt{\kappa _2^2} \\
				\sqrt{\kappa _1 \kappa _3} & \sqrt{\kappa _2 \kappa _3} & \kappa _3 & \sqrt{\kappa _3 \kappa _4} & \sqrt{\kappa _3^2} & \sqrt{\kappa _2 \kappa _3} \\
				\sqrt{\kappa _1 \kappa _4} & \sqrt{\kappa _2 \kappa _4} & \sqrt{\kappa _3 \kappa _4} & \kappa _4 & \sqrt{\kappa _3 \kappa _4} & \sqrt{\kappa _2 \kappa _4} \\
				\sqrt{\kappa _1 \kappa _3} & \sqrt{\kappa _2 \kappa _3} & \sqrt{\kappa _3^2} & \sqrt{\kappa _3 \kappa _4} & \kappa _3 & \sqrt{\kappa _2 \kappa _3} \\
				\sqrt{\kappa _1 \kappa _2} & \sqrt{\kappa _2^2} & \sqrt{\kappa _2 \kappa _3} & \sqrt{\kappa _2 \kappa _4} & \sqrt{\kappa _2 \kappa _3} & \kappa _2 \\
			\end{array}.
			\right)
		\end{equation}
		Experimentally, we cannot access this matrix directly, but in the next section, we show how to infer it using the experimental data.

		\subsection{Microwave mode reconstruction}
		As mentioned above, a small disorder in the microwave frequency $\omega_0$ lifts the degeneracy of the degenerate mode pairs $\left|u_{\pm 1}\right\rangle$ and $\left|u_{\pm 2}\right\rangle$. Here we exploit the different values of microwave mode couplings to the feedline in order to reconstruct their modeshapes. The method consists of four main steps: 1- Introducing an additional part to the Hamiltonian wherein the frequency of the individual microwave resonators are weakly perturbed by additive Gaussian noise with zero mean; 2- Applying the original eigenvector matrix $U$ to the new Hamiltonian, resulting in a block-diagonal matrix (using the first order perturbation theory) with two $2\times2$ blocks corresponding to the subspaces of the previously degenerate modes; 3- Introducing a new diagonalization matrix with linear combinations of $\left|u_{\pm 1}\right\rangle$ and $\left|u_{\pm 2}\right\rangle$; and 4- Diagonalizing the full non-Hermitian Hamiltonian and determining the coefficient of such linear combination that satisfy the experimental values.
		
		Considering a small disorder on the resonators frequencies $\omega_0+\delta_i$ the system Hamiltonian reads 
		\begin{equation}
			{\hat H}/\hbar={\hat H_0}/{\hbar}+ \sum_n^N \delta_n\hat a_n^{\prime\dagger}\hat a_n' ={\hat H_0}/{ \hbar}+\hat D_\delta=\bm{ \hat a}^{\prime\dagger}\left(\bm{ H_0+D}\right)\bm{\hat a}',
		\end{equation}
		where the assumption $\delta_i/\omega_0\ll 1$ is justified by the small splitting of the microwave modes $\left|u_{\pm 1}\right\rangle$ and $\left|u_{\pm 2}\right\rangle$ (see \secref{sec:MWimperfections} for more information). Considering a Gaussian frequency fluctuation
		\begin{equation}
			\begin{split}
				&\left\langle u_n\right| \bm{D} \left|u_n\right\rangle=\frac{1}{6}\sum_n^6\delta_n= 0,\\
				& \left\langle u_n\right| \bm{D}\left|u_m\right\rangle\ll\bm{\omega}_n-\bm{\omega}_m \qquad \text{for} \quad |n-m|>1,\\
				& \left\langle u_{1}\right|\bm{D}\left|u_{-1}\right\rangle=\frac{1}{6}\sum_n^6\delta_ne^{-i n \pi/3}=r e^{i\phi},\\
				& \left\langle u_{2}\right| \bm{D}\left|u_{-2}\right\rangle=\frac{1}{6}\sum_n^6\delta_ne^{-2i n \pi/3}=r' e^{i\phi'},
			\end{split}
		\end{equation}
		the full system diagonalization can be done by simply diagonalizing the two blocks on the main diagonal related to the ideally degenerate modes. In matrix form we have
		\begin{equation}
			U^t \left(\bm{ H_0+D}\right) U\approx\left(
			\begin{array}{cccccc}
				\omega _{\rm c} & \cdot & \cdot & \cdot & \cdot & \cdot \\
				\cdot & \omega _2 & r e^{i\phi} & \cdot & \cdot & \cdot \\
				\cdot & r e^{-i\phi} & \omega _2 & \cdot &\cdot &\cdot \\
				\cdot & \cdot & \cdot & \omega _4 & r' e^{i\phi'} & \cdot \\
				\cdot & \cdot & \cdot & r' e^{-i\phi'} & \omega _4 & \cdot \\
				\cdot & \cdot & \cdot & \cdot & \cdot & \omega _{\rm aux} \\
			\end{array}
			\right),
		\end{equation}
		where dots are the higher order perturbation terms. 
		
		The unperturbed degenerate modes now split by $2r$ and $2r'$ respectively. Such splitting can be readily measured by single tone spectroscopy. By diagonalizing each block, the new non-degenerate eigenstates of the system are 
		\begin{equation}
			\begin{split}
				&\left|u_{1, \rm nd}\right\rangle=\frac{1}{\sqrt{2}}\left(\left|u_{1}\right\rangle+e^{-i\phi}\left|u_{-1}\right\rangle\right)\\
				&\left|u_{-1, \rm nd}\right\rangle=\frac{1}{\sqrt{2}}\left(\left|u_{1}\right\rangle-e^{-i\phi}\left|u_{-1}\right\rangle\right)\\
				&\left|u_{2, \rm nd}\right\rangle=\frac{1}{\sqrt{2}}\left(\left|u_{2}\right\rangle+e^{-i\phi'}\left|u_{-2}\right\rangle\right)\\
				&\left|u_{-2, \rm nd}\right\rangle=\frac{1}{\sqrt{2}}\left(\left|u_{2}\right\rangle-e^{-i\phi'}\left|u_{-2}\right\rangle\right).
			\end{split}
		\end{equation}
		The goal is now to find the parameters $\phi$ and $\phi'$. For this, we apply the new diagonalization operator $U_{\rm nd}$ to the non-Hermitian component of the Hamiltonian $\bm{H_{\rm nH}}$, giving rise to a system of 6 unknowns ($\phi, \phi', \kappa_1, \kappa_2, \kappa_3, \kappa_4$) in 6 experimentally measured modes couplings. However, such system does not have an analytical solutions and we calculate its solutions by minimizing the cost function
		\begin{equation}
			\sum_n^N\left(\text{Imag}\left[U_{\rm nd}^t \bm{ H_{\rm nH}}U_{\rm nd}\right]_{nn}-\kappa_n^{\rm exp}\right)^2.
		\end{equation}
		The resulting parameters $\phi$ and $\phi'$ are used to reconstruct the mode shape in Fig.~1 of the main text. The extracted coupling rates of each site to the feedline are found to be $\kappa_1/2\pi=4.45$\,MHz, $\kappa_2/2\pi=2.24$\,MHz, $\kappa_3/2\pi=1.34$\,MHz and $\kappa_4/2\pi=1.09$\,MHz.
		
		\subsection{Linearized optomechanical Hamiltonian}
		\label{sec:ham_lin}
		Considering the mechanical oscillators and their optomechanical interaction with the microwave resonators, the full optomechanical Hamiltonian reads
		\begin{equation}
			\hat H/\hbar= \bm{ \hat a}^{\prime\dagger} \bm{ H_0}\bm{\hat a}'+\sum_i\Omega_{{\rm m},i} \hat b_i^\dagger\hat b_i+\sum_i g_{0,i}'\hat a_i^{\prime\dagger}\hat a_i' (\hat b_i^{\dagger}+\hat b_i),
		\end{equation}
		where $\hat{b}_i$ represents the annihilation operator for the mechanical oscillators, and $g_{0,i}'$ represents the single-photon optomechanical coupling rate. The microwave resonators can now be diagonalized in their basis by using the matrix $U$ in \secref{sec:full_mw}. Diagonal microwave modes $\hat a_i$ are related to the individual microwave resonators $\hat a_i'$ as $\hat{a}_i = \sum_j U_{ij} \hat{ a}'_j $. The Hamiltonian in the diagonalized basis is then rewritten as
		\begin{equation}
			\begin{aligned}
				\hat H/\hbar&= \sum_i \omega_i\hat a_i^\dagger\hat a_i+\sum_i\Omega_{{\rm m},i} \hat b_i^\dagger\hat b_i+\sum_i g_{0,i}'\left(\sum_j U_{ji}\hat a^\dagger_j \right)\left(\sum_k U^*_{ki}\hat a_k\right)(\hat b_i^{\dagger}+\hat b_i)\\
				&=\sum_i \omega_i\hat a_i^\dagger\hat a_i+\sum_i\Omega_{{\rm m},i} \hat b_i^\dagger\hat b_i+\sum_{i,j} g_{0,i}' \left|U_{ji}\right|^2 \hat a^\dagger_j\hat a_j(\hat b_i^{\dagger}+\hat b_i),
				\label{eq:fullOptoHam}
			\end{aligned}
		\end{equation}
		where $\omega_i$s are the microwave eigenfrequencies (Eq.~\eqref{eq:ws}). Here we have assumed the rotating wave approximation and ignored the terms $a_i^\dagger a_j$ when $i\neq j$, which is valid as $|\omega_i-\omega_j|\gg g_{0,i}'$. The collective optomechanical coupling of the $j$th microwave mode to the $i$th mechanical oscillator is $g_{0,i}'\left|U_{ji}\right|^2$. In particular, for the primary and auxiliary cavities, this yields to $g_{0,i} = g_{0,i}'/6$ for each mechanical oscillator. 
		
		From now on, we restrict the Hamiltonian~\eqref{eq:fullOptoHam} to the primary cavity with frequency $\omega_{\rm c}$. By considering a strong pump with frequency $\omega_{\rm p}$ red-detuned from the cavity by $\Delta=\omega_{\rm c} - \omega_{\rm p}$, the Hamiltonian simplifies to the the beam-splitter interaction (in the rotating frame of the pump):
		\begin{equation}
			\hat H/\hbar= \Delta\hat a^\dagger\hat a+\sum_i\Omega_{{\rm m},i} \hat b_i^\dagger\hat b_i+ \sum_i {{g_i}\left( {\hat a\hat b_i^\dag  + {{\hat a}^\dag }{{\hat b}_i}} \right)}.
			\label{eq:ham}
		\end{equation}
		where $g_i=g_{0,i}\sqrt{n_{\rm p}}$, $n_{\rm p}$ is the number of photons inside the cavity due to the pump, and $\hat a$ is the annihilation operator of the primary cavity (we remove the subscript 3 for simplicity).

		\subsection{Degenerate mechanical oscillators}
		\label{sec:degen}
		In this section, we show how to diagonalize the basis of degenerate mechanical oscillators. We consider the case where all the mechanical oscillators are degenerate (${\Omega _{{\rm m},i}} = {\Omega_{\rm m} }\;\;\forall i \in \left\{ {1,...,N} \right\}$) with the same optomechanical coupling rates (${g_i} = g\;\;\forall i \in \left\{ {1,...,N} \right\}$). The Hamiltonian in Eq.~(\ref{eq:ham}) then reads:
		\begin{equation}
			\hat H/\hbar  = \Delta {{\hat a}^\dag }\hat a + {\Omega_{\rm m}} \sum\limits_i {\hat b_i^\dag {{\hat b}_i}}  + g_i\left( {\hat a\sum\limits_i {\hat b_i^\dag }  + {{\hat a}^\dag }\sum\limits_i {{{\hat b}_i}} } \right).
			\label{eq:ham_deg}
		\end{equation}
		As it can be seen from this Hamiltonian, the cavity is now coupled to a mechanical collective mode, where all of the individual mechanics have equal amplitude and phase. This can be better seen by considering the mechanical oscillators in their own basis:
		\begin{equation}
			\Omega_{\rm m} \sum\limits_i {\hat b_i^\dag {{\hat b}_i}}  = \Omega_{\rm m} \left( {\hat b_1^\dag \;\;\hat b_2^\dag \;\; \cdots \;\;\hat b_N^\dag } \right){{\bf{I}}_{N \times N}}\left( {\begin{array}{*{20}{c}}
					{{{\hat b}_1}}\\
					{{{\hat b}_2}}\\
					\vdots \\
					{{{\hat b}_N}}
			\end{array}} \right),
		\end{equation}
		where ${\bf I}$ is an identity matrix. As it is a scalar matrix, there is a freedom to choose the basis. This can be done with any unitary matrix $\bf{U}$ where $\bf{U}^{\dagger}\bf{U} = \bf{I}$. We choose the first row of matrix $\bf{U}$ (one of the eigenvectors) to be 
		\begin{equation}
			{{\bf{U}}_{(1,i)}} = \frac{1}{{\sqrt N }}\left( {1\;\;1\;\; \cdots \;\;1} \right)^t
		\end{equation}
		which is divided by $\sqrt{N}$ for normalization. The other rows of the matrix should be orthogonal to this row, meaning
		\begin{equation}
			\sum\limits_{j = 1}^N {{{\bf{U}}_{(i \ne 1,j)}}}  = 0.
		\end{equation}
		We now rewrite the Hamiltonian in the new basis: 
		\begin{equation}
			\hat H/\hbar  = \Delta {{\hat a}^\dag }\hat a + \Omega \sum\limits_{i = 1}^N {\hat \beta_i^\dag {{\hat \beta}_i}}  + g\sqrt{N}\left( {\hat a\hat \beta_1^\dag  + {{\hat a}^\dag }{{\hat \beta}_1}} \right),
		\end{equation}
		where
		\begin{equation}
			\begin{aligned}
				\left( {\begin{array}{*{20}{c}}
						{{{\hat \beta}_1}}\\
						{{{\hat \beta}_2}}\\
						\vdots \\
						{{{\hat \beta}_N}}
				\end{array}} \right) = {\bf{U}}\left( {\begin{array}{*{20}{c}}
						{{{\hat b}_1}}\\
						{{{\hat b}_2}}\\
						\vdots \\
						{{{\hat b}_N}}
				\end{array}} \right),\\
				{{\hat \beta}_1} = \frac{1}{{\sqrt N }}\sum\limits_{i = 1}^N {{{\hat b}_i}}.
			\end{aligned}
			\label{eq:ideal_mode}
		\end{equation}
		Now, it can be clearly seen that the microwave mode is only coupled to $\beta_{1}$, which will be referred as the bright collective mode, wherein all individual mechanical oscillators have the same amplitude and relative phase. The other collective modes are decoupled from the cavity, which is why we name them dark collective modes. In addition, we see an $\sqrt{N}$ enhancement in the coupling of the bright collective mode to the cavity.
		
		The optomechanical damping rate of the bright collective mode induced by the pump is given by~\cite{aspelmeyer2014cavity}:
		\begin{equation}
			\begin{aligned}
				\Gamma_{\rm{opt}} & = \frac{4(g\sqrt{N})^2}{\kappa}=N\frac{4g^2}{\kappa}\\
				&=N\Gamma_{{\rm opt},i},
			\end{aligned}
			\label{eq:gam_col}
		\end{equation}
		where $\Gamma_{{\rm opt},i}$ is the optomechanical damping rate of an individual mechanical oscillator. The linewidth of the bright collective mode is then increased by a factor of $N$. It should be noted that, here, $\Gamma_{{\rm{opt}},i}$ is the optomechanical damping rate of one of the $N$ mechanical oscillators coupled to the primary cavity. In \secref{sec:ham_lin}, it is shown that by coupling to the primary cavity, the optomechanical coupling rate of each of the mechanical oscillators reduces by a factor of $N$ compared to a single-mode optomechanical system.

		\subsection{Multimode optomechanically induced transparency (OMIT)}
		\label{sec:omit}
		In this section, we provide the theoretical details of the observed multi-mode OMIT response for the \textbf{general case of non-degenerate mechanical frequencies}. We consider a pump with frequency $\omega_{\rm p}$ red-detuned by the average of the bare mechanical frequencies ${\bar \Omega_{\rm m}}$ from the primary cavity with frequency $\omega_{\rm c}$ ($\Delta = \omega_{\rm c} - \omega_{\rm p} = {\bar \Omega_{\rm m}}$) and a weak probe around its center frequency (Fig.~2\textbf{D} inset of the main text). The linearized Hamiltonian in the rotating frame of the pump is given by Eq.~\eqref{eq:ham}.
		
		The quantum Langevin equations, in the presence of a weak probe $\hat s_{\rm{in}}(t)$ and quantum noise, can be written as:
		\begin{equation}
			\begin{aligned}
				\frac{d}{{dt}} { \bf{\hat b}}  &= {\bf{ M}} {\bf{\hat b}}  +  {{{\bf{\hat b}}_{{\rm{in}}}}}, \\
				{\bf{\hat b}}  &= \left( {\begin{array}{*{20}{c}}
						{\hat a(t)}\\
						{{{\hat b}_1}(t)}\\
						\vdots \\
						{{{\hat b}_N}(t)}
				\end{array}} \right),\\
				{\bf{M}} &= \left( {\begin{array}{*{20}{c}}
						{ - i\Delta  - \kappa /2}&{ - i{g_1}}& \cdots &{ - i{g_N}}\\
						{ - i{g_1}}&{ - i{\Omega _{{\rm m},1}} - {\Gamma _{{\rm m},1}}/2}&0&0\\
						\vdots &0& \ddots &0\\
						{ - i{g_N}}&0&0&{ - i{\Omega _{{\rm m},N}} - {\Gamma _{{\rm m},N}}/2}
				\end{array}} \right),\\
				{{{\bf{\hat b}}_{{\rm{in}}}}}  &= \left( {\begin{array}{*{20}{c}}
						{\sqrt {{\kappa _{{\rm{ex}}}}} \hat s_{\rm{in}}(t) + \sqrt {{\kappa _{{\rm{ex}}}}} {{\hat a}_{{\rm{ex}}}}(t) + \sqrt {{\kappa _{\rm{0}}}} {{\hat a}_{\rm{0}}}(t)}\\
						{\sqrt {{\Gamma _{{\rm m},1}}} \hat b_{0,1}^{{\rm{in}}}(t)}\\
						\vdots \\
						{\sqrt {{\Gamma _{{\rm m},N}}} \hat b_{0,N}^{{\rm{in}}}(t)}
				\end{array}} \right).
			\end{aligned}
			\label{eq:lang_omit}
		\end{equation}
		Here, $\hat b_{0,i}^{{\rm{in}}}(t)$ is the noise operator for the internal loss of the $i$th mechanical oscillator, ${{\hat a}_{{\rm{0}}}}(t)$ is the noise operator for the internal loss of the primary cavity, and ${{\hat a}_{{\rm{ex}}}}(t)$ is the noise operator for the external field. These operators satisfy the following relations:
		\begin{equation}
			\begin{aligned}
				&\left\langle {\hat a_{{\rm{ex}}}^\dag (t){{\hat a}_{{\rm{ex}}}}(t')} \right\rangle  = 0,\\
				&\left\langle {\hat a_{{\rm{ex}}} (t){{\hat a}_{{\rm{ex}}}}^\dag (t')} \right\rangle  = \delta (t - t'),\\
				&\left\langle {\hat a_{\rm{0}}^\dag (t){{\hat a}_{\rm{0}}}(t')} \right\rangle  = n_{\rm{c}}^{{\rm{th}}}\delta (t - t'),\\
				&\left\langle {\hat b_{0,i}^{{\rm{in}}\dag}(t)\hat b_{0,i}^{{\rm{in}}}(t')} \right\rangle  = n_{\rm{m,i}}^{{\rm{th}}}\delta (t - t'),\\
				&\left\langle {\hat b_{0,i}^{{\rm{in}}\dag}(t)\hat b_{0,j\neq i}^{{\rm{in}}}(t')} \right\rangle  = \left\langle {\hat b_{0,i}^{{\rm{in}}}(t)\hat b_{0,j\neq i}^{{\rm{in}}\dagger}(t')} \right\rangle = 0.
			\end{aligned}
			\label{eq:noise_cor}
		\end{equation}
		Here, we have assumed zero temperature for the incident microwave field and finite temperature for the internal microwave and mechanical baths with $n_{\rm{c}}^{\rm{th}}$ and $n_{\rm{m},i}^{\rm{th}}$ occupations, respectively.

		The solution to Eq.~(\ref{eq:lang_omit}) is found by taking a Fourier transform, leading to:
		\begin{equation}
			\begin{aligned}
				&\left( {\begin{array}{*{20}{c}}
						{ - i(\omega  - \Delta ) + \kappa /2}&{  i{g_1}}& \cdots &{ i{g_N}}\\
						{ i{g_1}}&{ -i(\omega  - {\Omega _{{\rm m},1}}) + {\Gamma _{{\rm m},1}}/2}&0&0\\
						\vdots &0& \ddots &0\\
						{ i{g_N}}&0&0&{ - i(\omega  - {\Omega _{{\rm m},N}}) + {\Gamma _{{\rm m},N}}/2}
				\end{array}} \right)\left( {\begin{array}{*{20}{c}}
						{\hat a(\omega )}\\
						{{{\hat b}_1}(\omega )}\\
						\vdots \\
						{{{\hat b}_N}(\omega )}
				\end{array}} \right) =\\& \left( {\begin{array}{*{20}{c}}
						{\sqrt {{\kappa _{{\rm{ex}}}}} \hat s_{\rm{in}}(\omega ) + \sqrt {{\kappa _{{\rm{ex}}}}} {{\hat a}_{{\rm{ex}}}}(\omega ) + \sqrt {{\kappa _{\rm{0}}}} {{\hat a}_{\rm{0}}}(\omega )}\\
						{\sqrt {{\Gamma _{{\rm m},1}}} \hat b_{0,1}^{{\rm{in}}}(\omega )}\\
						\vdots \\
						{\sqrt {{\Gamma _{{\rm m},N}}} \hat b_{0,N}^{{\rm{in}}}(\omega )}
				\end{array}} \right),
			\end{aligned}
			\label{eq:matrix_omit}
		\end{equation}
		where the following convention is used for the Fourier transform:
		\begin{equation}
			\begin{aligned}
				{\cal F}\{ f(t)\}  &= F(\omega ) = \frac{1}{{\sqrt {2\pi } }}\int_{ - \infty }^{ + \infty } {f(t){e^{ + i\omega t}}dt} \\
				{{\cal F}^{ - 1}}\{ F(\omega )\}  &= f(t) = \frac{1}{{\sqrt {2\pi } }}\int_{ - \infty }^{ + \infty } {F(\omega ){e^{ - i\omega t}}d\omega }.
			\end{aligned}
		\end{equation}
		
		By inverting this matrix, we can find the solution of $\hat a$ and $\hat b_{1,...,N}$. To find the linear response of the system, we ignore the quantum noise and consider only the probe field, $\hat s_{\rm{in}}(t)$. In this case, using input-output relation, the output field, and consequently the reflection frequency response $S_{11}(\omega)$, will be given by:
		\begin{equation}
			\hat s_{\rm{out}}(\omega) = \hat s_{\rm{in}}(\omega) - \sqrt{\kappa_{\rm{ex}}}\hat a(\omega),
			\label{eq:in_out}
		\end{equation}
		and
		\begin{equation}
			\begin{aligned}
				S_{11}(\omega ) &\equiv \frac{{\left\langle {{{\hat s}_{{\rm{out}}}}(\omega )} \right\rangle }}{{\left\langle {{{\hat s}_{{\rm{in}}}}(\omega )} \right\rangle }}\\
				&= 1 - \frac{{{\kappa _{{\rm{ex}}}}}}{{\sum\limits_{i = 1}^N {{\chi _{\rm{m},i}(\omega)}g_i^2}  + 1/{\chi _{\rm{c}}(\omega)}}},
			\end{aligned}
			\label{eq:ref_omit}
		\end{equation}
		where $\chi_{\rm{m},i}(\omega)^{-1}=-i(\omega-\Omega_{{\rm m},i})+\Gamma_{{\rm m},i}/2$ and $\chi_{\rm{c}}(\omega)^{-1}=-i(\omega-\Delta)+\kappa/2$ are the bare mechanical and microwave susceptibilities, respectively. Equation (\ref{eq:ref_omit}) can be considered as the generalized OMIT response in the presence of $N$ mechanical oscillators. For $N=1$, we recover the well-known OMIT response \cite{weis2010optomechanically}.

		\subsection{Incoherent response}
		\label{sec:incoherent}
		In this part, we derive the output spectrum in the presence of a red detuned pump and absence of any resonant probe. Equation~(\ref{eq:matrix_omit}) is then rewritten in the following form:
		\begin{equation}
			\left( {\begin{array}{*{20}{c}}
					{\hat a(\omega )}\\
					{{{\hat b}_1}(\omega )}\\
					\vdots \\
					{{{\hat b}_N}(\omega )}
			\end{array}} \right) = \left( {\begin{array}{*{20}{c}}
					{{A_0}(\omega )}&{{A_1}(\omega )}& \cdots &{{A_N}(\omega )}\\
					{{B_{1,0}}(\omega )}&{{B_{1,1}}(\omega )}& \cdots &{{B_{1,N}}(\omega )}\\
					\vdots & \vdots & \vdots & \vdots \\
					{{B_{N,0}}(\omega )}&{{B_{N,1}}(\omega )}& \cdots &{{B_{N,N}}(\omega )}
			\end{array}} \right)\left( {\begin{array}{*{20}{c}}
					{\sqrt {{\kappa _{{\rm{ex}}}}} {{\hat a}_{{\rm{ex}}}}(\omega ) + \sqrt {{\kappa _{\rm{0}}}} {{\hat a}_{\rm{0}}}(\omega )}\\
					{\sqrt {{\Gamma _{{\rm m},1}}} \hat b_{0,1}^{{\rm{in}}}(\omega )}\\
					\vdots \\
					{\sqrt {{\Gamma _{{\rm m},N}}} \hat b_{0,N}^{{\rm{in}}}(\omega )}
			\end{array}} \right),
			\label{eq:matrix_omit_inv}
		\end{equation}
		where
		\begin{equation}
			\begin{aligned}
				&{A_p}(\omega ) = \frac{{ - i{g_p}{\chi _p} + (1 + i{g_p}{\chi _p}){\delta _{p,0}}}}{{D(\omega )}},\;\;p \in \{ 0,...,N\} \\\\
				&{B_{k,p}}(\omega ) = \frac{{ - i{g_k}{\chi _k}}}{{D(\omega )}}{\delta _{p,0}} + \left( {{\chi _k} - \frac{{{g_k}^2{\chi _k}^2}}{{D(\omega )}}} \right){\delta _{k,p}} + \frac{{ - {g_k}{g_p}{\chi _k}{\chi _p}}}{{D(\omega )}}(1 - {\delta _{k,p}})(1 - {\delta _{p,0}}),\\
				&k \in \{ 1,...,N\} ,\;\;p \in \{ 0,...,N\} \\\\
				&{\chi _p} = {\chi _{\rm{c}}}(\omega ){\delta _{p,0}} + {\chi _{{\rm{m}},p}}(\omega )(1 - {\delta _{p,0}}),\;\;p \in \{ 0,...,N\} \\\\
				&D(\omega ) = \sum\limits_{i = 1}^N {{\chi _{{\rm{m}},i}}(\omega )g_i^2}  + \chi _{\rm{c}}^{ - 1}(\omega ).
			\end{aligned}
			\label{eq:mat_elem}
		\end{equation}
		Here, $\delta_{i,j}$ is Kronecker delta and $\delta_{i,j}=1$ if $i=j$ and 0 if $i\neq j$. By using Eq.~(\ref{eq:matrix_omit_inv}), we find the output spectra from the cavity, the occupation of the individual mechanical oscillators, and the occupation of the collective mode. 
		
		\subsubsection{Microwave output spectrum}
		The output field from the cavity is found using input-output relation, by considering an input vacuum microwave field to the cavity $\hat a_{\rm{ex}}(\omega)$:
		\begin{equation}
			\begin{aligned}
				{{\hat a}_{{\rm{out}}}}(\omega ) &= {{\hat a}_{{\rm{ex}}}}(\omega ) - \sqrt {{\kappa _{{\rm{ex}}}}} \hat a(\omega )\\
				&= \left( {1 - {\kappa _{{\rm{ex}}}}{A_0}(\omega )} \right){{\hat a}_{{\rm{ex}}}}(\omega ) + \sqrt {{\kappa _{{\rm{ex}}}}{\kappa _{\rm{0}}}} {A_0}(\omega ){{\hat a}_{\rm{0}}}(\omega ) + \sum\limits_{i = 1}^N {\sqrt {{\kappa _{{\rm{ex}}}}{\Gamma _{{\rm m},i}}} {A_i}(\omega )\hat b_{0,i}^{{\rm{in}}}(\omega )}.
			\end{aligned}
		\end{equation}
		Using the noise correlators given in Eq.~(\ref{eq:noise_cor}), we find the the symmetrized output spectrum:
		\begin{equation}
			\begin{aligned}
				{{\bar S}}(\omega ) &= \frac{1}{2}\int_{ - \infty }^{ + \infty } {\left\langle {{{\hat a}_{\rm{out}}^\dagger }(\omega '){\hat a}_{\rm{out}}(\omega ) + {\hat a}_{\rm{out}}(\omega '){{\hat a}_{\rm{out} }^\dagger}(\omega )} \right\rangle d\omega '}\\
				& = \frac{1}{2}{\left| {1 - {\kappa _{{\rm{ex}}}}{A_0}(\omega )} \right|^2} + \left( {n_{\rm{c}}^{{\rm{th}}} + \frac{1}{2}} \right){\kappa _{{\rm{ex}}}}{\kappa _{\rm{0}}}{\left| {{A_0}(\omega )} \right|^2} + \sum\limits_{i = 1}^N {\left( {n_{{\rm{m,}}i}^{{\rm{th}}} + \frac{1}{2}} \right){\kappa _{{\rm{ex}}}}{\Gamma _{{\rm m},i}}{{\left| {{A_i}(\omega )} \right|}^2}}.
			\end{aligned}
			\label{eq:spec_a_unsimp}
		\end{equation}
		This equation can be further simplified by writing the first term as follows:
		\begin{equation}
			\begin{aligned}
				{\left| {1 - {\kappa _{{\rm{ex}}}}{A_0}(\omega )} \right|^2} &= 1 + \kappa _{{\rm{ex}}}^2{\left| {{A_0}(\omega )} \right|^2} - 2{\kappa _{{\rm{ex}}}}\Re \left\{ {{A_0}(\omega )} \right\}\\
				&= 1 + \kappa _{{\rm{ex}}}^2{\left| {{A_0}(\omega )} \right|^2} - 2{\kappa _{{\rm{ex}}}}\frac{1}{{{{\left| {D(\omega )} \right|}^2}}}\Re \left\{ {{D^*}(\omega )} \right\}\\
				&= 1 + \kappa _{{\rm{ex}}}^2{\left| {{A_0}(\omega )} \right|^2} - 2{\kappa _{{\rm{ex}}}}\frac{1}{{{{\left| {D(\omega )} \right|}^2}}}\left( {\frac{\kappa }{2} + \Re \left\{ {\sum\limits_{i = 1}^N {{\chi _{{\rm{m}},i}}(\omega )g_i^2} } \right\}} \right)\\
				&= 1 + \kappa _{{\rm{ex}}}^2{\left| {{A_0}(\omega )} \right|^2} - 2{\kappa _{{\rm{ex}}}}\frac{1}{{{{\left| {D(\omega )} \right|}^2}}}\left( {\frac{\kappa }{2} + \sum\limits_{i = 1}^N {{{\left| {{\chi _{{\rm{m}},i}}(\omega )} \right|}^2}g_i^2} \Re \left\{ {1/{\chi _{{\rm{m}},i}}(\omega )} \right\}} \right)\\
				&= 1 + \kappa _{{\rm{ex}}}^2{\left| {{A_0}(\omega )} \right|^2} - 2{\kappa _{{\rm{ex}}}}\frac{1}{{{{\left| {D(\omega )} \right|}^2}}}\left( {\frac{\kappa }{2} + \sum\limits_{i = 1}^N {{{\left| {{\chi _{{\rm{m}},i}}(\omega )} \right|}^2}g_i^2} \frac{{{\Gamma _{{\rm m},i}}}}{2}} \right),
			\end{aligned}
		\end{equation}
		where $\Re$ stands for the real part. We now pair all the 1/2 contributions in Eq.~(\ref{eq:spec_a_unsimp}) as:
		\begin{equation}
			\begin{aligned}
				&\frac{1}{2}\left( {{{\left| {1 - {\kappa _{{\rm{ex}}}}{A_0}(\omega )} \right|}^2} + {\kappa _{{\rm{ex}}}}{\kappa _{\rm{0}}}{{\left| {{A_0}(\omega )} \right|}^2} + \sum\limits_{i = 1}^N {{\kappa _{{\rm{ex}}}}{\Gamma _{{\rm m},i}}{{\left| {{A_i}(\omega )} \right|}^2}} } \right)\\\\
				&= \frac{1}{2}\left( 1 + \kappa _{{\rm{ex}}}^2{{\left| {{A_0}(\omega )} \right|}^2} - {\kappa _{{\rm{ex}}}}\frac{1}{{{{\left| {D(\omega )} \right|}^2}}}\left( {\kappa  + \sum\limits_{i = 1}^N {{{\left| {{\chi _{{\rm{m}},i}}(\omega )} \right|}^2}g_i^2} {\Gamma _{{\rm m},i}}} \right) + {\kappa _{{\rm{ex}}}}{\kappa _{\rm{0}}}{{\left| {{A_0}(\omega )} \right|}^2}\right.\\& \left.+ \sum\limits_{i = 1}^N {{\kappa _{{\rm{ex}}}}{\Gamma _{{\rm m},i}}{{\left| {{A_i}(\omega )} \right|}^2}}  \right)\\\\
				&= \frac{1}{2}\left( {1 + {\kappa _{{\rm{ex}}}}\kappa {{\left| {{A_0}(\omega )} \right|}^2} - {\kappa _{{\rm{ex}}}}\frac{1}{{{{\left| {D(\omega )} \right|}^2}}}\left( {\kappa  + \sum\limits_{i = 1}^N {{{\left| {{\chi _{{\rm{m}},i}}(\omega )} \right|}^2}g_i^2} {\Gamma _{{\rm m},i}}} \right) + \sum\limits_{i = 1}^N {{\kappa _{{\rm{ex}}}}{\Gamma _{{\rm m},i}}{{\left| {{A_i}(\omega )} \right|}^2}} } \right)\\\\
				&= \frac{1}{2}\left( {1 + {\kappa _{{\rm{ex}}}}\kappa \frac{1}{{{{\left| {D(\omega )} \right|}^2}}} - {\kappa _{{\rm{ex}}}}\frac{1}{{{{\left| {D(\omega )} \right|}^2}}}\left( {\kappa  + \sum\limits_{i = 1}^N {{{\left| {{\chi _{{\rm{m}},i}}(\omega )} \right|}^2}g_i^2} {\Gamma _{{\rm m},i}}} \right) + \sum\limits_{i = 1}^N {{\kappa _{{\rm{ex}}}}{\Gamma _{{\rm m},i}}\frac{{{g_i}^2{{\left| {{\chi _{{\rm{m}},i}}} \right|}^2}}}{{{{\left| {D(\omega )} \right|}^2}}}} } \right)\\\\
				&= \frac{1}{2},
			\end{aligned}
		\end{equation}
		where we used the explicit forms introduced in Eq.~(\ref{eq:mat_elem}). The simplified spectrum then reads:
		\begin{equation}
			\bar S(\omega ) = \frac{1}{2} + \frac{{{\kappa _{{\rm{ex}}}}{\kappa _{\rm{0}}}n_{\rm{c}}^{{\rm{th}}}}}{{{{\left| {\sum\limits_{i = 1}^N {{\chi _{{\rm{m}},i}}(\omega )g_i^2}  + \chi _{\rm{c}}^{ - 1}(\omega )} \right|}^2}}} + \frac{{\sum\limits_{i = 1}^N {{\kappa _{{\rm{ex}}}}{\Gamma _{{\rm m},i}}{g_i}^2{{\left| {{\chi _{{\rm{m}},i}}} \right|}^2}n_{{\rm{m,}}i}^{{\rm{th}}}} }}{{{{\left| {\sum\limits_{i = 1}^N {{\chi _{{\rm{m}},i}}(\omega )g_i^2}  + \chi _{\rm{c}}^{ - 1}(\omega )} \right|}^2}}}.
			\label{eq:spec_a}
		\end{equation}
		In this equation, the first term is the vacuum noise, the second term is the contribution of the cavity heating, and the last term is the contribution of each of the mechanical oscillators.

		\subsubsection{Individual and cross mechanical spectrums}
		Although we do not have direct access to each individual/physical mechanical oscillator (with $\hat b_i$ operator), we can find their occupations and their cross-correlations, respectively, using Eq.~(\ref{eq:matrix_omit_inv}):
		\begin{subequations}
			\begin{equation}
				\begin{aligned}
					{S_{\hat b_i^\dag {{\hat b}_i}}}(\omega ) =& \int_{ - \infty }^{ + \infty } {\left\langle {\hat b_i^\dag (\omega '){{\hat b}_i}(\omega )} \right\rangle d\omega '} \\
					=& \frac{{g_i^2{{\left| {{\chi _{{\rm{m}},i}}} \right|}^2}}}{{|D(\omega ){|^2}}}{\kappa _0}n_{\rm{c}}^{{\rm{th}}} + {\left| {{\chi _{{\rm{m}},i}} - \frac{{{g_i}^2{\chi _{{\rm{m}},i}}^2}}{{D(\omega )}}} \right|^2}{\Gamma _{{\rm m},i}}n_{{\rm{m}},i}^{{\rm{th}}} + \sum\limits_{\scriptstyle j = 1\hfill\atop
						\scriptstyle j \ne i\hfill}^N {\frac{{{g_i}^2{g_j}^2{{\left| {{\chi _{{\rm{m}},i}}{\chi _{{\rm{m}},j}}} \right|}^2}}}{{{{\left| {D(\omega )} \right|}^2}}}{\Gamma_{{\rm m},j}}n_{{\rm{m}},j}^{{\rm{th}}}},
				\end{aligned}
				\label{eq:spec_b_ind}
			\end{equation}
		%
		\begin{equation}
			\begin{aligned}
				{S_{\hat b_i^\dag {{\hat b}_j}}}(\omega ) =& \int_{ - \infty }^{ + \infty } {\left\langle {\hat b_i^\dag (\omega '){{\hat b}_j}(\omega )} \right\rangle d\omega '} \\
				=& \frac{{{g_i}{g_j}\chi _{{\rm{m}},i}^*{\chi _{{\rm{m}},j}}}}{{|D(\omega ){|^2}}}{\kappa _0}n_{\rm{c}}^{{\rm{th}}}\\
				&- {\left| {{\chi _{{\rm{m}},i}}} \right|^2}\left( {1 - \frac{{{g_i}^2\chi _{{\rm{m}},i}^*}}{{{D^*}(\omega )}}} \right)\left( {\frac{{{g_i}{g_j}{\chi _{{\rm{m}},j}}}}{{D(\omega )}}} \right){\Gamma _{{\rm m},i}}n_{{\rm{m}},i}^{{\rm{th}}}\\
				&- {\left| {{\chi _{{\rm{m}},j}}} \right|^2}\left( {1 - \frac{{{g_j}^2\chi _{{\rm{m}},j}}}{{{D}(\omega )}}} \right)\left( {\frac{{{g_i}{g_j}{\chi _{{\rm{m}},i}^*}}}{{D^*(\omega )}}} \right){\Gamma _{{\rm m},j}}n_{{\rm{m}},j}^{{\rm{th}}}\\
				&+ \sum\limits_{k = 1\hfill\atop
					k \ne i,j\hfill}^N {\frac{{{g_i}{g_j}{g_k}^2\chi _{{\rm{m}},i}^*{\chi _{{\rm{m}},j}}{{\left| {{\chi _{{\rm{m}},k}}} \right|}^2}}}{{{{\left| {D(\omega )} \right|}^2}}}{\Gamma _{{\rm m},k}}n_{{\rm{m}},k}^{{\rm{th}}}} 
			\end{aligned}
		\end{equation}
		\label{eq:spec_b_cross}
	\end{subequations}
	
	In the individual occupation expression (Eq.~\eqref{eq:spec_b_ind}), the first term is the effect of the cavity heating on the mechanical oscillator, the second term is the effect of its thermal bath, and the last term is the effect of other mechanical baths. The last term appears as the cavity is mediating between different mechanical oscillators due to the optomechanical coupling of each mechanical oscillator to the cavity. By taking the integral of the above spectra, we can find the occupation of each mechanical oscillator and their cross correlations.

	\subsubsection{Occupation of individual mechanical oscillators in an ideal system}
	Here we consider an ideal system where all the oscillators have the same frequency $(\Omega_{{\rm m},i}=\Omega_{\rm m})$, linewdith $(\Gamma_{{\rm m},i}=\Gamma_{\rm m})$, optomechanical coupling rate $(g_{i}=g)$, and thermal bath occupation ($n_{{\rm m},i}^{\rm th}=n_{{\rm m}}^{\rm th}$). Substituting these values into Eq.~\eqref{eq:spec_b_ind} yields:
	\begin{equation}
		\begin{aligned}
			&{S_{\hat b_i^\dag {{\hat b}_i}}}(\omega ) = \frac{{{g^2}{{\left| {{\chi _{\rm{m}}}} \right|}^2}}}{{|D(\omega ){|^2}}}{\kappa _0}n_{\rm{c}}^{{\rm{th}}} + {\left| {{\chi _{\rm{m}}} - \frac{{{g^2}{\chi _{\rm{m}}}^2}}{{D(\omega )}}} \right|^2}{\Gamma _{\rm{m}}}n_{\rm{m}}^{{\rm{th}}} + \frac{{(N - 1){g^4}{{\left| {{\chi _{\rm{m}}}} \right|}^4}}}{{{{\left| {D(\omega )} \right|}^2}}}{\Gamma _{\rm{m}}}n_{\rm{m}}^{{\rm{th}}}\\
			&= \frac{{{g^2}{{\left| {{\chi _{\rm{m}}}} \right|}^2}}}{{|D(\omega ){|^2}}}{\kappa _0}n_{\rm{c}}^{{\rm{th}}} + \left( {{{\left| {{\chi _{\rm{m}}}} \right|}^2} + \frac{{{g^4}{{\left| {{\chi _{\rm{m}}}} \right|}^4}}}{{{{\left| {D(\omega )} \right|}^2}}} - \frac{{2{g^2}{{\left| {{\chi _{\rm{m}}}} \right|}^2}}}{{{{\left| {D(\omega )} \right|}^2}}}\Re \left\{ {{\chi _{\rm{m}}}{D^*}(\omega )} \right\}} \right){\Gamma _{\rm{m}}}n_{\rm{m}}^{{\rm{th}}}\\
			& + \frac{{(N - 1){g^4}{{\left| {{\chi _{\rm{m}}}} \right|}^4}}}{{{{\left| {D(\omega )} \right|}^2}}}{\Gamma _{\rm{m}}}n_{\rm{m}}^{{\rm{th}}}\\
			&= \frac{{{g^2}{{\left| {{\chi _{\rm{m}}}} \right|}^2}}}{{|D(\omega ){|^2}}}{\kappa _0}n_{\rm{c}}^{{\rm{th}}} + \left( {{{\left| {{\chi _{\rm{m}}}} \right|}^2} + \frac{{{g^4}{{\left| {{\chi _{\rm{m}}}} \right|}^4}}}{{{{\left| {D(\omega )} \right|}^2}}} - \frac{{2{g^2}{{\left| {{\chi _{\rm{m}}}} \right|}^4}}}{{{{\left| {D(\omega )} \right|}^2}}}\left( {N{g^2} + \Re \left\{ {\chi _{\rm{m}}^{ - 1}\chi _{\rm{c}}^{ - 1}} \right\}} \right)} \right){\Gamma _{\rm{m}}}n_{\rm{m}}^{{\rm{th}}}\\&+ \frac{{(N - 1){g^4}{{\left| {{\chi _{\rm{m}}}} \right|}^4}}}{{{{\left| {D(\omega )} \right|}^2}}}{\Gamma _{\rm{m}}}n_{\rm{m}}^{{\rm{th}}}\\
			&= \frac{{{g^2}{{\left| {{\chi _{\rm{m}}}} \right|}^2}}}{{|D(\omega ){|^2}}}{\kappa _0}n_{\rm{c}}^{{\rm{th}}} + \left( {{{\left| {{\chi _{\rm{m}}}} \right|}^2} - \frac{{N{g^4}{{\left| {{\chi _{\rm{m}}}} \right|}^4}}}{{{{\left| {D(\omega )} \right|}^2}}} - \frac{{2{g^2}{{\left| {{\chi _{\rm{m}}}} \right|}^4}}}{{{{\left| {D(\omega )} \right|}^2}}}\left( {{\Gamma _{\rm{m}}}\kappa /4 - {{(\omega  - {\Omega _{\rm{m}}})}^2}} \right)} \right){\Gamma _{\rm{m}}}n_{\rm{m}}^{{\rm{th}}},
		\end{aligned}
		\label{eq:spec_b_cross_deg}
	\end{equation}
	where now
	\begin{equation}
		D(\omega ) = N{\chi _{\rm{m}}}(\omega ){g^2} + \chi _{\rm{c}}^{ - 1}(\omega ).
	\end{equation}
	The optomechanical coupling rate of the collective mechanical mode is enhanced by the number of mechanical oscillators coupled to the cavity, given by $\Gamma_{\rm opt}=N\frac{4g^2}{\kappa}$. Now, assuming a weak coupling regime $\Gamma_{\rm opt}\ll\kappa$, the microwave cavity susceptibility can be approximated as $\chi _{\rm{c}}^{ - 1}(\omega )\simeq \kappa/2$. This simplification allows us to express $D(\omega )$ as:
	\begin{equation}
		D(\omega ) \simeq \frac{{{\chi _{\rm{m}}}(\omega )\kappa }}{2}\left( { - i(\omega  - {\Omega _{\rm{m}}}) + \frac{{{\Gamma _{{\rm{opt}}}}}}{2} + \frac{{{\Gamma _{\rm{m}}}}}{2}} \right).
	\end{equation}
	Using this approximation, we derive the analytical form of individual mechanical occupations by integrating Eq.~\eqref{eq:spec_b_cross_deg}:
	\begin{equation}
		\begin{aligned}
			\left\langle {\hat b_i^\dag {{\hat b}_i}} \right\rangle  &= \frac{{\int_{ - \infty }^{ + \infty } {{S_{\hat b_i^\dag {{\hat b}_i}}}(\omega )d\omega } }}{{2\pi }}\\
			&= \frac{{{\Gamma _{{\rm{opt}}}}{\kappa _0}n_{\rm{c}}^{{\rm{th}}}}}{{N\kappa ({\Gamma _{{\rm{opt}}}} + {\Gamma _{\rm{m}}})}} \\
			&+ \left( {1 - N{{\left( {\frac{{{\Gamma _{{\rm{opt}}}}}}{{N({\Gamma _{{\rm{opt}}}} + {\Gamma _{\rm{m}}})}}} \right)}^2} - \frac{2}{N}\frac{{{\Gamma _{\rm{m}}}}}{{{\Gamma _{{\rm{opt}}}} + {\Gamma _{\rm{m}}}}}\frac{{{\Gamma _{{\rm{opt}}}}}}{{{\Gamma _{{\rm{opt}}}} + 2{\Gamma _{\rm{m}}}}}\frac{{\kappa  - {\Gamma _{{\rm{opt}}}} - {\Gamma _{\rm{m}}}}}{\kappa }} \right)n_{\rm{m}}^{{\rm{th}}}.
		\end{aligned}
	\end{equation}
	Considering that microwave cavity heating is given by $n_{\rm c }=\frac{\kappa_0n_{\rm c}^{\rm th}}{\kappa}$, and under the condition $\Gamma_{\rm m}\ll \Gamma_{\rm opt}\ll\kappa$, the expression simplifies to:
	\begin{equation}
		\left\langle {\hat b_i^\dag {{\hat b}_i}} \right\rangle  \simeq \frac{{{n_{\rm{c}}}}}{N} + \left( {1 - \frac{1}{N}} \right)n_{\rm{m}}^{{\rm{th}}}.
	\end{equation}
	The resulting equation illustrates that individual oscillator cooling becomes unattainable in the limit $N\gg1$. This observation aligns with the existence of a single bright collective mode and $(N-1)$ dark collective modes. In other words, the bright collective mode can only extract $1/N$ of thermal occupation $n_{\rm m}^{\rm th}$ from each physical oscillator. Notably, the influence of the microwave bath $n_{\rm c}$ on each oscillator diminishes by a factor of $N$ compared to a single optomechanical system.

	\subsubsection{Occupation of collective mechanical modes based on the covariance matrix}
	\label{sec:cov}
	The occupation of the bright and dark collective modes can be defined based on the covariance matrix of the system in the basis of the mechanical oscillators. As the initial state is Gaussian (thermal state) and the Hamiltonian contains only quadratic terms, the state remains Gaussian as time evolves, so the covariance matrix is sufficient to describe the system. The covariance matrix in the basis of the mechanical oscillators is given by:
	
	\begin{equation}
		{{\rm{COV}} = \left( {\begin{array}{*{20}{c}}
					{\left\langle {\hat b_1^\dag {{\hat b}_1}} \right\rangle }&{\left\langle {\hat b_1^\dag {{\hat b}_2}} \right\rangle }& \cdots &{\left\langle {\hat b_1^\dag {{\hat b}_N}} \right\rangle }\\
					{\left\langle {\hat b_2^\dag {{\hat b}_1}} \right\rangle }&{\left\langle {\hat b_2^\dag {{\hat b}_2}} \right\rangle }& \cdots &{\left\langle {\hat b_2^\dag {{\hat b}_N}} \right\rangle }\\
					\vdots & \vdots & \ddots & \vdots \\
					{\left\langle {\hat b_N^\dag {{\hat b}_1}} \right\rangle }&{\left\langle {\hat b_N^\dag {{\hat b}_2}} \right\rangle }& \cdots &{\left\langle {\hat b_N^\dag {{\hat b}_N}} \right\rangle }
			\end{array}} \right)}.
		\label{eq:cov}
	\end{equation}
	In the above equation,
	\begin{equation}
		\left\langle {\hat b_i^\dag {{\hat b}_j}} \right\rangle  = \frac{{\int_{ - \infty }^{ + \infty } {{S_{\hat b_i^\dag {{\hat b}_j}}}(\omega )d\omega } }}{{2\pi }},
	\end{equation}
	where ${S_{\hat b_i^\dag {{\hat b}_j}}}(\omega )$ are introduced in Eqs.~(\ref{eq:spec_b_cross}). To find the occupations of the collective mechanical modes, we can diagonalize the covariance matrix. The eigenvalues, in this case, provide us with the occupations of the collective modes. We attribute the lowest eigenvalue to the occupation of the bright collective mode, as it has the maximum coupling to the cavity. The rest of the eigenvalues correspond to the occupations of the dark modes.

	\subsection{Extraction of collective mode linewidths and frequencies}
	\label{sec:eigens}
	To find the collective linewidths and frequencies reported in Fig.~2 of the main text, we fit Eq.~(\ref{eq:ref_omit}) to OMIT response data by taking $g_i$, $\kappa$, and $\Delta$ as free parameters. The mechanical frequencies and linewidths are found independently and are given as fixed parameters; however, microwave frequency and linewidth are taken as free parameters, as we observe they slightly shift by changing the pump power. Upon finding these values, we construct the matrix $\bf M$ in Eq.~\eqref{eq:lang_omit}. Imaginary parts and real parts of the eigenvalues of this matrix provide us with the collective mode frequencies and the collective mode linewidths, respectively.

	\subsection{Direct extraction of collective mode linewidths and frequencies}	
	Here, we provide another method to extract collective linewidths and frequencies directly using the OMIT experiment. We start by diagonalizing the matrix $\bf M$ in Eq.~(\ref{eq:lang_omit}). This matrix is not unitary, but it is still diagonalizable, i.e. ${\bf{M}} = {\bf{S}}\Lambda {{\bf{S}}^{ - 1}}$, where $\bf{\Lambda}$ is a diagonal matrix of eigenvalues of the matrix $\bf{M}$ (with $\lambda_i$ as eigenvalues). Equation~(\ref{eq:lang_omit}) is then written as follows:
	\begin{equation}
		\begin{aligned}
			\frac{d}{{dt}} {\boldsymbol{\hat \alpha }}  &= \Lambda  {\boldsymbol{\hat \alpha }} +  {{{\boldsymbol{\hat \alpha }}_{{\rm{in}}}}} \\
			{{{\boldsymbol{\hat \alpha }}_{{\rm{in}}}}}  &= {{\bf{S}}^{ - 1}} {{{\bf{\hat b}}_{{\rm{in}}}}} \\
			{\boldsymbol{\hat \alpha }}  &= {{\bf{S}}^{ - 1}} {\bf{\hat b}} 
		\end{aligned}
		\label{eq:eigen_lang}
	\end{equation}
	In the above equation, ${\boldsymbol{\hat \alpha }}$ are the eigenvectors of the system. We take the Fourier transform of the above equation to find its steady-state solution:
	\begin{equation}
		{\boldsymbol{\hat \alpha }} (\omega ) = {\left( { - i\omega {\bf{I}} - {\boldsymbol{\Lambda }}} \right)^{ - 1}} {{{\boldsymbol{\hat \alpha }}_{{\rm{in}}}}}(\omega ).
	\end{equation}
	The advantage of working in the collective basis is that the matrix ${ - i\omega {\bf{I}} - {\bf{\Lambda }}}$ is diagonal, so its inverse is diagonal as well. To find $\hat s_{\rm{out}}(\omega)$, we need to first find $\hat a(\omega)$ in terms of eigenvectors. Using Eq.~(\ref{eq:eigen_lang}), we can write:
	\begin{equation}
		\begin{aligned}
			\hat a(\omega ) &= \sum\limits_{j = 1}^{N + 1} {{s_{(1,j)}}{\hat \alpha _j}(\omega )}, \\
			{\hat \alpha _j}(\omega ) &= \frac{1}{{ - i\omega  - {\lambda _j}}}{\hat \alpha _{{\rm{in}},j}}(\omega )\\
			&= \frac{{{s_{r,(j,1)}}}}{{ - i\omega  - {\lambda _j}}}\sqrt {{\kappa _{{\rm{ex}}}}} {s_{{\rm{in}}}}(\omega ),
		\end{aligned}
		\label{eq:col}
	\end{equation}
	where $s_{(i,j)}$ is the $(i,j)$th element of the matrix $\bf{S}$, $s_{r,(i,j)}$ is the $(i,j)$th element of the matrix $\bf{S}^{-1}$, $\hat \alpha_{j}(\omega)$ is the $j$th eigenvector, and $\hat \alpha_{\text{in},j}(\omega)$ is the $j$th element of $ {{{\boldsymbol{\hat \alpha }}_{{\rm{in}}}}} (\omega )$.
	
	Using input-output relation Eq.~(\ref{eq:in_out}) and Eq.~(\ref{eq:col}), we find the reflection frequency response $S_{11}(\omega)$ in the collective basis:
	\begin{equation}
		S_{11}(\omega ) = 1 - {\kappa _{{\rm{ex}}}}\sum\limits_{j = 1}^{N + 1} {\frac{{{s_{(1,j)}}{s_{r,(j,1)}}}}{{ - i\omega  - {\lambda _j}}}}.
		\label{eq:col_ref}
	\end{equation}
	For fitting purpose, we can consider ${{s_{(1,j)}}{s_{r,(j,1)}}}$ as a single complex-valued parameter. The advantage of Eq.~(\ref{eq:col_ref}) is that we do not have multiplication of fitting parameters, which is an advantage, especially for large number of free parameters. In addition, by finding $\lambda_{j}$, which is a complex-valued parameter, we can directly find the eigenvalues of the system, where $\Re \{ {\lambda _j}\}$ gives half of the linewidth of the eigenmode and $\Im \{ {\lambda _j}\} $ gives its eigenfrequency. We confirm that this method reproduces the same results we obtained using the method introduced in \secref{sec:eigens}.

	\section{Numerical Analysis}
	\label{sec:Numerical}
	In this section, we provide the numerical analysis of the system in the presence/absence of disorder.
	
	\subsection{Effects of frequency disorder of microwave resonators}
	\label{sec:MWimperfections}
	The frequency of the microwave resonators is designed to be $\omega_0/2\pi=5.45$\,GHz  with large inter-sites coupling (see design simulation section \secref{sec:simulation}). Due to fabrication imperfection, e.g. in the gap-size of the vacuum gaps, such target frequency may fluctuate by a small amount, breaking the symmetry of the Hamiltonian~\eqref{eq:fullMW_Hamiltonian} and perturbing the ideal microwave mode shape. We study this effect by numerically simulating the distribution of the microwave eigenfrequencies and eigenmodes for fluctuations in  $\omega_0$ ranging from 0 to 0.25\%.

	\begin{figure*}[hbt!]%
		\includegraphics[width=\linewidth]{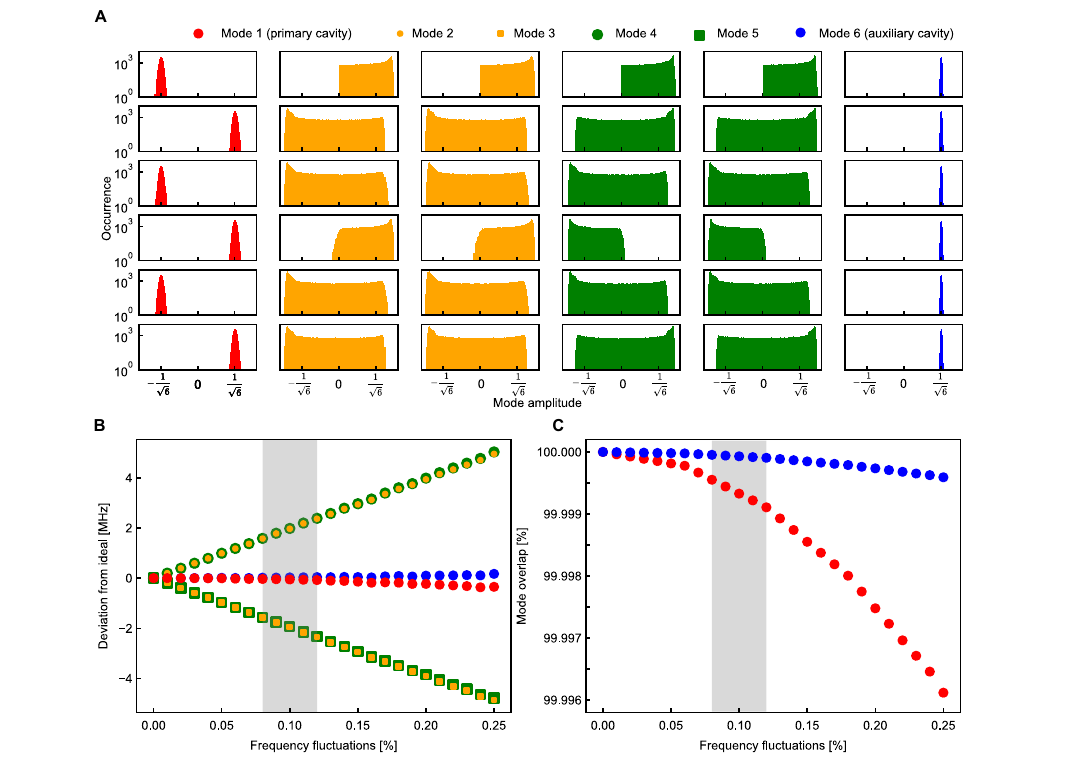}
		\caption{\textbf{Statistical analysis of microwave frequency disorder.} \textbf{A}, Eigenmode distribution for $10^5$ realization of Eq.~\eqref{eq:ham_unper} with frequency fluctuations sampled with a zero-mean Gaussian distribution and a variance of 0.26\%. Each panel corresponds to the distribution of the normalized amplitude in the particular microwave mode (indicated by the color) on the individual microwave resonator (each row). For example, for the primary cavity (the anti-symmetric mode, red color), the amplitudes alternate sign and cluster around $\pm 1/\sqrt{6}$. For the auxiliary cavity (the symmetric mode, blue color), the amplitudes keep the sign and cluster around $+1/\sqrt{6}$. \textbf{B}, Average eigenfrequencies of $10^5$ realization of  Eq.~\eqref{eq:ham_unper}, where the microwave resonators frequency is randomly selected by a Gaussian distribution. It is plotted versus the Gaussian variance which represents the fabrication fluctuation. \textbf{C}, Average inner product (fidelity) of the mode shape of  $10^5$ realization of Eq.~\eqref{eq:ham_unper} and the ideal case for the primary and auxiliary cavities. 
		}
		\label{fig:MW_disorder}
	\end{figure*}
	We sample $10^5$ realization of Eq.~\eqref{eq:ham_unper} by drawing $\omega_0$ from a Gaussian distribution for each value of frequency disorder. The average of the microwave mode frequencies as a function of the individual resonator frequency fluctuation is shown in \figref{fig:MW_disorder}\textbf{B}. Small perturbations break the degeneracy of the modes, and the difference between the new eigenfrequencies increases linearly with the frequency disorder. However, the frequencies of the primary and auxiliary cavities shift 1 order of magnitude less compared with the aforementioned modes. By taking into account that the separation of modes -1 and 1 (-2 and 2) is 3 (4)\,MHz, from this study we can infer that the frequency fluctuations in the fabrication process are $\approx 0.1\%$. It is worth noticing that such low disorder implies a very precise control of the vacuum gap capacitors. In addition, we can use the average value of mode -1 and 1 (-2 and 2) to infer the frequency of such modes in absence of noise. 
	
	In \figref{fig:MW_disorder}\textbf{C}, we study the fidelity of the mode shape of the primary and auxiliary cavities compared to the ideal case, defined as $F = {\left| {\left\langle {{{\psi _{{\rm{ideal}}}}}}\mathrel{\left | {\vphantom {{{\psi _{{\rm{ideal}}}}} {{\psi _{{\rm{sim}}}}}}}\right. \kern-\nulldelimiterspace}{{{\psi _{{\rm{sim}}}}}} \right\rangle } \right|^2}$, where $\left| {{\psi _{{\rm{ideal}}}}} \right\rangle $ and $\left| {{\psi _{{\rm{sim}}}}} \right\rangle $ stand for the ideal and the simulated modeshapes, respectively. We verify that these two modes  are not affected significantly by frequency fluctuation. We sample $10^5$ realization of Eq.~\eqref{eq:ham_unper} by drawing $\omega_0$ from a Gaussian distribution for each value of frequency disorder and calculate the fidelity. For fluctuation of 0.1\%, we expect that such modes are $>99.9\%$ overlapping with the ideal case. 
	
	\subsection{Degenerate mechanical oscillators}
	
	In section~{\ref{sec:degen}}, we provided the theoretical grounds for appearance of the bright collective mode. Here, we show numerically the linewidth of the collective modes as a function of the average of cooperativities. The eigenvalues of the system and their corresponding eigenvectors can be found by diagonalizing the matrix $\bf{M}$ in Eq.~(\ref{eq:lang_omit}). Similar to our device, we consider $N=6$. The simulation parameters are similar to the actual device parameters given in Tables~\ref{tab:glossary} and \ref{tab:mechanics}. The result can be seen in Fig.~\ref{fig:disorder}\textbf{A}.

	In Fig.~{\ref{fig:disorder}}\textbf{A}, the collective modes' linewidths (twice of the real part of the eigenvalues of the matrix $\bf M$) are plotted versus the average of the cooperativities $\mathcal{\bar C}$. We see an increase in the effective linewidths only for one of the eigenmodes. The rest (5 modes) are dark, as their linewidths do not change by increasing the power.  In this figure, we also plot with a dashed line 6 times of the average of the optomechanical damping rates of individual mechanics. As it is elaborated in \secref{sec:degen}, it can predict the behavior of the bright collective mode. 
	
	When the linewidth of the bright collective mode becomes comparable with the linewidth of the cavity, the two enter the strong coupling regime \cite{teufel2011circuit}. In this regime, we have hybridization between the cavity and the bright collective mode. By further increasing the power, the linewidth of the both remain constant ($\kappa/2$) and their frequencies split (Fig.~\ref{fig:disorder}\textbf{A} inset). The modeshape of the eigenmodes are found using the eigenvectors of the matrix $\bf M$ (see \secref{sec:mode_col} for more details). The shaded region in this plot represents 90\% certainty range. 
	
	\subsection{Nearly-degenerate mechanical oscillators}
	Due to inevitable fabrication imperfections, we cannot make perfectly degenerate mechanical oscillators. In this section, we consider disorder in mechanical frequencies and linewidths, given by:
	\begin{equation}
		\begin{aligned}
			{\Omega _{{\rm m},i}} &= \Omega_{\rm m} \left( {1 + {\cal N}\left( {{x_i}|\mu  = 0,\sigma  = {\sigma _{\Omega_{\rm m}}/\Omega_{\rm m} }} \right)} \right),\\
		\end{aligned}
	\end{equation}
	where we consider a Gaussian distribution with zero mean and $\sigma_{\Omega_{\rm m}}$ standard deviation of the mechanical frequencies. The result of simulations for two values $\sigma_{\Omega_{\rm m}}/\Omega_{\rm m}=0.001\%$ and $1\%$ can be seen in \figsref{fig:disorder}\textbf{B} and \textbf{C}, respectively. Considering these figures, three main regions can be defined: 1- When mechanical oscillators behave as individual mechanics, where their linewidth increase by $\Gamma_{\rm{opt},i}$, given in Eq.~(\ref{eq:gam_col}). 2- When bright collective mode appears, evidenced by linewidth increase given by $\Gamma_{\rm{opt}}=6\bar \Gamma_{\rm{opt,i}}$ and appearance of dark modes. 3- When strong coupling between the bright collective mode and the primary cavity occurs. This happens when linewidths of the both become equal. 
	
	The transition between regions 1 and 2 happens when the disorder between the mechanical frequencies become comparable with the optomechanical damping rates of individual mechanics. When the disorder is much larger than the optomechanical damping rates, they behave as independent mechanical oscillators. By increasing the pump power, the optomechanical damping rates increase. At a certain power, the linewidths of the mechanical oscillator become comparable to their frequency difference. This is where they interact and form the collective modes. Such condition can be written as:
	\begin{equation}
		\bar \Gamma_{\rm{opt,i}}=\frac{\overline {4g_i^2} }{\kappa}\simeq\frac{\sigma_{\Omega_{\rm m}}}{N}.
	\end{equation}
	Considering $g_i=g_{0,i}\sqrt{n_{\rm p}}$, where $n_{\rm p}$ is the number of photons inside the cavity, we can write:
	\begin{equation}
		n_{\rm p}^{\rm{(1)}} \simeq \frac{\kappa}{4\overline{g_{0,i}^2} N}\sigma_{\Omega_{\rm m}},
		\label{eq:nd1}
	\end{equation}
	where $n_{\rm p}^{(1)}$ stands for the required number of photons inside the cavity to observe the bright collective mode. 
	
	A transition from region 2 to region 3 occurs when the linewidth of the collective mode becomes comparable with the cavity linewidth, meaning:
	\begin{equation}
		N\bar \Gamma_{\rm{opt,i}}\simeq\kappa,
	\end{equation}
	so
	\begin{equation}
		n_{\rm p}^{\rm{(2)}} \simeq \frac{\kappa^2 }{4\overline{g_{0,i}^2} N},
		\label{eq:nd2}
	\end{equation}
	where $n_{\rm p}^{(2)}$ stands for the required number of photons inside the cavity to observe strong coupling between the bright collective mode and the microwave cavity. 
	
	We aim to observe the bright collective mode before the onset of strong coupling, so it requires $n_{\rm p}^{(1)}<n_{\rm p}^{(2)}$, leading to:
	\begin{equation}
		\sigma_{\Omega_{\rm m}}<\kappa.
	\end{equation}
	This means the frequency fluctuations should be less than the linewidth of the cavity. 
	
	\begin{figure*}[hbt!]%
		\includegraphics[width=\linewidth]{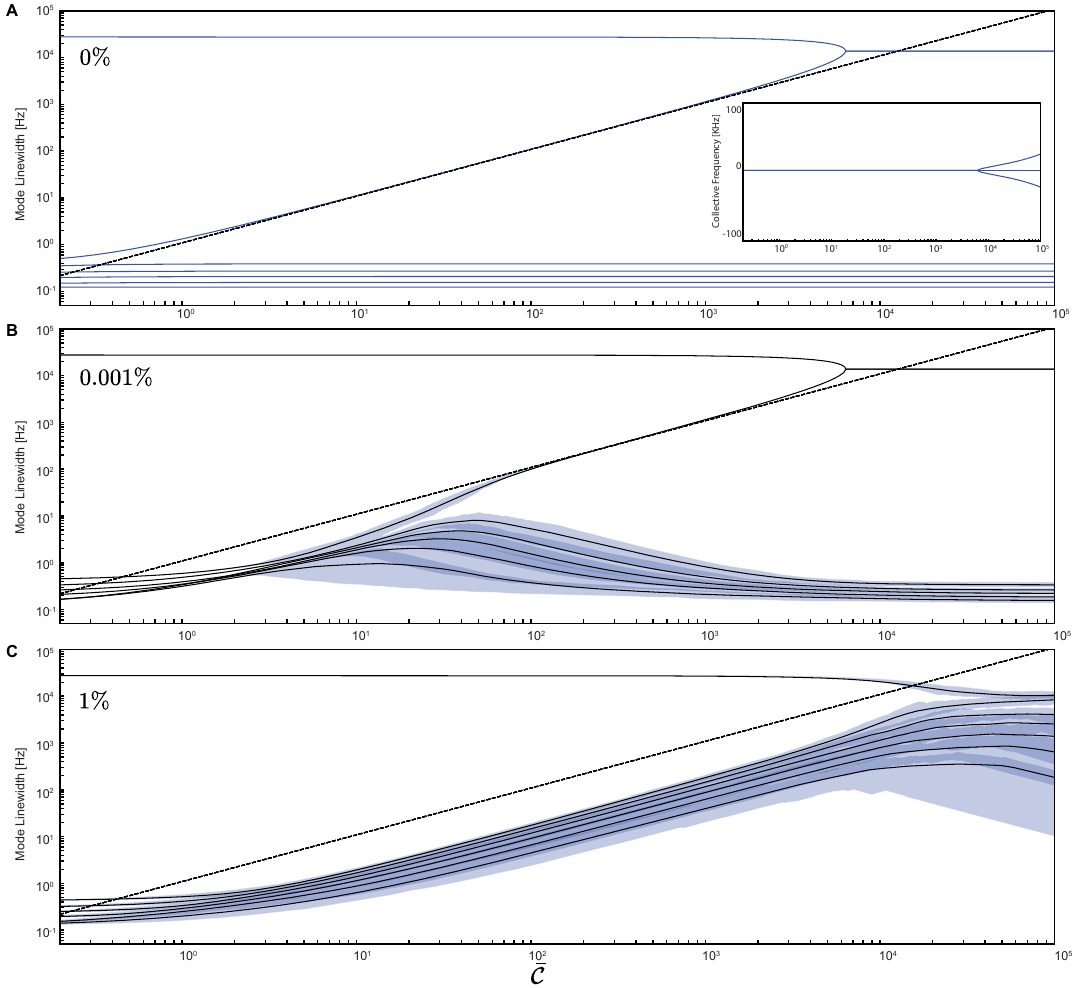}
		\caption{\textbf{Effect of mechanical disorder}. \textbf{A}-\textbf{C} Collective mode linewidths by finding real parts of eigenvalues of matrix $\mathbf{M}$ for mechanical disorders $\sigma_{\Omega_{\rm{m}}}/{\bar \Omega_{\rm m}}=0\%,\;0.001\%\text{, and }1\%$, respectively, versus average of cooperativities. The dashed line shows 6 times of the average of optomechanical damping rates of individual oscillators. The shaded region represents 90\% certainty region. The inset shows the collective mode frequencies(subtracted from $\Omega_{\rm m}$).
		}
		\label{fig:disorder}
	\end{figure*}

	\section{Fabrication}
	\subsection{Using DUV stepper instead of direct laser writer}
	In contrast to our previous works~\cite{youssefi2023squeezed,youssefiTopological2022}, here we use DUV stepper (ASML PAS 5500/350C) instead of direct laser writer (Heidelberg MLA 150) for all the lithography steps. The main motivation is the better uniformity of the defined patterns as well as the higher resolution of the tool. The stepper offers better uniformity $<15$~nm ($<120$~nm for MLA), better resolution 200~nm (1~$\mu \rm{m}$ for MLA). Importantly, MLA suffers from 500~nm Field Stitching Accuracy (FSA) and 1~$\mu\rm{m}$ Blanket Stage Accuracy (BSA), which may generate some jagged boundaries when defining patterns. As we are aiming for making identical trenches (identical frequencies) for mechanical oscillators, we decide to use DUV for lithography steps. As a result, we achieve 0.1\% disorder among the mechanical frequencies, an order of magnitude less than our previous work~\cite{youssefiTopological2022}.

	\subsection{Details of the nano-fabrication process}
	
	The full fabrication process flow is reported in \figref{fig:fabProcess} and can be divided in four main lithography steps: patterning of the trenches, definition of bottom electrodes, planarization and vias,  and finally patterning of top electrode.
	\begin{figure*}[hbt!]%
		\includegraphics[width=\linewidth]{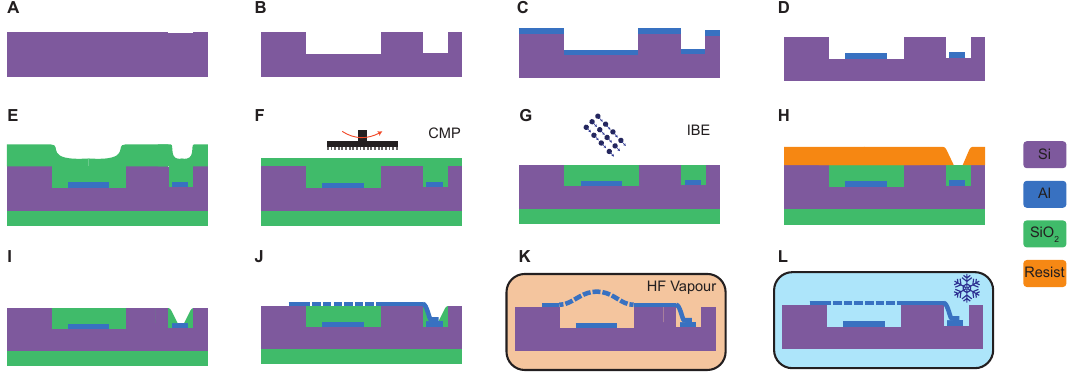}
		\caption{\textbf{Main steps in the fabrication process.} \textbf{A}, The fabrication starts with a high-resistivity silicon wafer. We then etch a trench \textbf{B} and deposit aluminum \textbf{C}. The aluminum is patterned with a wet etching process \textbf{D} followed by deposition of the $\rm{SiO_2}$ sacrificial layer \textbf{E}. The topography is removed by chemical mechanical polishing (CMP, \textbf{F}). We reach to the surface of the silicon using ion beam etching \textbf{G}. The top and bottom layer are galvanically connected by re-flowing the resist \textbf{H} and etching the sacrificial layer with a 1:1 DRIE etching method \textbf{I}. The top aluminum electrode is deposited and then patterned with a wet etching method \textbf{J}. The sacrificial layer is removed using HF vapor \textbf{K}. At cryogenic temperatures, the top layer becomes flat \textbf{L} due to the tensile stress of the material. }
		\label{fig:fabProcess}
	\end{figure*}
	
	\textit{Patterning of trenches}. The fabrication starts by measuring the bow (Toho Technology FLX 2320-S) of a 4" intrinsic Silicon wafer (Topsil GlobalWafers). After logging the average bow value  $<\pm 1\,\mu$m the wafer goes through standard rinse dry (SRD) with deionized (DI) water. The wafer is loaded in an automatic coater/developer (Tokyo Electron Cleantrack ACT 8) followed by a vapour deposition of HMDS and a spin-coating of 1\,$\mu$m of M35G. After baking the resist at 130 degrees Celsius for 90s, the trenches and alignment markers are exposed with intensity of 21~mJ/cm$^2$ and 0 focus offset in a DUV stepper (ASML PAS 5500/350C) through a commercially produced reticle (Compugraphics). Due to the limited space on the mask, only 4 different chip layouts can be realized. The resist goes though a post exposure backing at $130^\circ$~C for 90~s and is automatically developed in a TMAH-based solution (JSR TMA238WA) for 60~s followed by SRD. To ensure complete removal of the resist from the exposed zones, the wafer is exposed to 200W-generated oxygen plasma (Tepla GiGAbatch) for 20~s, removing $\approx 80$~nm of resist. 
	The exposed pattern is transferred to 350~nm trenches in the Silicon by Deep Reactive Ion Etching (Alcatel AMS 200 SE) using $\rm C_4F_8$ gas for 26~s. After such process the top layer of the resist is removed with 2 minutes of 200W-generated oxygen plasma. Any possible additional organic residue is removed by two baths of Remover 1165 at 70 degree Celsius for 5 minutes each, followed by two 5 mins dip in $100^\circ$~C Piranha solution, rinsing in DI water, and 10~s Oxygen plasma at 200~W to ensure all the resist is removed. A sketch of the wafer cross section at this stage is shown in \figref{fig:fabProcess}\textbf{B}.
	
	\textit{Definition of bottom electrodes}. The wafer is loaded in an electron beam evaporator (Alliance-Concept EVA 760) where 100~nm of high purity aluminum is deposited on the surface of the wafer (see \figref{fig:fabProcess}\textbf{C}). After a SRD, the wafer is loaded in the automatic coater/developer where it is dehydrated in air on a hot plate at $160^\circ$~C and a $1\,\mu$m  layer of DUV resist (M35G) is spin-coated and baked at 140 degrees Celsius for 90s. The pattern for the bottom layer is exposed in the DUV stepper with a dose of 25~mJ/cm$^2$ and 0 focus offset. After a post-exposure baking step at  $140^\circ$~C for 90~s, it is automatically developed in a TMAH-based solution (JSR TMA238WA) for 60~s followed by SRD. To ensure complete removal of the resist from the exposed zones and to make the surface hydrophilic, the wafer is exposed to 200W-generated oxygen plasma for 20~s, removing $\approx 80$~nm of resist. The aluminum is then etched with a commercial solution of Phosphoric, Nitric, and Acetic acid (TechniEtch Al80) at $35^\circ$~C for 35~s, and the resist is chemically removed by dipping the wafer for 5 mins in two baths of Remover 1165 at 70 degree Celsius, followed by dipping it in DI water, and 10~s Oxygen plasma at 200~W to ensure all the resist is removed from the metal surface, as shown in \figref{fig:fabProcess}\textbf{D}. 
	
	Next, we need to define the labels for each chip in the wafer. This cannot be done with DUV. Instead, we use MLA for defining the labels. The wafer is loaded in a automatic coater/developer (Süss ACS200 GEN3), vapor coated with HMDS, spin coated with $1.5\,\mu$m of positive photoresist (AZ ECI 3007), and finally soft-baked at 100 degrees Celsius for 60~s. The wafer is then mounted on the vacuum stage of a Mask-less Aligner (Heidelberg MLA 150) and chip label patterns are exposed to the resist with an intensity of 170~mJ/cm$^2$ and zero focus offset. After a post exposure backing at $110^\circ$~C, the resist is developed with TMAH-based solution (AZ 726 MIF) for 47~s. After an oxygen plasma step of 20~s at 200~W, the chip labels are etched into the aluminum with wet etching. 
	
	\textit{Planarization and vias}. At this stage $2\,\mu$m layer of $\rm SiO_2$ is deposited using low thermal oxide (LTO) deposition at $425^\circ$\,C with a mixture of silane ($\rm SiH_4$) and oxygen in a furnace (Centrotherm). The oxide maps the same topography seeded by the wafer, resulting in trenches on its surface; notice that a similar layer of oxide is deposited on the backside of the wafer (see \figref{fig:fabProcess}\textbf{E}). In order to planarise the surface of the oxide, we use chemical mechanical polishing (CMP) for 8 mins (ALPSITEC MECAPOL E 460), with a low rate of PH-neutral slurry (ultra-sol 7A) removing $\approx1.2\,\mu$m of the oxide as highlighted in \figref{fig:fabProcess}\textbf{F}. To reach the silicon surface, we physically etch the remaining 800\,nm of silicon oxide by argon ion beam etching (IBE,Veeco Nexus IBE350) for 12 minutes (see \figref{fig:fabProcess}\textbf{G}). The following steps requires to galvanically connect top and bottom electrodes; after a SRD, the wafer is loaded in the DUV automatic coater. It is first coated with HMDS vapor followed by spin-coating of $1\,\mu$m layer of DUV resist (M35G) and baking at 140 degrees Celsius for 90s. The pattern for the bottom layer is exposed in the DUV stepper with a dose of 21~mJ/cm$^2$ and 0 focus offset. After a post-exposure baking step at  $140^\circ$~C for 90~s, it is automatically developed in a TMAH-based solution (JSR TMA238WA) for 60~s followed by SRD. In order to avoid resist residues in the developed areas, the sample is exposed to oxygen plasma at 200~W for 20~s. In order to obtain a smooth surface as depicted in \figref{fig:fabProcess}\textbf{H}, the resist is re-flowed for 90~s at $170^\circ$~C in a pre-heated oven (Heraeus T6060). We then use a 1:1 deep reactive ion etching (DRIE, SPTS APS) to transfer the pattern of the resist to the oxide (\figref{fig:fabProcess}\textbf{I}). After etching, the resist is removed similar to the previous steps.

	\textit{Patterning of top electrode}. The sample is now loaded in the load lock chamber of an ultrahigh-vacuum electron evaporator (Plassys MEB550SL3 UHV) reaching a base pressure of $5\cdot10^{-8}$\,mTorr. At this point, a Kaufman gun is turned on for 4 mins at 400~V to Argon-mill the native oxide of the aluminum. After transferring the wafer to the deposition chamber and reaching a base vacuum of $5\cdot10^{-9}$~Torr, 200~nm of aluminum is deposited on the surface of the wafer. The top layer is then patterned with DUV (similar to the bottom layer electrodes) and etched using the same wet etching method. After etching, the resist is removed with the standard recipe.
	
	\textit{Dicing}. 
	The wafer is now ready for dicing. We coat 1.5~$\mu m$ resist AZ 3007 for protecting the wafer during dicing. We then dice the wafer into $5\;\rm{mm}\times 5\;\rm{mm}$ chips (Disco DAD321). 
	
	\textit{HF releasing}
	After inspecting the samples, we remove the resist by dipping them in the wet remover followed by dipping in DI water and 2 mins of oxygen plasma. The sacrificial layer ($\rm{SiO_2}$) is then removed by Hydrofluoric acid (HF) vapor (SPTS uEtch), which is a dedicated tool for suspending MEMS structures.

	\textit{Bonding}
	The chips are glued to the sample holder with a butyral-phenol based glue and bonded with aluminum wires (F\&S Bondtec 56i). The sample is then connected to the mixing chamber of a dilution refrigerator. Due to the tensile stress of the material at cryogenic temperatures, the top layer will be flat (\figref{fig:fabProcess}\textbf{J}).

	\section{Characterization}

	\subsection{Measurement of the single photon optomechanical coupling rates ($g_{0,i}$)}
	\label{sec:g0}
	To measure the single photon optomechanical coupling rates ($g_{0,i}$) of all the mechanical oscillators coupled to the primary cavity, we use a resonant weak pump on the primary cavity (\figref{fig:g0}\textbf{A}) to measure the upper and lower motional sidebands using a spectrum analyzer \cite{bernier2019multimode,youssefi2023squeezed}. The weak pump ensures that its backaction on the mechanical oscillators is negligible, i.e. $n_{{\rm{ba}},i} \equiv \frac{{4g_{0,i}^2}}{{\kappa \Gamma _{{\rm{m}},i}}}\frac{{(4{\kappa _{{\rm{ext}}}}/{\kappa ^2}){P_{{\rm{MW}}}}}}{{1 + {{(2\Omega _{{\rm{m}},i}/\kappa )}^2}}} \ll n_{{\rm{m}},i}^{{\rm{th}}}$, where $n_{{\rm ba},i}$ is the equivalent backaction noise in the unit of quanta and $P_{\rm MW}$ is the microwave input power to the device at the cavity center frequency \cite{bowen2015quantum}. 
	
	\begin{figure*}[hbt!]%
		\includegraphics[width=\linewidth]{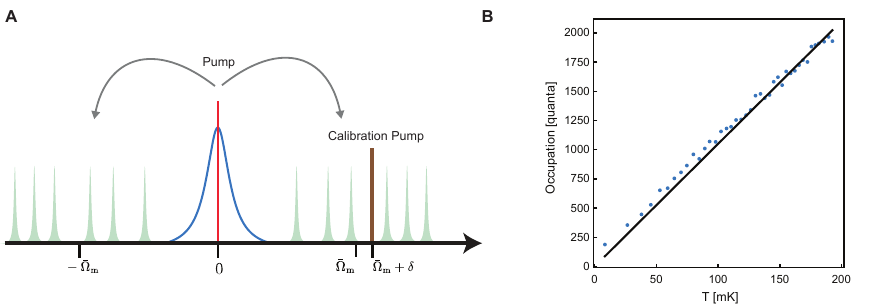}
		\caption{\textbf{$g_0$ Measurement}. \textbf{A}, Measurement scheme for finding $g_0$ of all the mechanical oscillators. \textbf{B}, Thermalization of the lowest frequency mechanical oscillator when we fit linearly to high temperatures $T>100$~mK.}
		\label{fig:g0}
	\end{figure*}
	
	The pump signal passes through attenuation lines in the fridge to reach the device, and then it is amplified using a HEMT at 4K stage followed by more amplifications using room temperature amplifiers (see \figref{fig:fullSetup} for the details of the experimental setup). The total attenuation and amplification in the measurement chain is unknown; therefore, we send an additional weak calibration pump located at the upper motional sideband, which passes through the same lines as the resonant pump. Using nominal values for the resonant/calibration pump powers at sources, $P_{\rm MW}^{\rm src}$ and $P_{\rm cal}^{\rm src}$ respectively, together with the measured calibration pump power at the spectrum analyzer and the power of the upper motional sideband of the $i$th oscillator, $P_{\rm cal}^{\rm meas}$ and $P_{{\rm SB},i}^{\rm meas}$ respectively, we can write:
	\begin{equation}
		\frac{P_{{\rm SB},i}^{\rm meas}}{P_{\rm MW}^{\rm src}}\frac{P_{\rm cal}^{\rm src}}{P_{\rm cal}^{\rm meas}} = 4g_{0,i}^2n_{{\rm m},i}\frac{(\kappa_{\rm ext}/\kappa)^2}{\Omega_{{\rm m},i}^2+((\kappa_{\rm ext} - \kappa_0)/2)^2} \frac{\omega_{\rm c}}{\omega_{\rm c}+ \Omega_{{\rm m},i}}.
		\label{eq:g0}
	\end{equation}
	
	By sweeping the fridge temperature, we can approximate $n_{{\rm m},i} \simeq \frac{k_{B}T}{\hbar \Omega_{{\rm m},i}}$, where $k_B$ is the Boltzmann constant. At each fridge temperature, we measure the sidebands of all the mechanics. Now, with a linear fit to the power of the sideband versus temperature and using Eq.~\eqref{eq:g0}, we find $g_0$ of all the mechanical oscillators. All the values for the mechanical oscillators are given in \tabref{tab:mechanics}. The thermalization of the lowest frequency mechanical oscillator is given in \figref{fig:g0}\textbf{B} when we fit linearly to $T>100$~mK, where the thermalization is more accurate. It is worth mentioning that the measured $g_0$ is less than that of a single optomechanical system by a factor of 6, as it is elaborated in \secref{sec:ham_lin}.  
	
	\subsection{Microwave cavity heating}
	\label{sec:heating}
	
	\begin{figure*}[hbt!]%
		\includegraphics[width=\linewidth]{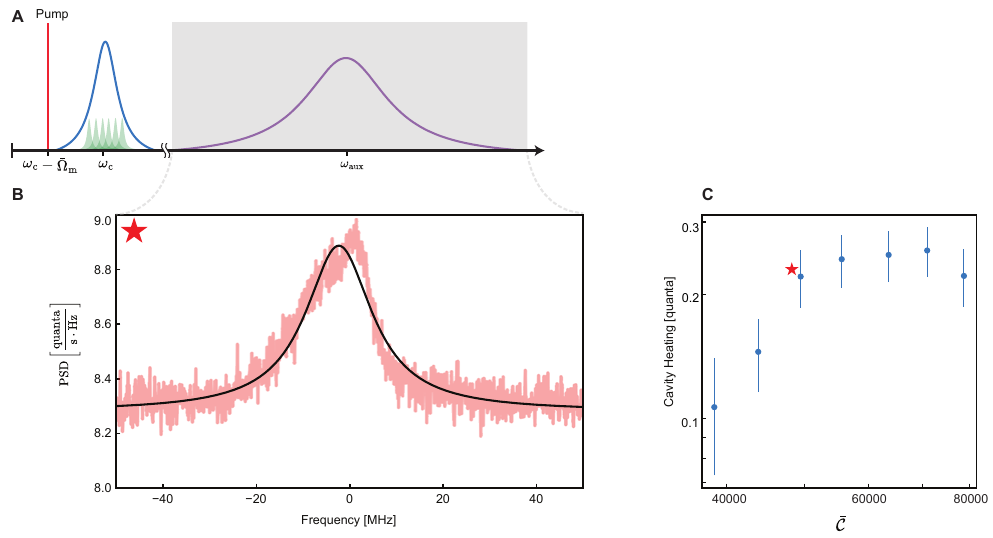}
		\caption{\textbf{Cavity heating of the auxiliary cavity}. \textbf{A}, Measurement scheme for finding cavity heating of the auxiliary cavity by pumping on the red sideband of the primary cavity. \textbf{B}, Detected spectrum by subtracting HEMT background and calibrating it. \textbf{C}, Inferred cavity heating by sweeping pump power. 
		}
		\label{fig:heating}
	\end{figure*}
	
	By applying a strong pump on the red sideband of the primary cavity, we observe heating of the auxiliary cavity. Characterizing such heating is necessary for the sideband asymmetry experiment \cite{youssefi2023squeezed} (\secref{sec:asym}). To measure it, we pump on the red sideband of the primary cavity while there is no pump at the auxiliary cavity (\figref{fig:heating}\textbf{A}). Using a spectrum analyzer around the auxiliary cavity, we can find its spectrum $\bar S_{\rm c}(\omega)$ and subtract it from the HEMT background (which is found by turning off all the pumps). An example is provided in \figref{fig:heating}\textbf{B}. The background of the spectrum is calibrated as elaborated in \secref{sec:hemt}. The heating of the auxiliary cavity, in the units of quanta, is then obtained \cite{youssefi2023squeezed}:
	\begin{equation}
		{n_{\rm{aux}}} = \frac{{\int_{ - \infty }^{ + \infty } {{{\bar S}_{\rm{c}}}(\omega )d\omega } }}{{2\pi {\kappa_{\rm{ex}}^{\rm aux}}}},
	\end{equation}
	where $\kappa_{\rm ex}^{\rm aux}$ is the external coupling rate of the auxiliary cavity. The result for a range of cooperativities is given in \figref{fig:heating}\textbf{C}.

	It is worth mentioning that in the sideband asymmetry experiment \secref{sec:asym}, another pump is present at the auxiliary cavity, which may generate additional heating. However, we do not observe a  background change around the sidebands. Hence, we used only data in \figref{fig:heating}\textbf{C} for analysing the data of sideband asymmetry experiment.

	\section{Experimental details}
	\subsection{Design considerations}
	The simplest design for performing this experiment would require only a single microwave mode equally coupled to six mechanical oscillators (Fig.~1\textbf{A}). However, our design involved six microwave resonators coupled in a rotation-symmetric way, each coupled to a mechanical oscillator, leading to six microwave modes, where the lowest and highest frequency ones satisfy our Hamiltonian in Eq.~\eqref{eq:ham}. One of our main goals was to conduct sideband asymmetry on the collective mode. This design allowed us to use two cavities simultaneously: one very narrow, suitable for ground state cooling of the collective mode, and an auxiliary cavity for demonstrating sideband asymmetry. This setup enabled us to perform the two experiments separately.
	
	An alternative design would involve coupling a single microwave cavity with a much larger linewidth ($\approx 300$~kHz) to six mechanical oscillators. The linewidth could be increased as long as we remain in the resolved sideband regime $\kappa/4\Omega_{\rm m} \ll 1$. In this regime, the optomechanical damping rate is independent of the cavity linewidth (with an over-coupled microwave cavity), so we would not sacrifice the cooling rate of the collective mode:
	\begin{equation}
		\begin{aligned}
			\Gamma_{\rm opt}&=\frac{4g_0^2}{\kappa}n_{\rm p}\\
			&=\frac{4g_0^2}{\kappa}\frac{\kappa_{\rm ex}}{\Omega_{\rm m}^2 + \kappa^2/4}\frac{P_{\rm in}}{\hbar (\omega_{\rm c}-\Omega_{\rm m})}\\
			&\approx \frac{4g_0^2}{\kappa} \frac{\kappa_{\rm ex}}{\Omega_{\rm m}^2}\frac{P_{\rm in}}{\hbar \omega_{\rm c}}\\
			&=\left( \frac{2g_0}{\Omega_{\rm m}}\right)^2{\color{red}\frac{\kappa_{\rm ex}}{\kappa}}\frac{P_{\rm in}}{\hbar \omega_{\rm c}}.
		\end{aligned}
	\end{equation}
	Here, $P_{\rm in}$ is the power of the red-detuned pump right before the device. We could then benefit from the large linewidth of the cavity and perform sideband asymmetry using the same cavity with some detuned pumps.
	
	However, increasing the linewidth of the main cavity would have some drawbacks. First, for a fixed pump power, a larger cavity linewidth would increase the number of photons inside the cavity, while the optomechanical damping rate would remain nearly constant (in the resolved sideband regime). This results in a technical issue of higher cavity heating, which is proportional to the number of photons inside the cavity. Looking at our current data (Fig.~4), the minimum occupation of the collective mode is limited by the cavity heating ($\approx 0.4$ quanta). If we were to increase the cavity linewidth by a factor of ten ($\approx 200$ kHz), for the same power, the number of photons inside the cavity would increase by a factor of ten, resulting in $\approx 1.1$ quanta of cavity heating, which means we could not achieve the ground state. This is also consistent with our previous data in \cite{youssefi2023squeezed}.
	
	Additionally, we were interested in entering the strong coupling regime with the microwave cavity, which occurs when $\Gamma_{\rm opt} \approx \kappa$. Given that the optomechanical damping rate is independent of $\kappa$, increasing the cavity linewidth tenfold would require increasing the input power by 10 dB, which is beyond the capability of our instrument.
	
	Overall, we decided to use a narrow linewidth cavity for the ground state cooling experiment and an auxiliary (wide) cavity for the sideband asymmetry. Although this increased the complexity of the design,this solution meets all the requirements. 
	\subsection{Design and Simulations}
	\label{sec:simulation}
	The design of the microwave resonators and the feedline is carried out by simulating the electromagnetic response with a commercial method-of-moments solver (SONNET\textregistered). Setting the purely real dielectric constant to $\epsilon_r=11.9$ for the silicon substrate, the simulated phase of the scattering parameter $S_{11}(\omega)$ is fitted to extract the electromagnetic mode frequencies and external coupling rates (see \figref{fig:MW_simulations}). From the simulation of individual microwave resonators (as shown in \figref{fig:stray_capacitance}\textbf{A} for the resonator closest to the feedline), we expect an average resonance frequency of $\omega_0^{\rm sim}/2\pi=5.45\pm0.03$\,GHz.
	\begin{figure*}[hbt!]%
		\includegraphics[width=\linewidth]{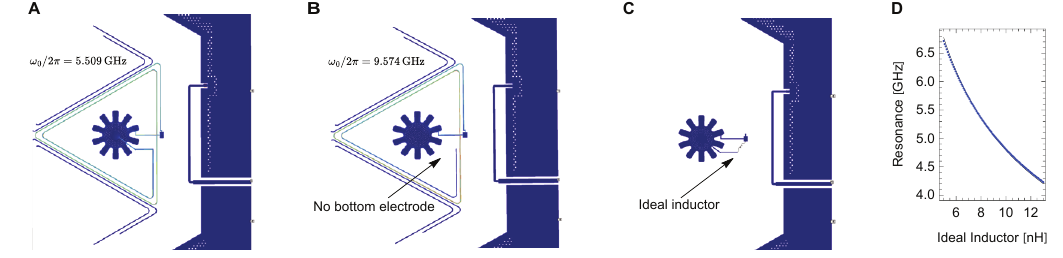}
		\caption{\textbf{Simulation of the microwave response of a single resonator}. \textbf{A}, Simulation of individual microwave resonator, where only the close sections of the inductors are left, giving a resonance frequency close to $\omega_0^{\rm sim}/2\pi=5.5$~GHz. \textbf{B}, In order to estimate the stray capacitance, the same simulation is repeated after removing the bottom electrode, yielding a larger resonance frequency around $\omega_{0,{\rm nodrum}}^{\rm sim}/2\pi=9.6$~GHz. \textbf{C} The estimation of the drum capacitor is performed by simulating the resonance frequency as a function of the inductance of an ideal inductor. The results are reported in \textbf{D}.
		}
		\label{fig:stray_capacitance}
	\end{figure*}
	To extract the value of stray capacitance we take the ratio between the resonator with and without bottom electrode (as shown in \figref{fig:stray_capacitance}\textbf{B}): 
	\begin{equation}
		\left(\frac{\omega_0^{\rm sim}}{\omega_{0,\rm no drum}^{\rm sim}}\right)^2=\frac{C_{\rm s}}{C_{\rm d }+C_{\rm s}}\approx 0.32, 
	\end{equation}
	so 68\% of the total capacitance is localized between the top drum and bottom electrode. In addition, by simulating the resonance frequency of an ideal inductor shunted by a vacuum gap capacitor, we can determine the value of both stray capacitance $C_{\rm s}=37.3$~fF  and vacuum gap capacitance $C_{\rm d}=75.2$~fF (see \figref{fig:stray_capacitance}\textbf{C} and \textbf{D}). 
	
	We also simulate the full system to ensure the equal distribution of the symmetric and anti-symmetric modes. The phase of the reflection coefficient  is reported in \figref{fig:MW_simulations}\textbf{A}.
	\begin{figure*}[hbt!]%
		\includegraphics[width=\linewidth]{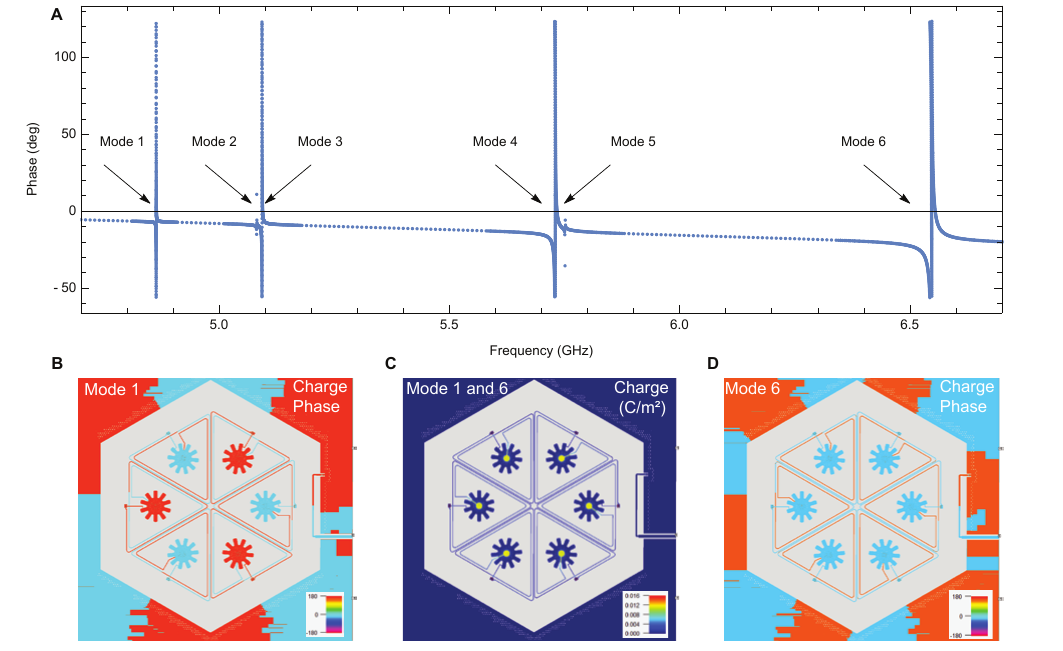}
		\caption{\textbf{Microwave simulation of the full circuit}. \textbf{A}, Simulated phase of the $S_{11}$ reflection coefficient as a function of driving frequency to the microwave feedline. The arrows highlight the frequencies of each modes. \textbf{B}-\textbf{D}, Choosing the driving frequency of modes 1 or 6, the simulation shows the preserved symmetry in charge distribution and its phase.}
		\label{fig:MW_simulations}
	\end{figure*}
	The reflection remarkably shows 6 individual modes. The splitting of the modes 2 and 3 (4 and 5), that would be degenerate in the ideal case, is 40\,MHz, which is one order of magnitude larger than the experimentally measured one. We attribute this deviation to discretization error while meshing the metallic layout, where $2\times2\,\mu\rm m^2$ elements are used. Despite this large numerical errors, we qualitatively asses the mode shape of the primary and auxiliary cavities, by plotting the charge phase in \figsref{fig:MW_simulations}\textbf{B} and \textbf{D} respectively. Finally, the charge distribution, that represents the microwave occupation of each individual resonators, is reported in  \figref{fig:MW_simulations}\textbf{C} and it identical for both modes.

	\subsection{Experimental setup}
	\label{sec:setup}
	The schematic of the full setup, room temperature (RT) and cryogenic, is illustrated in \figref{fig:fullSetup}. 
	\begin{figure*}[hbt!]%
		\includegraphics[width=\linewidth]{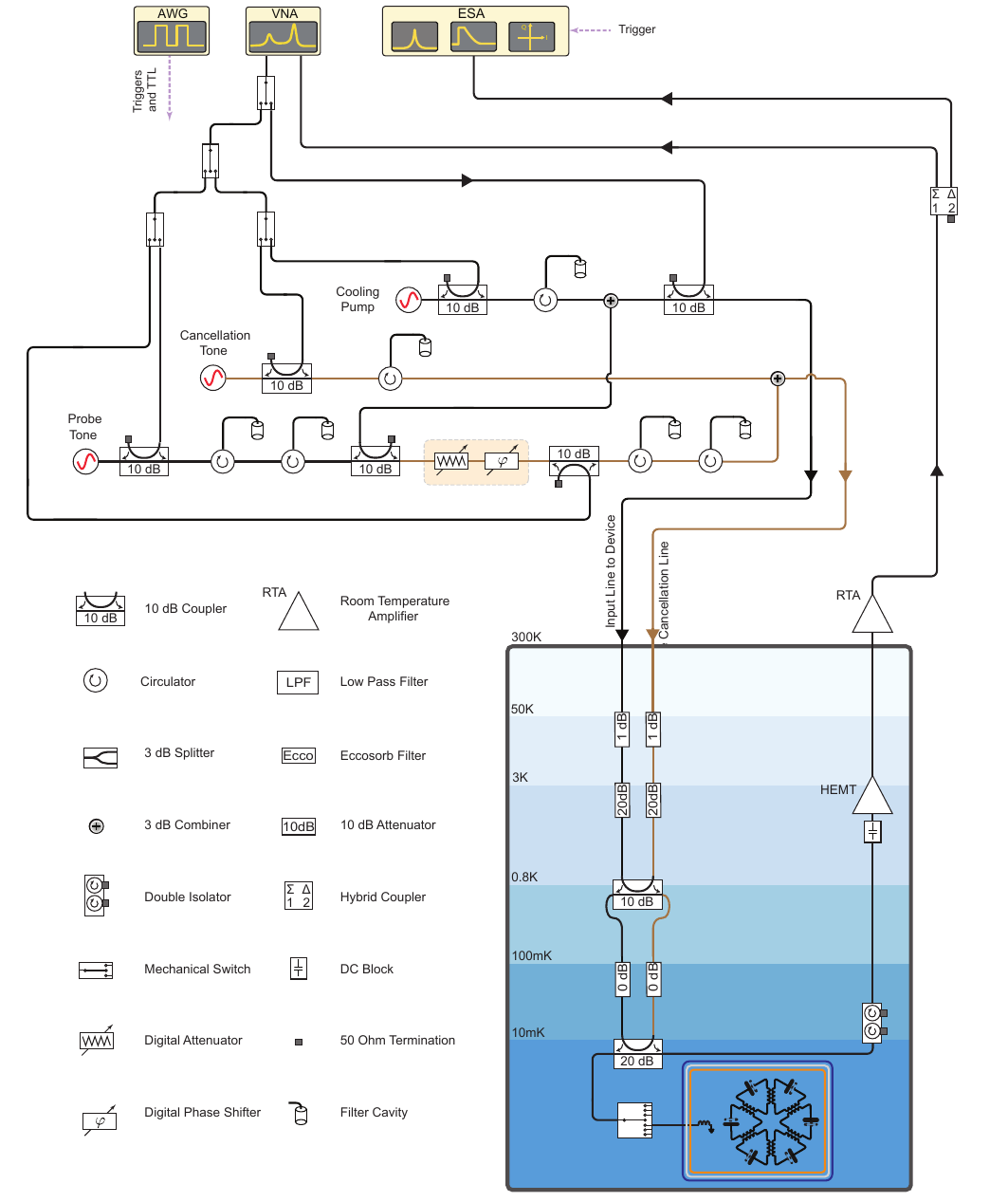}
		\caption{\textbf{Complete cryogenic and room temperature setup.} Black lines are the main signal path to the device and brown lines are the tone cancellation lines. Solid-state switches are not shown in this schematic.}
		\label{fig:fullSetup}
	\end{figure*}
	
	For OMIT and ground state cooling experiment, we use a VNA (Rohde \& Schwartz, ZNB~20) together with a strong pump (\textit{Cooling Pump} on the figure, Rohde \& Schwartz, SMA100B), which is red-detuned from the primary cavity. The pump is routed to the device through an attenuated line in the fridge and a cryo-switch (Radiall, R583-423-141) in the mixing chamber. The reflected signal (together with the sidebands) passes through a dual-junction isolator (Low Noise Factory, LNF-ISISC4 8A) and is amplified at 3K stage of the fridge using a high-electron-mobility transistor (HEMT: Low Noise Factory, LNF-LNC4-8C). Finally, it is amplified at RT (Low Noise Factory, LNF-LNR1-15A-SV) and goes to a VNA and an electrical spectrum analyzer (ESA: Rohde \& Schwartz, FSW) for detection. Frequency-tunable motorized filter cavities~\cite{joshi2021automated} are used to remove phase noise of the source at anti-Stokes sideband, as it may increase the mechanical occupation~\cite{rabl2009phase}. The power of the cooling pump can reach very high values, $\approx 27$~dBm at RT, which may saturate the HEMT. To avoid it, the cooling pump signal needs to be canceled before the HEMT. In our experiment, we use another source, (\textit{Cancellation Tone} on the figure, Rohde \& Schwartz, SMA100B), which interferes destructively with the cooling pump signal after the device. The frequency of this source is similar to the cooling pump, and its phase and power are swept to maximize the cancellation. We do not use a digital phase shifter and an attenuater~\cite{youssefi2023squeezed}, as these devices have $\approx 13$~dB loss, preventing us from reaching high cooperativities necessary for this experiment. 
	
	For performing the sideband asymmetry experiment, we use another probe tuned on the auxiliary cavity center frequency (\textit{Probe Tone} on the figure, Rohde \& Schwartz, SMF100A). The power of this tone is moderate, hence we use a digital phase shifter and attenuater to cancel it after the device. Because we measure both the Stokes and anti-Stokes sidebands, we use two filter cavities tuned on the upper and lower motional sidebands. We use another pair of filter cavities after the tone cancellation branch to remove any added noise due to this branch. To monitor the filter cavities and the device, we rout the VNA signal through all the lines using mechanical switches (Mini-Circuits, MSP2T-18XL+).
	
	For the modeshape measurement experiment, we change the experimental setup by removing digital attenuators/phase shifters and using the probe tone as the excitation signal (see Fig.~3 of the main text). For making pulses, we use solid-state switches (Mini-Circuits, ZFSWA2R-63DR+) after the cooling pump, cancellation tone, and probe tone which have $\approx 35$~ns rising/falling time and can be controlled with TTL generated by an arbitrary wave generator (AWG: Tektronix, AFG3252C). Considering the switching time of these switches, the uncertainties of the modeshape phase is $\Delta \varphi\simeq 2\pi\times5\;\rm{kHz}\times35\;{\rm ns} = 2\;\rm{mRad}$, which is acceptable in our experiment. 
	
	\begin{figure*}[hbt!]%
		\includegraphics[width=\linewidth]{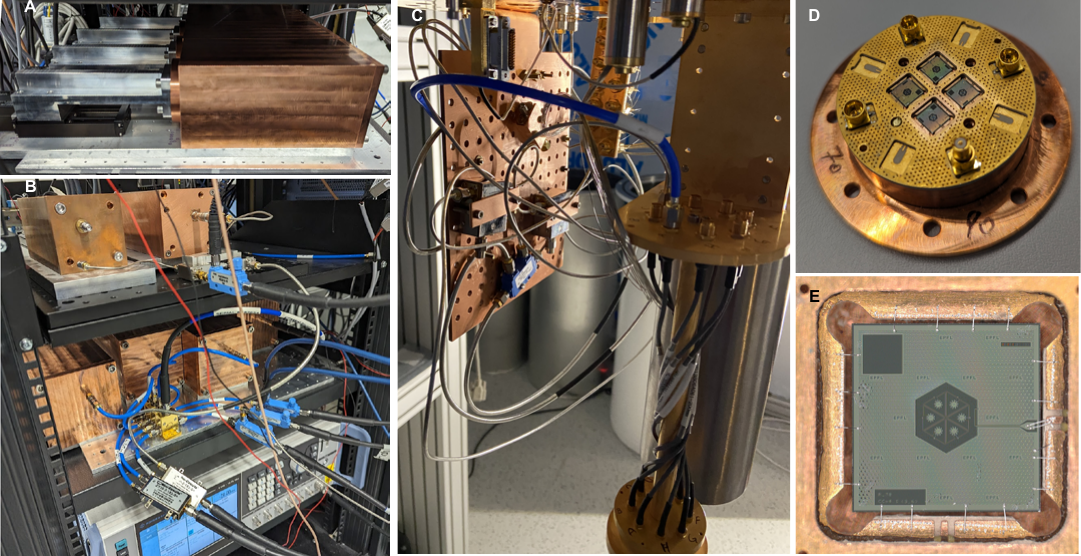}
		\caption{\textbf{Details of the experimental setup.} \textbf{A} and \textbf{B}, filter cavities for removing the phase noise of the sources. \textbf{C}, cold fingers for connecting the sample to the mixing chamber. \textbf{D}, sample holder with four bounded chips. \textbf{E}, bounded chip to the sample holder.
		}
		\label{fig:sample_setup}
	\end{figure*}
	
	As shown in \figref{fig:fullSetup} the experiments require careful phase noise cancellation that is achieved by employing 6 filter cavities whose frequencies can be tuned remotely using stepper motors (see \figref{fig:sample_setup}\textbf{A}).  A photograph of their inclusion in the experimental setup is shown in \figref{fig:sample_setup}\textbf{B}. 
	
	The sample is thermally and mechanically anchored at the mixing chamber stage of a dilution fridge (BLUEFORS, LD250) with an average base temperature of 10\,mK. The cold fingers which connect the device to the mixing chamber (see \figref{fig:sample_setup}\textbf{C}) are enclosed on a triple layer shield of $\mu-$metal, eccosorb, and copper, which protect the sample microwave environment~\cite{kono2023mechanically}. The cryogenic microwave components are placed on a copper plate. The sample is bonded to a trilayer goldplated PCB, which is mounted on a copper plate that hosts four chips as shown in~\figref{fig:sample_setup}\textbf{D}. We keep the number of bonds per side to four to minimize the possible effect of the ultrasonic vibration of bonding on the metallic drums (\figref{fig:sample_setup}\textbf{E}). 
	
	For the thermal photon calibration (see \secref{sec:hemt}), the radiation emitter is the 50\,$\Omega$ termination of the isolator, placed right after the sample. The photons travel to the sample and after their reflection continue along the amplification chain.
	
	\subsection{Extraction of loss between the source and the device}
	\begin{figure*}[hbt!]
		\centering
		\includegraphics{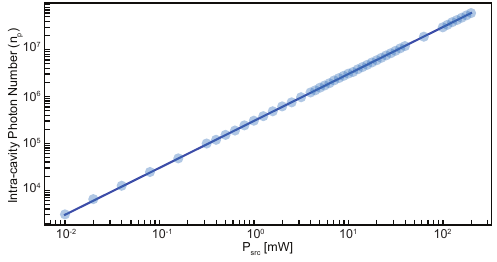}
		\caption{\textbf{Intra-cavity photon number versus source pump power}. Intra-cavity photon number vs the room-temperature source power, which is linear. The loss between the source and the device can be found from the slope. Experimental data is shown with circles and linear fit is shown with solid blue line. We used data for the lowest frequency mechanical oscillator (see \tabref{tab:mechanics}).}
		\label{fig:loss_g}
	\end{figure*}
	After finding single photon optomechanical coupling rates ($g_{0,i}$) as described in \secref{sec:g0}, we can find the intra-cavity photons for each pump power. For each pump power red-detuned from the primary cavity, we find optomechanical coupling rates ($g_i$) from the OMIT response, as explained in \secref{sec:eigens}. Optomechanical coupling rates are related to the intra-cavity photons ($n_{\rm p}$) according to $g_i=g_{0,i}\sqrt{n_{\rm p}}$. The number of photons inside the cavity is related to the pump power at room temperature ($P_{\rm src}$) according to:
	\begin{equation}
		{n_{\rm{p}}} = {\left( {\frac{{{g_i}}}{{{g_{0,i}}}}} \right)^2} \approx \eta_{\rm tot} \frac{{{\kappa _{{\rm{ex}}}}}}{{{\kappa ^2}/4 + {{\bar \Omega }_{\rm{m}}}^2}}\frac{{{P_{{\rm{src}}}}}}{{\hbar {\omega _{\rm{d}}}}},
	\end{equation}
	where $\eta_{\rm tot}$ is the loss between the source and the device (to be determined) and $\omega_{\rm d}=\omega_{\rm c}-\bar \Omega_{\rm m}$ is the frequency of the pump detuned from the primary cavity by the average of the mechanical frequencies. In \figref{fig:loss_g}, we plot the intra-cavity photon number versus the source power at room temperature. By fitting a line to data and using the device parameters (\tabref{tab:glossary} and \tabref{tab:mechanics}), we find the loss to be 60~dB. The loss at room temperature, considering filter cavities and cables, is estimated to be $\approx 5-10$~dB, meaning that the loss in the fridge is $\approx 50-55$~dB, consistent with a separate measurement (see \secref{sec:attenuation}).
	
	\subsection{Attenuation scheme and photon occupation}
	\label{sec:attenuation}
	Two attenuated lines are used to rout a driving tone to the device and to destructively interfere this tone to prevent the HEMT amplifier saturation. Importantly, attenuations in the last three stages (still, cold plate, and mixing chamber plate) have been removed to avoid  their thermal overload when a strong pump is sent. However, the photonic bath of such lines is kept cold by adding directional couplers instead of attenuators (\figref{fig:fullSetup}). While a more optimized case would dump the excess driving power in a third line towards higher stages, the limited lines available in our dilution refrigerator constrained our experiment to use only these two. 
	
	In the presented scheme, the estimation of the total attenuation as well as the photon bath occupation slightly deviates from the simple case of attenuators, because the signal in a single branch travels back and forth between the stages. 
	
	In order to find the total transmission from this system, we use transfer matrix method~\cite{yeh2006optical}, mainly used for analyzing cascaded systems. This matrix describes the relation between the waves (incident and reflected) on either side of each system. For example, the transfer matrix of a directional coupler with coupling $c$ and insertion loss $t$ (\figref{fig:couplers}\textbf{A}) is given by
	\begin{equation}
		\begin{aligned}
			\left( {\begin{array}{*{20}{c}}
					{{a_1}}\\
					{{a_3}}\\
					{{b_1}}\\
					{{b_3}}
			\end{array}} \right) &= {{\bf{T}}_{c{\rm{dB}}}}\left( {\begin{array}{*{20}{c}}
					{{b_2}}\\
					{{b_4}}\\
					{{a_2}}\\
					{{a_4}}
			\end{array}} \right)\\
			&= \left( {\begin{array}{*{20}{c}}
					{1/c}&0&0&{t/c}\\
					0&{1/c}&{t/c}&0\\
					0&{t/c}&{1/c}&0\\
					{t/c}&0&0&{1/c}
			\end{array}} \right)\left( {\begin{array}{*{20}{c}}
					{{b_2}}\\
					{{b_4}}\\
					{{a_2}}\\
					{{a_4}}
			\end{array}} \right),\;\;\;{t^2} + {c^2} = 1,\;\;\;t,c > 0.\\
		\end{aligned}
	\end{equation}
	In our case, the cascaded system is composed of a directional coupler, two transmission lines (phase delay), some attenuation (from cables and connectors), and another directional coupler (\figref{fig:couplers}\textbf{A}). The transfer matrix of the transmission line with phase delay $\varphi$ and equivalent attenuation $\eta$ is given by:
	\begin{equation}
		{{\bf{T}}_{\varphi ,\eta }} = \left( {\begin{array}{*{20}{c}}
				{\frac{1}{\eta }{e^{ - j\varphi }}}&0&0&0\\
				0&{\frac{1}{\eta }{e^{ - j\varphi }}}&0&0\\
				0&0&{\eta {e^{ + j\varphi }}}&0\\
				0&0&0&{\eta {e^{ + j\varphi }}}
		\end{array}} \right).
	\end{equation}
	The full transmission of the cascaded system can be found by multiplying the transfer matrices, i.e.:
	\begin{equation}
		{\bf{T}} = {{\bf{T}}_{10{\rm{dB}}}}{{\bf{T}}_{\varphi ,\eta  = 2{\rm{dB}}}}{{\bf{T}}_{20{\rm{dB}}}}.
	\end{equation}
	Using this matrix, we can find the transmission:
	\begin{equation}
		{S_{21}} = \frac{{ - {{\bf{T}}_{2,2}}}}{{{{\bf{T}}_{2,1}}{{\bf{T}}_{1,2}} - {{\bf{T}}_{2,2}}{{\bf{T}}_{1,1}}}},
	\end{equation}
	where ${\bf{T}}_{i,j}$ is the $(i,j)$th element of the transfer matrix ${\bf{T}}$.
	
	The estimated loss between two couplers is at least 2~dB for each transmission line. We arrived at this number by measuring the transmission for a similar cascaded system of couplers and cables and comparing it with different loss values $\eta$ (\figref{fig:couplers_exp}\textbf{A} and \textbf{B}). The transmission in this configuration is given in \figref{fig:couplers}\textbf{B} for varying loop phase ($2\varphi$). As it can be seen from this figure (and \figref{fig:couplers_exp}\textbf{B}), the total loss varies between 24~dB and 36~dB for the destructive ($2\varphi=\pi$) and constructive ($2\varphi=0$) interference, respectively. The estimated attenuation in our setup is then 45-57~dB. For 50~dB attenuation, we arrive at the total photon number of 0.09 and 0.07 at 4.8~GHz and 6.2\,GHz, respectively.
	
	\begin{figure*}[hbt!]
		\includegraphics[width=\linewidth]{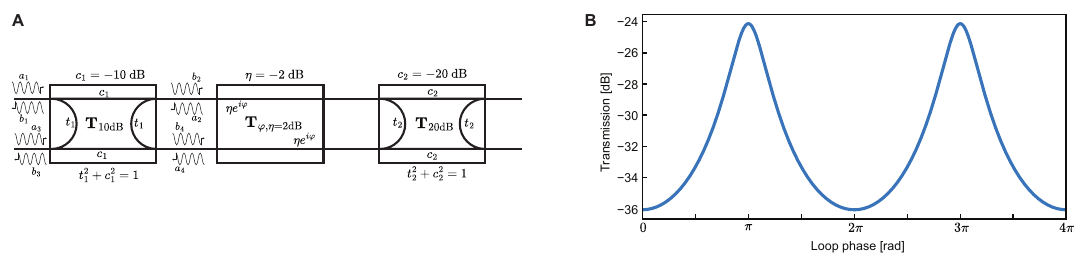}
		\caption{\textbf{Transmission from two cascaded directional couplers}. \textbf{A}, System of two cascaded directional couplers with 20~dB and 10~dB loss mediated by transmission lines. The transmission lines in the middle make phase delay and loss. Total transmission is found by multiplying all the transfer matrices. \textbf{B}, Total transmission as a function of the loop phase ($2\varphi$) by considering 2~dB loss for each transmission line.}
		\label{fig:couplers}
	\end{figure*}
	
	\begin{figure*}[hbt!]
		\includegraphics[width=\linewidth]{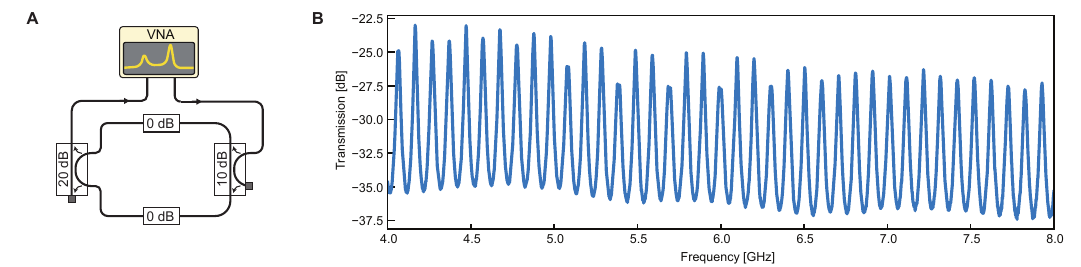}
		\caption{\textbf{Measured transmission from two cascaded directional couplers}. \textbf{A}, Measurement scheme for finding transmission from two cascaded directional couplers with 20~dB and 10~dB coupling connected via coaxial cables and connectors used in the main setup. \textbf{B}, Transmission as a function of the frequency.}
		\label{fig:couplers_exp}
	\end{figure*}
	
	\subsection{Microwave modes characterization}
	\label{sec:mm_modes}
	The values reported in \secref{sec:params} are obtained by fitting the microwave reflection $S_{11}(\omega)$ for all the microwave modes. In particular, we remove the microwave background due to imperfection in our experimental setup (see \secref{sec:setup} for more details). 
	\begin{figure*}[hbt!]
		\includegraphics[width=\linewidth]{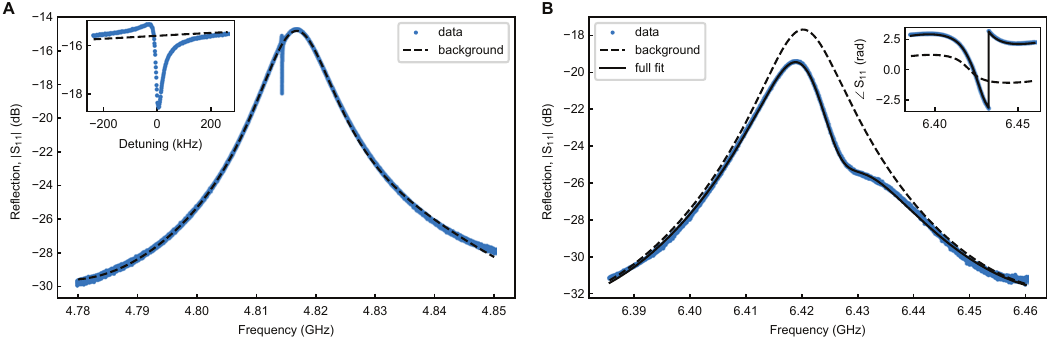}
		\caption{\textbf{Characterization of the microwave modes}. \textbf{A}, Amplitude of the reflection coefficient for the primary cavity. The fit to the background (dashed line) clearly capture the behaviour of the main feature of the signal. In the inset, we show the comparison between the estimated background near the mode resonance. \textbf{B}, Amplitude of the reflection coefficient for the auxiliary cavity; due to a linewidth comparable to the background feature, the background (dashed line) and the microwave mode are simultaneously fitted (solid line). In the inset we show the phase of $S_{11}$ and the fits.}
		\label{fig:MW_fit}
	\end{figure*}
	
	In \figref{fig:MW_fit}, we report the raw reflection data from a VNA, specifically for the primary cavity in \textbf{A} and the auxiliary cavity in \textbf{B}. For the primary cavity (and for all other modes except the auxiliary cavity), the linewidth is much narrower than the background fluctuation, so the background can be removed by fitting only the amplitude part to 
	\begin{equation}
		|S_{11}^{\rm bg}|=\frac{A_{\rm bg}}{\pi\gamma_{\rm bg}(1+(\omega-\omega_{\rm bg})/\gamma_{\rm bg})^2}+B_{\rm bg}\cos(C_{\rm bg}\omega+\phi_0)+D_{\rm bg}
		\label{eq:bg_fit}
	\end{equation} 
	which takes into account the Lorentzian-like resonance due to the reflection from two cascaded couplers as well as possible impedance mismatch. 
	The result of the fits (dashed line) well reproduces the background that is removed to obtain each microwave mode reflection data (see inset in \figref{fig:MW_fit}\textbf{A}).
	Such procedure is repeated for all the microwave modes except for the auxiliary cavity that requires a different fitting scheme. After subtracting the background and applying a normalization, the reflection $S_{11}(\omega)$ is fitted to 
	\begin{equation}
		S_{11}=\exp{\left(i \omega \tau-i \alpha_i\right)}\left(
		1-\frac{2\kappa_{\rm ex}^{i}e^{i\phi}}{\kappa_{\rm ex}^{i}+\kappa_{0}^{i}+2i (\omega-\omega_{i})}\right)
\end{equation}
and the results are reported in Figure 1 in the main text. 

For the auxiliary cavity, due to its large linewidth, it is not possible to fit independently the background. In order to extract this mode parameters, we fit the complex reflection coefficient: 
\begin{equation}
	S_{11}^{\rm combined}=\left(
	\frac{A_{\rm bg}\exp{\left(i \omega \tau-i \alpha_{\rm bg}\right)}\sqrt{\kappa_{\rm i, bg}\kappa_{\rm ext, bg}}}
	{\kappa_{\rm tot, bg}+2 i(\omega-\omega_{\rm bg})}+
	B_{\rm bg}\cos(C_{\rm bg}\omega+\phi_0)+D_{\rm bg}\right)S_{11},
	\label{eq:aux_fit}
\end{equation} 
where the background in this case is modeled as a resonator in trasmission cofiguaration, hence the three coupling terms $\kappa_{\rm i, bg}$, $\kappa_{\rm ext, bg}$ and $\kappa_{\rm tot, bg}$ . Figure~\ref{fig:MW_fit}\textbf{B} display the magnitude and the phase (inset) as well as the fit (in solid line). Using the fitted values, the reconstructed background is shown in the dashed line.

\subsection{Benchmark of the microwave characterization and Fano effects}
\label{sec:benchmark}
In this section, we validate the fitting protocol described in \secref{sec:mm_modes} for removing the background (due to Fabry-Perot resonances of the two cascaded couplers) by simulating the cryogenic setup below the Still plate using Cadence AWR\textsuperscript{\textcopyright}, and comparing the extracted internal and external decay rates with actual values. In details, we implement two circuits: (i) the complete wiring diagram beneath the Still plate, inductively coupled to an RLC resonator (see \figref{fig:SPICE_full}A), and (ii) the same RLC resonator, but without two directional couplers (see \figref{fig:SPICE_full}B).

\begin{figure*}[hbt!]
	\includegraphics[width=\linewidth]{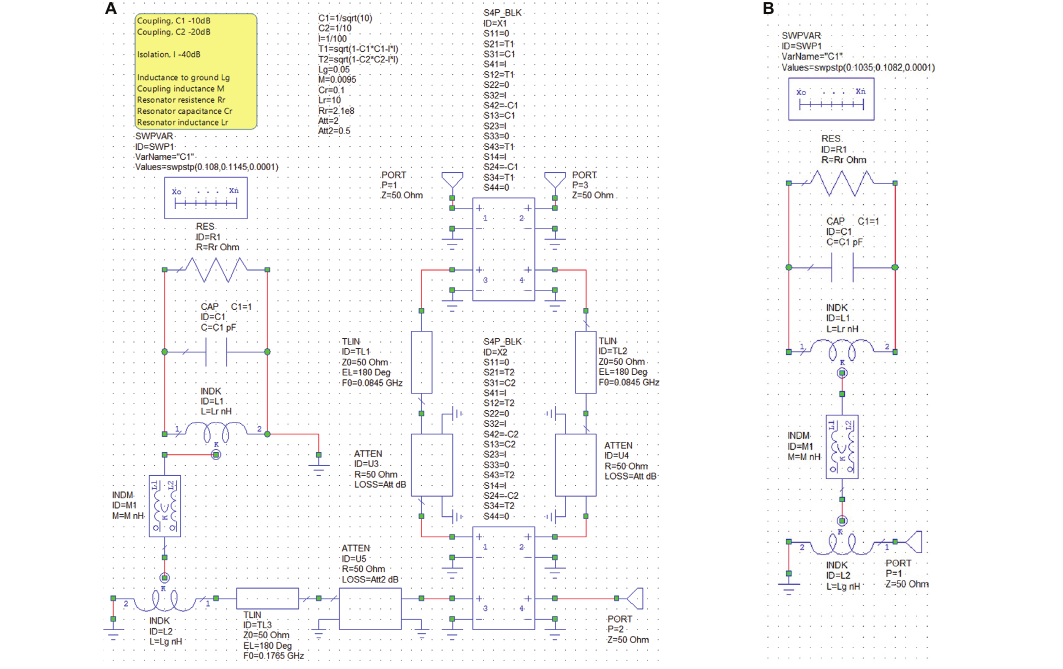}
	\caption{\textbf{Microwave SPICE circuits for the benchmark of the microwave background removal procedure.} \textbf{A}, Full circuit setup, including the directional coupler loop and a RLC resonator whose resonance is varied by chnging the value of its capacitor. \textbf{B}, RLC resonators inductively coupled to the microwave feedline. 
	}
	\label{fig:SPICE_full}
\end{figure*}

The circuit (i) has three free parameter that we determine from measurements: First, the length of each branch in the couplers' loop, which is determined by fixing the free spectral range of the interference pattern to 84.5\,MHz (the measure value). Second, the attenuation in each branch of the circulator loop, which is set to $A=-2$\,dB and is determined from an identical replica of the setup measured at room temperature (see \secref{sec:attenuation} for more details). Finally, the attenuation and the length of the branch connecting the circulator, which is set according to the cryogenic setup. 

Due to imperfect isolation of the second coupler, Fano interference is present in the circuit~\cite{rieger2023fano}, as the leakage signal from the second isolator ($S_{41}\neq0$) interferes with the signal reflected form the sample. By initially neglecting the circulator leakage (in the absence of Fano interference), our background removing procedure faithfully recovers the internal and external decay rates, both for the primary and auxiliary cavities.

\begin{figure*}[hbt!]
	\includegraphics[width=\linewidth]{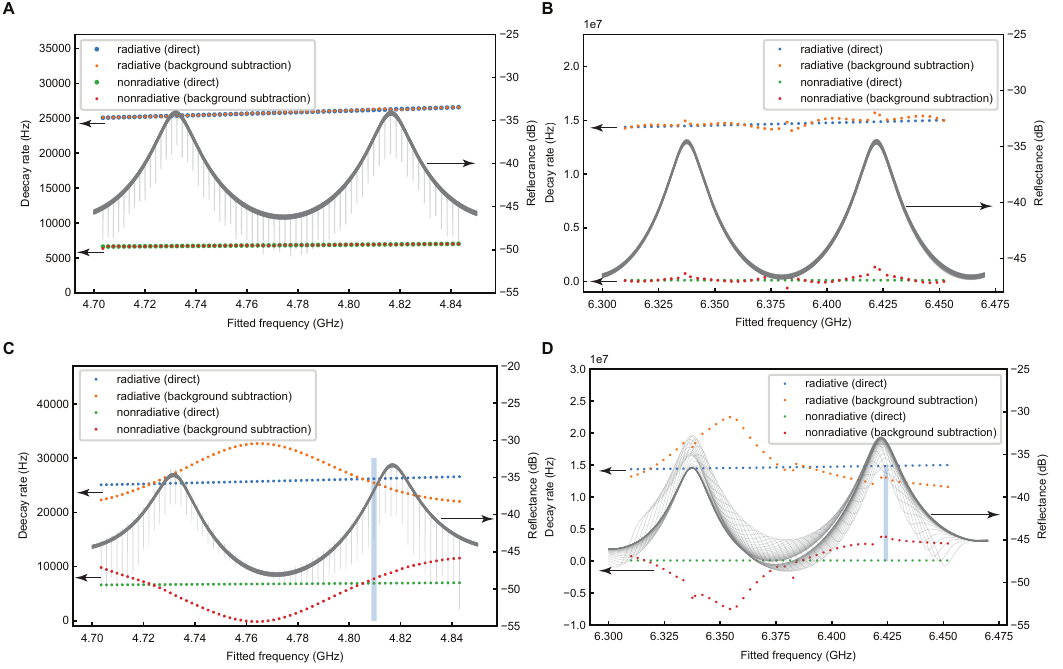}
	\caption{\textbf{Benchmark of the microwave background removal procedure.} \textbf{A}, Internal and external decay rates extracted by fitting the complex $S_{11}$ to the model in Eq.~\eqref{eq:simpleS11} as a function of the resonance frequency and using the full background removal in Eq.~\eqref{eq:bg_fit}. \textbf{B}, Comparison of the internal and external decay rates extracted for a much broader resonance and using the complex $S_{11}$ model in Eq.~\eqref{eq:aux_fit}. \textbf{C}, Comparison of the extracted decay rates for the main and auxiliary (\textbf{D}) cavities, with the Fano interference present in the circuit.
	}
	\label{fig:MW_fit_benchmark}
\end{figure*}

In \figsref{fig:MW_fit_benchmark}\textbf{A} and \textbf{B}, we report the extracted decay rates for multiple resonances obtained by varying the capacitance of the RLC resonator and considering perfect isolation. We fit the resonance simulated by the circuit depicted in \figref{fig:SPICE_full}A with the complex $S_{11}$ in Eq.~\eqref{eq:aux_fit}. The ones simulated in the simple RLC circuit (\figref{fig:SPICE_full}B) are fitted by the complex function 
\begin{equation}
	S_{11}=
	1-\frac{2\kappa_{\rm ex}}{\kappa_{\rm ex}+\kappa_{0}+2i (\omega-\omega_{0})}.
	\label{eq:simpleS11}
\end{equation}
The results show well agreement between the extracted decay rates both for resonances around the primary (\figref{fig:MW_fit_benchmark}A) and the auxiliary cavities (\figref{fig:MW_fit_benchmark}B). The extracted decay rates are chosen to be close to the ones measured in the cryogenic setup. 

Finally, we simulate the same resonances, but in the presence of a finite isolation in the coupler ($S_{41}=S_{14}=-40$\,dB) and report the comparison in \figref{fig:MW_fit_benchmark}\textbf{C} and \textbf{D}. In this case, the Fano interference produces less reliable estimations of the decay rates.

In the experiments, the resonance of the primary cavity lies within the blue-shaded region. For the primary cavity, when the Fano interference produces a peak in the reflectance, the fit yields a large discrepancy between the extracted decay rates and the one set by the resonator (\figref{fig:MW_fit_benchmark}\textbf{C}). The same behavior is shown in \figref{fig:MW_fit_benchmark}\textbf{D} for resonances around the auxiliary cavity; in this case, resonances that appear as peak cannot be fitted, as they converge towards negative decay rates for our automated routine. However, in our experimental realization, both primary and auxiliary cavities appear as a dip. To take into account the uncertainty of fitting, we conclude a 10\% error in the estimation of the microwave external decay rates and propagate it in our data analysis.

The internal decay rate of the auxiliary and primary cavity should be the same, at least in theory. Our numerical simulations reveal that, for the auxiliary cavity, the fitted internal decay rate is over-estimated by one order of magnitude. While this is a large error, it does not account completely for the discrepancy between the decay rates of the primary and auxiliary cavities. The inconsistency can be partially explained by tangent loss, which is frequency dependent.  One of our goals in the near future is to investigate more this observation.

\subsection{Microwave nonlinearities}
In our optomechanical model, we assume that fourth order Kerr nonlinearity, $K$, in the microwave modes has negligible effect in the photon number regime produced by the pumping powers used in all experiment. To confirm this assumption, here we characterize the Kerr effect as a function of intra-cavity photon number for the primary cavity mode. We use a VNA and measure the amplitude and phase of the reflection coefficient $S_{11}(\omega)$ reported in \figsref{fig:nonlinearities}\textbf{A} and \textbf{B}, respectively.

\begin{figure*}[hbt!]
	\includegraphics[width=\linewidth]{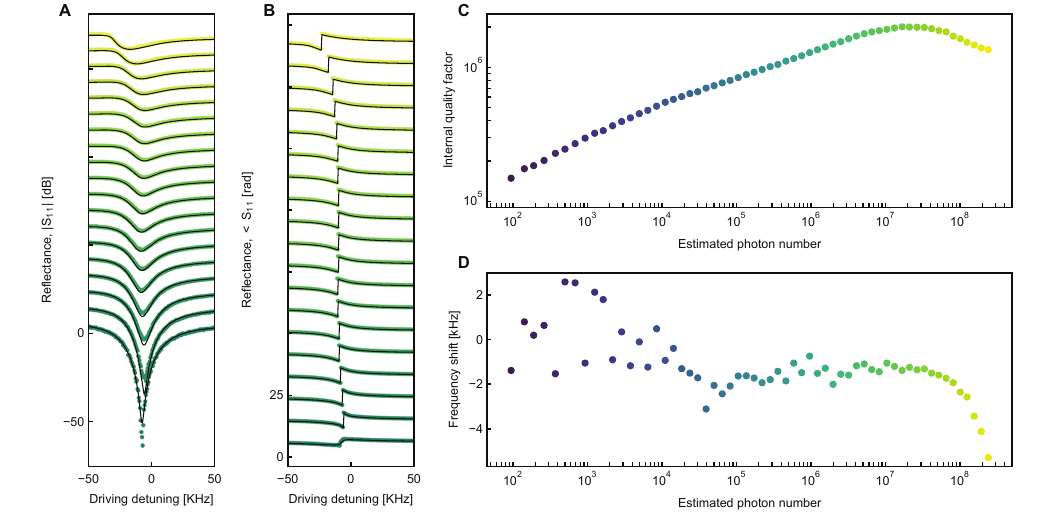}
	\caption{\textbf{Microwave response of the primary cavity as a function of driving power.} \textbf{A}, Magnitude and \textbf{B} phase of the reflection coefficient $S_{11}$ of the primary cavity mode as a function of power drive (each trace is shifted by 9\,dB and 9\,rad to make them visible). \textbf{C}, Internal quality factor and \textbf{D} resonance frequency extracted by fitting the complex $S_{11}$ to the model in Eq.~\eqref{eq:nonlin_model} as a function of the inferred intra-cavity photon number. The color code shows the intra-cavity photon number.
	}
	\label{fig:nonlinearities}
\end{figure*}
Following the treatment in \cite{eichler2014controlling} and assuming a total cryogenic attenuation of 50~dB, we fit all traces with a modified expression for the reflection coefficient,
\begin{equation}
	S_{11}=\exp{\left(i \omega \tau-i \alpha_i\right)}\left(
	1-\frac{2\kappa_{\rm ex}e^{i\phi}}{\kappa_{\rm ex}+\kappa_{\rm 0}+2i (\omega-\omega_{\rm c})+i  \frac{|\alpha_{\rm in}|^2 K}{\kappa_0+\kappa_{\rm ex}} n}\right),
	\label{eq:nonlin_model}
\end{equation}
that takes into account the nonlinear relation between the driving power $P_{\rm in}$ and the intra-cavity photon number due to the Kerr term. Here, by using the driving field $|\alpha_\textrm{in}|^2 =P_{\rm in}/\hbar \omega$ and the intra-cavity coherent field $\alpha$, the parameter $n=(\kappa_0+\kappa_{\rm ex})^2|\alpha|^2/(\kappa_{\rm ex}|\alpha_{\rm in}|^2)$ is obtained by solving the third order algebraic equation for each driving power 
\begin{equation}
	\left[\left(\frac{\omega_{\rm c}-\omega}{\kappa_{0}+\kappa_{\rm ex}}\right)^2  +\frac{1}{4}\right]n-2\frac{\omega_{\rm c}-\omega}{(\kappa_0+\kappa_{\rm ex})^2}K n^2+\left(\frac{K}{\kappa_0+\kappa_{\rm ex}}\right)^2 n^3 =\frac{\kappa_0}{\kappa_0+\kappa_{\rm ex}}.
\end{equation}

The model well reproduces the experimental data. The internal quality factors and the resonance frequencies extracted from the fit are reported in \figsref{fig:nonlinearities}\textbf{C} and \textbf{D}, respectively. The strong power dependence of the quality factor suggests that at low intra-cavity photon numbers, the primary cavity mainly dissipates via the two-level systems bath located between the capacitor electrodes; such behavior has been already observed in the literature for vacuum gap capacitors~\cite{zemlicka2023compact, cicak2009vacuum}. At intermediate powers, the resonance frequency is constant within the experimental errors, while the Lorentzian dip shifts (compare  \figref{fig:nonlinearities}\textbf{B} and \textbf{D}). This is due to the presence of small Kerr nonlinearity ($K/2\pi< 1\;{\rm Hz}$) whose effect appears for large photon numbers. Finally above $10^8$ photons, the resonance frequency shifts and the quality factor slightly degrades. We attribute such behavior to quasi-particles generation in the inductors that results in a lossy resonator.

\subsection{Microwave thermal response}
In this section, we study the single microwave tone response of the primary cavity as a function of the mixing chamber temperature, to which the sample is thermally anchored. We drive the cavity by a single coherent tone generated by a VNA and measure the amplitude and phase of reflection coefficient $S_{11}$ reported in \figsref{fig:MW_temperature}\textbf{A} and \textbf{B}, respectively.
\begin{figure*}[hbt!]
	\includegraphics[width=\linewidth]{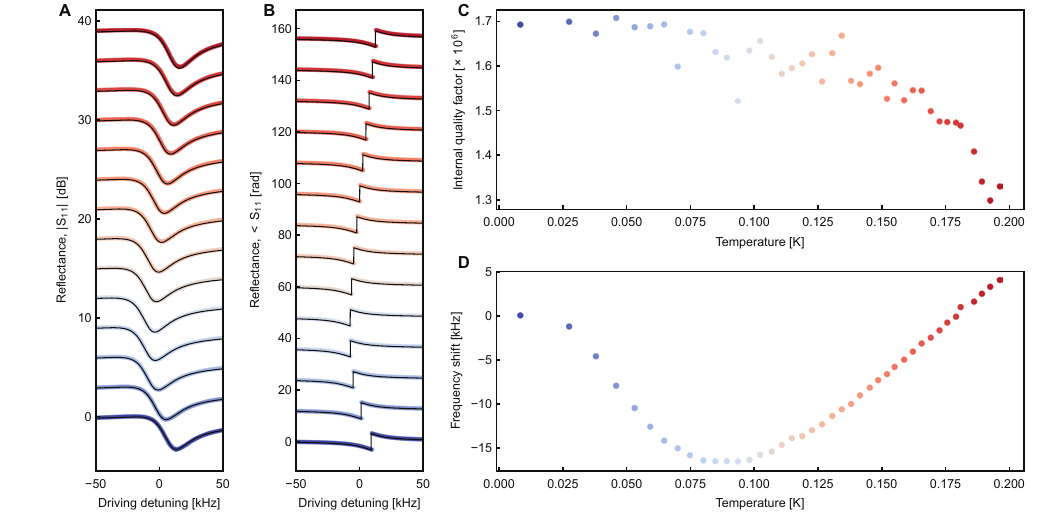}
	\caption{\textbf{Microwave response of the primary cavity as a function of temperature.}  \textbf{A}, Magnitude and \textbf{B} phase of the reflection coefficient $S_{11}$ of the primary cavity mode as a function of mixing chamber temperature for a driving power at room temperature $P_{\rm in}=-20$\,dBm corresponding to $\approx10^7$ photons (each trace is shifted by 9\,dB and 9\,rad to make them visible). \textbf{C}, Internal quality factor and \textbf{D} resonance frequency extracted by fitting the complex $S_{11}$ to the linear model as a function of the temperature. The color code shows the mixing chamber temperature.
	}
	\label{fig:MW_temperature}
\end{figure*}
As expected~\cite{scigliuzzo2020phononic}, the quality factor of aluminum resonators degrades due to quasi-particle poisoning when the temperature becomes a considerable fraction of the superconducting critical temperature, see \figref{fig:MW_temperature}\textbf{C}. In addition, for temperatures below 100\,mK, the frequency of the primary cavity decreases due to an increase in broken cooper pairs, which increases the superconducting films kinetic inductance (see~\figref{fig:MW_temperature}\textbf{D}). However, for higher temperatures, the frequency shift changes sign and increases with temperature. We attribute this effect to the relaxation of the strain in the film, resulting in the increase of the gap size, and consequently reducing the capacitance.

\subsection{Noise floor calibration}
\label{sec:hemt}
To extract the occupation of the bright collective mode, we need to calibrate the noise floor of the output spectrum. We  model the experimental setup after the device as given in Fig.~{\ref{fig:hemt}}\textbf{A}. In this model, $\eta$ is the loss between the device under test (DUT) and HEMT and can be modeled with a beam-splitter, $n_{\rm{add}}^{\rm{H}}$ is the added noise of HEMT, and $G_{\rm{H}}$ is the gain of HEMT. The added noise of the measurement chain after HEMT can be neglected due to its high gain. The detected signal at room temperature is then given by:

\begin{equation}
	{G_{\rm{H}}}\left( {\eta (s + \frac{1}{2})+(1-\eta)\frac{1}{2}}+ n_{{\rm{add}}}^{\rm{H}} + \frac{1}{2} \right) = {G_{\rm{H}}}\eta \left( {s + \frac{{1 + n_{{\rm{add}}}^{\rm{H}}}}{\eta }} \right).
\end{equation}
Here, $s$ is the signal leaving the device together with its background (vacuum noise), which enters the first port of the beam-splitter, the second $1/2$ is the vacuum noise coming from the second port of the beam-splitter, $n_{\rm add}^{\rm H}$ is the added noise of HEMT, and the second $1/2$ is the added noise due to phase-insensitive amplification. The background of the signal referred to the device is $1+n_{\rm add}=\frac{1+n_{{\rm{add}}}^{\rm{H}}}{\eta}$. Note that in an ideal phase-insensitive measurement $n_{\rm add}=0$.

We need then to reliably find the added noise of HEMT. To this end, we sweep the temperature of mixing chamber and wait until all the lines and devices connected to it thermalize to this temperature. In this case, the thermal input signal to HEMT is given by:
\begin{equation}
	n_{\rm{th}} = \frac{\exp(-\hbar \omega/k_{\rm{B}}T)}{1-\exp(-\hbar \omega/k_{\rm{B}}T)},
\end{equation}
where $k_{\rm{B}}$ is the Boltzmann constant and $T$ is temperature of the mixing chamber. The detected signal is given by: 
\begin{equation}
	{G_{\rm{H}}}\left(n_{\rm{th}} + n_{\rm{add}}^{\rm{H}}+1 \right).
	\label{eq:hemt}
\end{equation}
The signal in a wide span is detected using a spectrum analyzer (4-8~GHz) with 1~MHz resolution bandwidth. We then divide the entire frequency span into sections, calculate the average noise power for each section, and plot these average values versus temperature. By fitting this curve to Eq.~\eqref{eq:hemt} with $n_{\rm{add}}^{\rm{H}}$ and $G$ as free parameters, we can extract $n_{\rm{add}}^{\rm{H}}$ at each frequency section. Calibration data around the primary and auxiliary cavity center frequencies are given in \figsref{fig:hemt}\textbf{B} and \textbf{C}, respectively. Between DUT and HEMT we have a cryo switch, which has 0.2~dB attenuation, a dual-junction isolator with 0.2~dB attenuation, and cables with 0.8~dB attenuation. The numbers are considered from datasheet. We consider 0.5~dB systematic error for the estimated loss. The extracted added noised (referred to DUT) are then $1+n_{\rm add}=9.0\pm1.0$ quanta and $8.3\pm1.0$ quanta around primary and auxiliary cavity frequencies, respectively. 

\begin{figure*}[hbt!]
	\includegraphics[width=\linewidth]{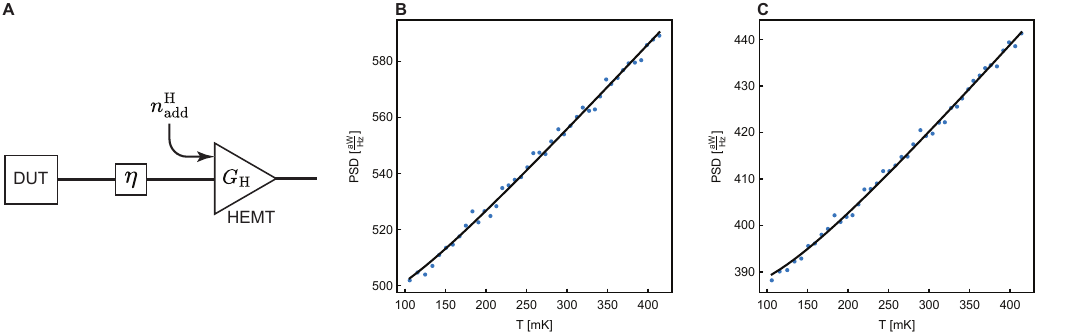}
	\caption{\textbf{Noise floor calibration}. \textbf{A}, Simplified measurement setup after the device. \textbf{B} and \textbf{C}, Calibration data for HEMT around the primary and auxiliary cavity frequencies, respectively. 
	}
	\label{fig:hemt}
\end{figure*}

\subsection{Modeshape of collective modes}
\label{sec:mode_col}
In order to measure the modeshapes of the collective modes, we leverage the slight non-degeneracy of the mechanical oscillator frequencies. For each pump power, we first take a VNA trace around the cavity frequency, which provides us with the multi-mode OMIT response (\secref{sec:omit}). We then fit Eq.~(\ref{eq:ref_omit}) to data with $g_i$, $\kappa$, and $\omega_{\rm{c}}$ as free parameters. Using these parameters, we constitute the matrix $\bf M$ in Eq.~(\ref{eq:lang_omit}). Then, we find eigenvalues of this matrix, which are complex valued parameters. The real parts are half of the mode linewidths and the imaginary parts are the mode frequencies. Afterwards, for each pump power, we send an excitation signal that is blue detuned from the pump by each mode frequency. The beating of these two tones will excite selectively the target mode. In the basis of the individual mechanical oscillators, each oscillator will be excited based on its participation ratio in the excited collective mode. To measure these amplitudes, we turn off both the pump and the excitation signal and send a weak pump which is red detuned from the cavity by the average of the bare mechanical frequencies. The power of the pump is weak enough to ensure that $\left| {{\Omega _{{\rm{m}},i}} - {\Omega _{{\rm{m}},j}}} \right| \gg {\Gamma _{{\rm{opt,}}i}},{\Gamma _{{\rm{opt,}}j}}$ and high enough to ${\Gamma _{{\rm{opt,}}i}} \gg {\Gamma _{{\rm{m,}}i}}$. This is to assure that the oscillator sidebands do not interfere, but at the same time, the rates are more than the bare damping rates. We then record both the in-phase ($I$) and out-of-phase ($Q$) quadratures of the emitted signal around the cavity frequency with sampling rate 50~kHz using the I-Q mode of a spectrum analyzer. The sampling rate is chosen so that all the ringdowns of the mechanical oscillators can be recorded (they all fall in a span as low as 10~kHz). Having both $I$ and $Q$ of the signal enable us to extract the phase of the collective mode. Now, to find the participation ratio of each mechanical oscillator in the excited mode, we take a Fourier transform of the recorded signal $z(t) = I(t)+iQ(t)$. As we can see from Fig.~3 of the main text, this contains 6 peaks related to the bare mechanical oscillator frequencies. Next, we fit the following complex Lorentzian function to each of these peaks,
\begin{equation}
	f(\omega) = \frac{A_j^ke^{i\varphi_j}(\Gamma_{{\rm{m}},j}/2 - j (\omega - \Omega_{{\rm{m}},j}) )}{\Gamma_{{\rm{m}},j}^2/4 + (\omega - \Omega_{{\rm{m}},j}) ^2}+{\rm BG}_j
\end{equation}
which is Fourier transform of $A_j^ke^{i\varphi_j}e^{(-i(\omega-\Omega_{{\rm{m}},j})-\Gamma_{{\rm{m}},j}/2)t}\theta(t)$, where $A_j^k$ is ringdown amplitude of the $j$th mechanical oscillator of the $k$th excited collective mode, $\varphi_j^k$ is its phase, ${\rm BG}_j$ is the fitting background, and $\theta(t)$ is the Heaviside function. For each pump power, we have $N+1$ eigenmodes, for each of which we follow the aforementioned procedure. The normalized amplitude can be found as:
\begin{equation}
	\eta_i^k = \frac{{{A_i^k}}}{\sqrt{\sum\limits_{j = 1}^N {{{\left| {{A_j^k}} \right|}^2}} }}.
\end{equation}
Theoretical values for eigenvectors are determined by diagonalizing the matrix $\bf{M}$ in Eq.~(\ref{eq:lang_omit}), utilizing the fitting parameters to the multi-mode OMIT response Eq.~(\ref{eq:ref_omit}). The reconstructed modeshapes and the theoretical ones for pump cooperativities $\mathcal{\bar C}=328$, 2056, 13301, and 67182 are given in \figsref{fig:modeshape}\textbf{A}-\textbf{D}, respectively, where a great overlap is observed, both in phases and amplitudes. 

It is worth mentioning that after strong coupling ($\mathcal{\bar C}>65000$), modes $N$ and $N+1$ will be the result of the hybridization between the microwave cavity and the bright collective mode. The only difference between them is that in one of them, the cavity and the bright collective mode hybridize in-phase, while in the other they hybridize out-of-phase. As we are only able to reconstruct the mode in the reduced phase space of the mechanical oscillators, there will not be any difference between them. This is why we only show modeshapes of six modes rather than seven modes even in the strong coupling regime.

\begin{figure*}[hbt!]
	\includegraphics[width=\linewidth]{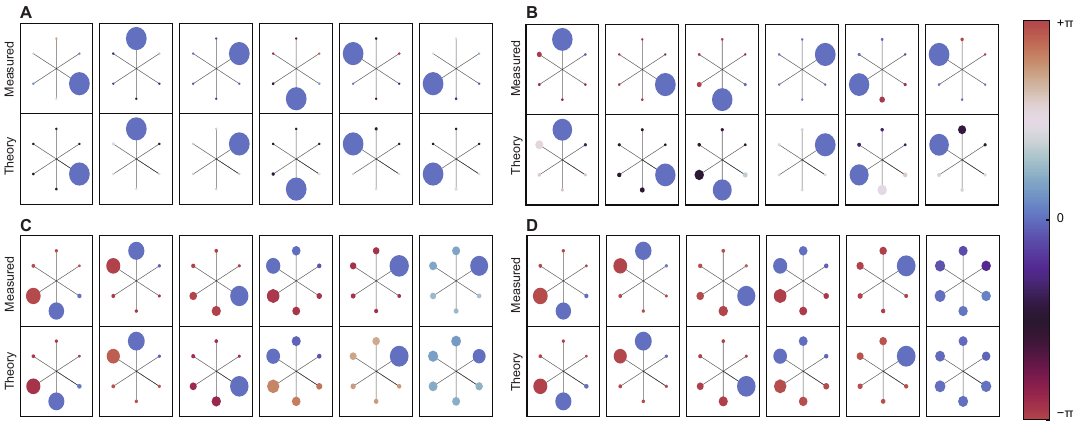}
	\caption{\textbf{Collective modeshapes}. \textbf{A}-\textbf{D}, Collective modeshape reconstruction for pump cooperativities $\mathcal{\bar C}=328, 2056, 13301,\;{\rm and }\;67182$, respectively. Normalized amplitudes are given by the circle radius and the phase is given by the color. 
	}
	\label{fig:modeshape}
\end{figure*}

\subsection{Occupation of bright and dark collective modes}
\label{sec:occupations}
\begin{figure*}[hbt!]
	\includegraphics{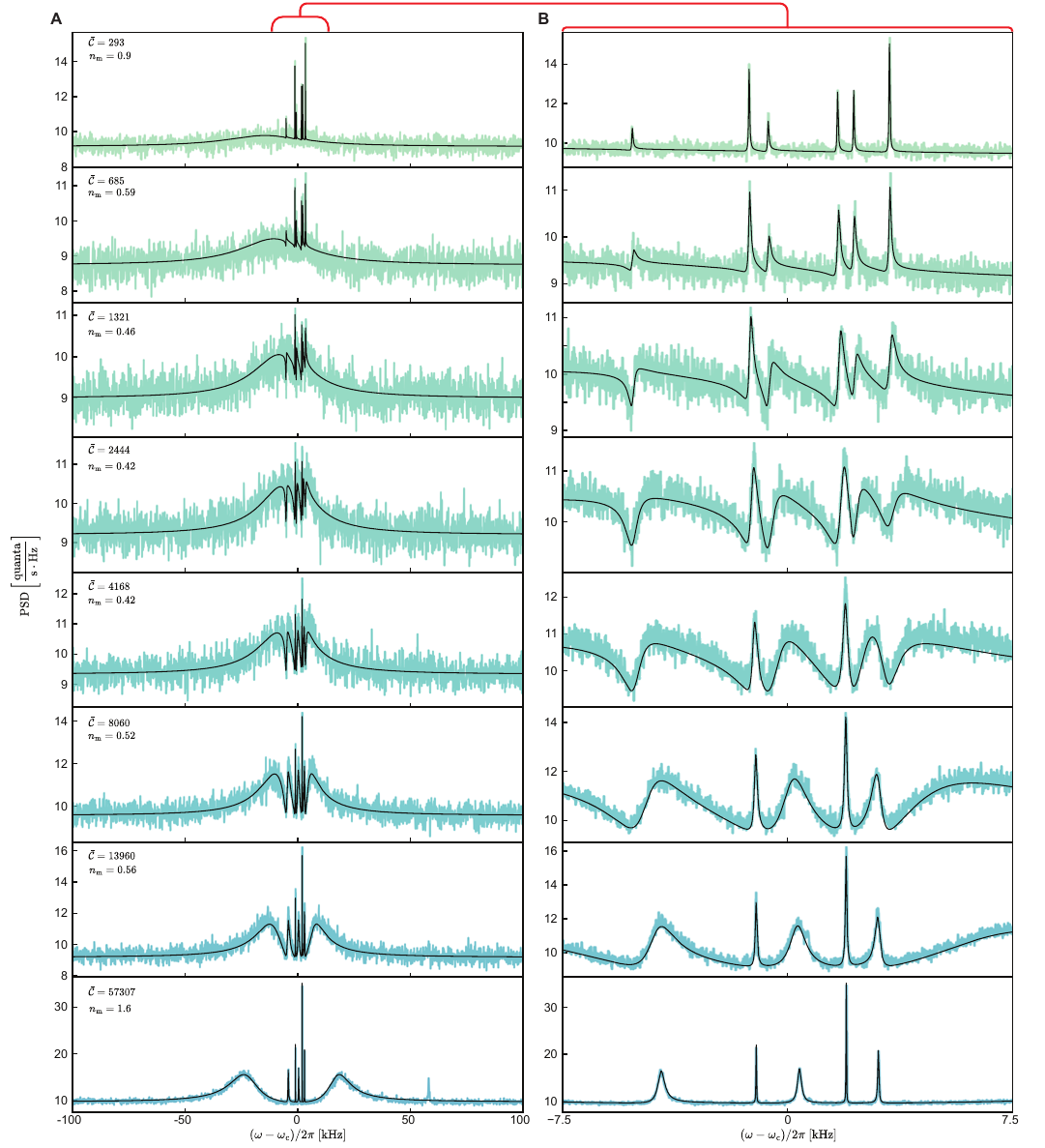}
	\caption{\textbf{Detected spectrum}. \textbf{A}, Wide span and \textbf{B}, zoomed-in spectrum for different rang of cooling powers and the corresponding occupation of the bright collective mode ($n_{\rm m}$).
	}
	\label{fig:specs}
\end{figure*}

To find the occupation of the dark and bright collective modes in the presence of a pump red-detuned from the primary cavity by $\bar \Omega_{\rm m}$, we need to first detect the power spectral density (Eq.~\eqref{eq:spec_a}). Considering gain, loss, and added noise of the measurement chain~\ref{sec:hemt}, the detected spectrum is given by:
\begin{equation}
	\bar S'(\omega ) = G\left( {\bar S(\omega ) + \frac{1}{2} + {n_{{\rm{add}}}}} \right).
	\label{eq:spec_a_det}
\end{equation}
Here, $G$ is the scaling factor induced by all the gain and loss of the measurement chain and $n_{\rm add}$ is the equivalent added noise of the chain referred to the device. For an ideal measurement chain (zero loss and ideal phase-insensitive measurement), $n_{\rm add}$ is zero. In our experiment, it is measured in \secref{sec:hemt}. By knowing the background of the detected spectrum, we can find the scaling factor $G$. In \figref{fig:specs}, we show the detected spectrum at different cooling pump powers. 

We then fit Eq.~\eqref{eq:spec_a} to the calibrated spectrum considering decoherence rates ($\Gamma_{{\rm th},i}=\Gamma_{{\rm m},i}n_{{\rm m},i}^{\rm th}$), microwave cavity frequency and linewidth, as well as coupling rates as free parameters. To propagate the uncertainties in the external coupling rate and the added noise of the measurement chain into the fitting parameters, we use the Monte Carlo method. We sample from a Gaussian distribution based on the known parameters' mean and variance, and then we fit Eq.~\eqref{eq:spec_a} to the spectrum. This process is repeated $N_{\rm rep}$ times, after which the mean and variance of the fitting parameters are obtained from the $N_{\rm rep}$ samples.

After finding these parameters, we plug them into Eq.~\eqref{eq:spec_b_cross} to find correlations among different mechanical oscillators. Having all these correlations, we can constitute the covariance matrix Eq.~\eqref{eq:cov}. Diagonalizing this matrix provides us with the occupations of the modes. Importantly, the Hermiticity of this matrix ensures that all the eigenvalues are real. The minimum eigenvalue corresponds to the occupation of the bright collective mode, as it has the maximum coupling to the cavity. The other eigenvalues, however, correspond to the occupation of the dark collective modes. The occupation of dark and bright collective modes together with the cavity heating versus cooperativity is plotted in \figref{fig:darks}\textbf{A}. 

The eigenvector corresponding to each eigenvalue provides us with the mechanical modeshape of that mode. In \figref{fig:darks}\textbf{B}, we plot the fidelity, defined as $F(\psi_i,\psi_i') = {\left| {\left\langle {{{\psi _i}}}\mathrel{\left | {\vphantom {{{\psi_i}} {{\psi_i'} }}}\right. \kern-\nulldelimiterspace}{{{\psi_i'} }} \right\rangle } \right|^2} $ for two vectors $\left| {{\psi _i}} \right\rangle$ and $\left| {{\psi _i'}} \right\rangle $, between the ideal bright collective mode (Eq.~\eqref{eq:ideal_mode}) and the mechanical modeshapes found using eigenvectors of the covariance matrix. As it can be seen from this figure, all the fidelities start from $1/6$, meaning that they are localized in one of the mechanical oscillators. This can be also seen in \figref{fig:modeshape} at low cooperativities. By increasing the pump power, the fidelity of the bright collective mode (blue circle) increases, and finally it reaches to one. The fidelity of dark collective modes, however, go to zero by increasing the pump power, meaning that they become orthogonal to the collective mode.

\begin{figure*}[hbt!]
	\includegraphics{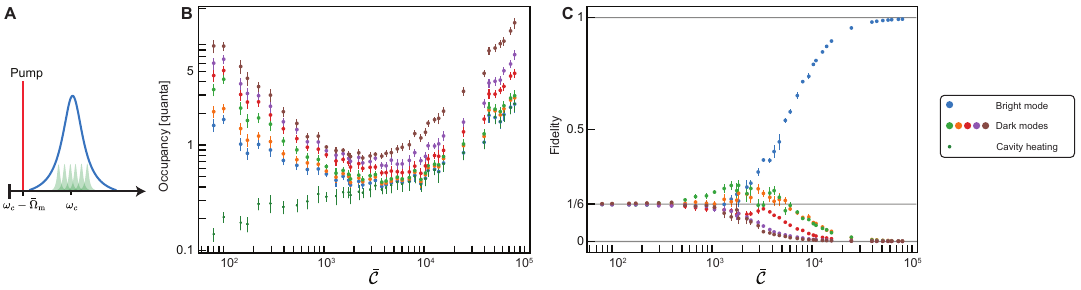}
	\caption{\textbf{Occupancy of collective modes}. \textbf{A}, Measurement scheme for finding the occupation of dark and bright collective modes. \textbf{B}, Occupation of dark and bright collective modes using the eigenvalues of the covariance matrix. Cavity heating is shown with small green circles. \textbf{C}, Fidelities between bright/dark collective modes and the ideal collective mode (Eq.~\eqref{eq:ideal_mode}).
	}
	\label{fig:darks}
\end{figure*}

\subsection{Sideband asymmetry}
\label{sec:asym}
To cross verify the occupation measured using the method mentioned in \secref{sec:occupations}, we use sideband asymmetry experiment, which is an out-of-loop measurement technique \cite{weinstein2014observation,clark2017sideband,youssefi2023squeezed}. The extracted occupation in \secref{sec:occupations} depends on the noise floor calibration, which is found using fridge base temperature sweep (\secref{sec:hemt}). The sideband asymmetry experiment, on the other hand, is independent of the noise floor and only depends on the difference between the Stokes and anti-Stokes motional sidebands. In this experiment, we use the auxiliary cavity which has a broad linewidth ($\kappa_{\rm aux}/2\pi\simeq 15\;\rm{MHz} > \bar \Omega_{\rm m}/2\pi\simeq 2\;\rm{ MHz}$). By pumping on-resonant (slightly red-detuned with detuning $\Delta<0$) on this cavity, we can generate two upper (anti-Stokes) and lower (Stokes) motional sidebands whose areas are proportional to the mechanical occupations. For cooling, we use a pump on the red sideband of the primary cavity (see Fig.~4\textbf{E} of the main text). 

Here, we perform sideband asymmetry experiment in two different regimes: 1- When the power of the cooling pump is low enough that the mechanical sidebands of individual mechanical oscillators do not interfere ($\Gamma_{{\rm{opt}},i},\Gamma_{{\rm{opt}},j}\ll \left| {{\Omega _{{\rm{m}},i}} - {\Omega _{{\rm{m}},j}}} \right|\ll \kappa$). In this case, we can consider the system as $N$ single optomechanical systems. 2- When power of the cooling pump is sufficiently high to induce the emergence of the bright collective mode and its strong coupling with the primary cavity. In this case, as we are only interested in the occupation of the bright collective mode, we can approximate the sidebands with the spectrum of a single optomechanical system in the strong coupling regime, where the mechanical oscillator is the bright collective mode. In both regimes, the cavity heating of the auxiliary cavity is considered, as it introduces a fake asymmetry \cite{youssefi2023squeezed}.

\subsubsection{Sideband asymmetry of individual mechanical oscillators (low power)}
In this regime, we perform the sideband asymmetry experiment on the mechanical oscillator that is furthest from the others, $\Omega_{\rm{m},1}-\bar \Omega_{\rm{m}} = -2\pi \times 5.19$~kHz. By using this mechanical oscillator, we can use a wider range of powers before it interferes with the other ones. By pumping on the auxiliary cavity slightly red-detuned $\Delta<0$, we generate two upper and lower motional sidebands whose areas are proportional to $\Gamma_+ (n_{{\rm{m}},1}-2n_{{\rm aux}})$ and $\Gamma_-(n_{{\rm{m}},1}+1+2n_{{\rm aux}})$, respectively. Here, $n_{{\rm{m}},1}$ is the occupation of the first mechanical oscillator and 
\begin{equation}
	\Gamma_{\pm} = \frac{4g_{1,{\rm aux}}^2\kappa_{\rm aux}}{\kappa_{\rm aux}^2+4(\Omega_{{\rm m},1}\pm\Delta)^2},
\end{equation}
where $g_{1,{\rm aux}}$ is the optomechanical coupling rate of the first mechanical oscillator to the auxiliary cavity. In the case of resonant pumping, $\Gamma_-=\Gamma_+$. We always pump slightly red-detuned ($\approx 100$~kHz) to avoid entering parametric instability regime \cite{aspelmeyer2014cavity}. In this case, to take into account the slight difference between $\Gamma_+$ and $\Gamma_-$, we turn off the cooling pump (red-detuned from the primary cavity by $\Omega_{{\rm m},1}$) and measure the upper and lower motional sidebands (with this, $n_{\rm{m}}\simeq n_{m}+1 \gg 1$). The ratio of the areas then provides us with $\Gamma_+/\Gamma_-$. It should be noted that using a resonant pump on the auxiliary cavity will introduce backaction on the mechanical oscillator, and its effective occupation will increase given by~\cite{youssefi2023squeezed}:
$\frac{{{\Gamma _ - }({n_{{\rm{aux}}}} + 1) + {\Gamma _ + }{n_{{\rm{aux}}}}}}{{{\Gamma _{{\rm{m}},1}} + {\Gamma _{{\rm{opt}},1}} + {\Gamma _{\rm{ + }}} - {\Gamma _ - }}}$. We confirm that in this regime of cooling powers, we did not observe heating of the auxiliary cavity. In addition, as the linewidth of the Lorentzian mechanical sideband is less than (or comparable with) the spectrum analyzer Gaussian resolution bandwidth, the mechanical sideband is fitted using a Voigt function. By finding the mechanical occupation using the calibrated background and comparing it with the sideband asymmetry experiment, we find an excellent overlap, confirming the quantum efficiency extracted in \secref{sec:hemt}.

\subsubsection{Sideband asymmetry of the bright collective mode (high power)}
In this regime, the linewidth of the bright collective mode becomes comparable with the linewidth of the primary cavity, so they enter the strong coupling regime \cite{teufel2011circuit}. It should be noted that the linewidth of the auxiliary cavity ($\simeq 15$~MHz) is much larger than that of the bright collective mode ($\simeq 30$~kHz). We again pump the auxiliary cavity slightly red-detuned ($\Delta<0$) and pump on the red-sideband of the primary cavity (with detuning $\bar \Omega_{\rm m}$). Here, due to the hybridization, the upper and lower motional sidebands are given by (ignoring the narrow peaks of dark modes)~\cite{teufel2011sideband}: 
\begin{equation}
	S(\delta ) = {\rm{BG}} + \frac{{A{\delta ^2}}}{{|4{g^2} - 4\delta (\delta  + \tilde \Delta ) + 2i\delta \kappa {|^2}}} + \frac{B}{{|4{g^2} - 4\delta (\delta  + \tilde \Delta ) + 2i\delta \kappa {|^2}}},
	\label{eq:sidebands}
\end{equation}
where $\delta = \omega +\omega_{\rm aux}\pm \bar \Omega_{\rm m}$, $\tilde \Delta = \omega_{\rm p}-\omega_{\rm c}+\bar \Omega_{\rm m}$, $\omega_{\rm p}$ is the frequency of the pump (red-detuned from the primary cavity), and BG is the background. The $\pm$ in $\delta$ refers to the anti-Stokes and Stokes sidebands, respectively. After fitting Eq.~\eqref{eq:sidebands} to the sidebands (as depicted in Fig.~4 of the main), we find area under each of them. Together with the cavity heating already found in \secref{sec:heating}, we can find the occupation of the bright collective mode, which are plotted in Fig.~4 of the main text. 

The sidebands in this regime have very low signal-to-noise ratio; therefore, given that we do not have a quantum-limited amplifier (such as JTWPA) in our measurement chain, we averaged for 90000 times with spectrum analyze resolution bandwidth RBW = 100~Hz per sideband.

\clearpage


\begin{thebibliography}{57}%
	\makeatletter
	\providecommand \@ifxundefined [1]{%
		\@ifx{#1\undefined}
	}%
	\providecommand \@ifnum [1]{%
		\ifnum #1\expandafter \@firstoftwo
		\else \expandafter \@secondoftwo
		\fi
	}%
	\providecommand \@ifx [1]{%
		\ifx #1\expandafter \@firstoftwo
		\else \expandafter \@secondoftwo
		\fi
	}%
	\providecommand \natexlab [1]{#1}%
	\providecommand \enquote  [1]{``#1''}%
	\providecommand \bibnamefont  [1]{#1}%
	\providecommand \bibfnamefont [1]{#1}%
	\providecommand \citenamefont [1]{#1}%
	\providecommand \href@noop [0]{\@secondoftwo}%
	\providecommand \href [0]{\begingroup \@sanitize@url \@href}%
	\providecommand \@href[1]{\@@startlink{#1}\@@href}%
	\providecommand \@@href[1]{\endgroup#1\@@endlink}%
	\providecommand \@sanitize@url [0]{\catcode `\\12\catcode `\$12\catcode
		`\&12\catcode `\#12\catcode `\^12\catcode `\_12\catcode `\%12\relax}%
	\providecommand \@@startlink[1]{}%
	\providecommand \@@endlink[0]{}%
	\providecommand \url  [0]{\begingroup\@sanitize@url \@url }%
	\providecommand \@url [1]{\endgroup\@href {#1}{\urlprefix }}%
	\providecommand \urlprefix  [0]{URL }%
	\providecommand \Eprint [0]{\href }%
	\providecommand \doibase [0]{https://doi.org/}%
	\providecommand \selectlanguage [0]{\@gobble}%
	\providecommand \bibinfo  [0]{\@secondoftwo}%
	\providecommand \bibfield  [0]{\@secondoftwo}%
	\providecommand \translation [1]{[#1]}%
	\providecommand \BibitemOpen [0]{}%
	\providecommand \bibitemStop [0]{}%
	\providecommand \bibitemNoStop [0]{.\EOS\space}%
	\providecommand \EOS [0]{\spacefactor3000\relax}%
	\providecommand \BibitemShut  [1]{\csname bibitem#1\endcsname}%
	\let\auto@bib@innerbib\@empty
	\bibitem [{\citenamefont {Anderson}(1972)}]{andersonMoreDifferent1972}%
	\BibitemOpen
	\bibfield  {author} {\bibinfo {author} {\bibfnamefont {P.~W.}\ \bibnamefont
			{Anderson}},\ }\bibfield  {title} {\bibinfo {title} {More is different},\
	}\href {https://doi.org/10.1126/science.177.4047.393} {\bibfield  {journal}
		{\bibinfo  {journal} {Science}\ }\textbf {\bibinfo {volume} {177}},\ \bibinfo
		{pages} {393} (\bibinfo {year} {1972})}\BibitemShut {NoStop}%
	\bibitem [{\citenamefont {Onuki}(2002)}]{onuki2002phase}%
	\BibitemOpen
	\bibfield  {author} {\bibinfo {author} {\bibfnamefont {A.}~\bibnamefont
			{Onuki}},\ }\href@noop {} {\emph {\bibinfo {title} {Phase transition
				dynamics}}}\ (\bibinfo  {publisher} {Cambridge University Press},\ \bibinfo
	{year} {2002})\BibitemShut {NoStop}%
	\bibitem [{\citenamefont {Anderson}\ \emph {et~al.}(1995)\citenamefont
		{Anderson}, \citenamefont {Ensher}, \citenamefont {Matthews}, \citenamefont
		{Wieman},\ and\ \citenamefont {Cornell}}]{anderson1995observation}%
	\BibitemOpen
	\bibfield  {author} {\bibinfo {author} {\bibfnamefont {M.~H.}\ \bibnamefont
			{Anderson}}, \bibinfo {author} {\bibfnamefont {J.~R.}\ \bibnamefont
			{Ensher}}, \bibinfo {author} {\bibfnamefont {M.~R.}\ \bibnamefont
			{Matthews}}, \bibinfo {author} {\bibfnamefont {C.~E.}\ \bibnamefont
			{Wieman}},\ and\ \bibinfo {author} {\bibfnamefont {E.~A.}\ \bibnamefont
			{Cornell}},\ }\bibfield  {title} {\bibinfo {title} {Observation of
			bose-einstein condensation in a dilute atomic vapor},\ }\href
	{https://doi.org/10.1126/science.269.5221.198} {\bibfield  {journal}
		{\bibinfo  {journal} {science}\ }\textbf {\bibinfo {volume} {269}},\ \bibinfo
		{pages} {198} (\bibinfo {year} {1995})}\BibitemShut {NoStop}%
	\bibitem [{\citenamefont {Dicke}(1954)}]{dicke1954coherence}%
	\BibitemOpen
	\bibfield  {author} {\bibinfo {author} {\bibfnamefont {R.~H.}\ \bibnamefont
			{Dicke}},\ }\bibfield  {title} {\bibinfo {title} {Coherence in spontaneous
			radiation processes},\ }\href {https://doi.org/10.1103/PhysRev.93.99}
	{\bibfield  {journal} {\bibinfo  {journal} {Physical review}\ }\textbf
		{\bibinfo {volume} {93}},\ \bibinfo {pages} {99} (\bibinfo {year}
		{1954})}\BibitemShut {NoStop}%
	\bibitem [{\citenamefont {Tavis}\ and\ \citenamefont
		{Cummings}(1968)}]{tavis1968exact}%
	\BibitemOpen
	\bibfield  {author} {\bibinfo {author} {\bibfnamefont {M.}~\bibnamefont
			{Tavis}}\ and\ \bibinfo {author} {\bibfnamefont {F.~W.}\ \bibnamefont
			{Cummings}},\ }\bibfield  {title} {\bibinfo {title} {Exact solution for an
			n-molecule—radiation-field hamiltonian},\ }\href
	{https://doi.org/10.1103/PhysRev.170.379} {\bibfield  {journal} {\bibinfo
			{journal} {Physical Review}\ }\textbf {\bibinfo {volume} {170}},\ \bibinfo
		{pages} {379} (\bibinfo {year} {1968})}\BibitemShut {NoStop}%
	\bibitem [{\citenamefont {Skribanowitz}\ \emph {et~al.}(1973)\citenamefont
		{Skribanowitz}, \citenamefont {Herman}, \citenamefont {MacGillivray},\ and\
		\citenamefont {Feld}}]{skribanowitz1973observation}%
	\BibitemOpen
	\bibfield  {author} {\bibinfo {author} {\bibfnamefont {N.}~\bibnamefont
			{Skribanowitz}}, \bibinfo {author} {\bibfnamefont {I.}~\bibnamefont
			{Herman}}, \bibinfo {author} {\bibfnamefont {J.}~\bibnamefont
			{MacGillivray}},\ and\ \bibinfo {author} {\bibfnamefont {M.}~\bibnamefont
			{Feld}},\ }\bibfield  {title} {\bibinfo {title} {Observation of dicke
			superradiance in optically pumped hf gas},\ }\href
	{https://doi.org/10.1103/PhysRevLett.30.309} {\bibfield  {journal} {\bibinfo
			{journal} {Physical Review Letters}\ }\textbf {\bibinfo {volume} {30}},\
		\bibinfo {pages} {309} (\bibinfo {year} {1973})}\BibitemShut {NoStop}%
	\bibitem [{\citenamefont {Fink}\ \emph {et~al.}(2009)\citenamefont {Fink},
		\citenamefont {Bianchetti}, \citenamefont {Baur}, \citenamefont {G{\"o}ppl},
		\citenamefont {Steffen}, \citenamefont {Filipp}, \citenamefont {Leek},
		\citenamefont {Blais},\ and\ \citenamefont {Wallraff}}]{fink2009dressed}%
	\BibitemOpen
	\bibfield  {author} {\bibinfo {author} {\bibfnamefont {J.}~\bibnamefont
			{Fink}}, \bibinfo {author} {\bibfnamefont {R.}~\bibnamefont {Bianchetti}},
		\bibinfo {author} {\bibfnamefont {M.}~\bibnamefont {Baur}}, \bibinfo {author}
		{\bibfnamefont {M.}~\bibnamefont {G{\"o}ppl}}, \bibinfo {author}
		{\bibfnamefont {L.}~\bibnamefont {Steffen}}, \bibinfo {author} {\bibfnamefont
			{S.}~\bibnamefont {Filipp}}, \bibinfo {author} {\bibfnamefont {P.~J.}\
			\bibnamefont {Leek}}, \bibinfo {author} {\bibfnamefont {A.}~\bibnamefont
			{Blais}},\ and\ \bibinfo {author} {\bibfnamefont {A.}~\bibnamefont
			{Wallraff}},\ }\bibfield  {title} {\bibinfo {title} {Dressed collective qubit
			states and the tavis-cummings model in circuit qed},\ }\href
	{https://doi.org/10.1103/PhysRevLett.103.083601} {\bibfield  {journal}
		{\bibinfo  {journal} {Physical review letters}\ }\textbf {\bibinfo {volume}
			{103}},\ \bibinfo {pages} {083601} (\bibinfo {year} {2009})}\BibitemShut
	{NoStop}%
	\bibitem [{\citenamefont {Aspelmeyer}\ \emph {et~al.}(2014)\citenamefont
		{Aspelmeyer}, \citenamefont {Kippenberg},\ and\ \citenamefont
		{Marquardt}}]{aspelmeyer2014cavity}%
	\BibitemOpen
	\bibfield  {author} {\bibinfo {author} {\bibfnamefont {M.}~\bibnamefont
			{Aspelmeyer}}, \bibinfo {author} {\bibfnamefont {T.~J.}\ \bibnamefont
			{Kippenberg}},\ and\ \bibinfo {author} {\bibfnamefont {F.}~\bibnamefont
			{Marquardt}},\ }\bibfield  {title} {\bibinfo {title} {Cavity optomechanics},\
	}\href {https://doi.org/10.1103/revmodphys.86.1391} {\bibfield  {journal}
		{\bibinfo  {journal} {Reviews of Modern Physics}\ }\textbf {\bibinfo {volume}
			{86}},\ \bibinfo {pages} {1391} (\bibinfo {year} {2014})}\BibitemShut
	{NoStop}%
	\bibitem [{\citenamefont {O’Connell}\ \emph {et~al.}(2010)\citenamefont
		{O’Connell}, \citenamefont {Hofheinz}, \citenamefont {Ansmann},
		\citenamefont {Bialczak}, \citenamefont {Lenander}, \citenamefont {Lucero},
		\citenamefont {Neeley}, \citenamefont {Sank}, \citenamefont {Wang},
		\citenamefont {Weides} \emph {et~al.}}]{o2010quantum}%
	\BibitemOpen
	\bibfield  {author} {\bibinfo {author} {\bibfnamefont {A.~D.}\ \bibnamefont
			{O’Connell}}, \bibinfo {author} {\bibfnamefont {M.}~\bibnamefont
			{Hofheinz}}, \bibinfo {author} {\bibfnamefont {M.}~\bibnamefont {Ansmann}},
		\bibinfo {author} {\bibfnamefont {R.~C.}\ \bibnamefont {Bialczak}}, \bibinfo
		{author} {\bibfnamefont {M.}~\bibnamefont {Lenander}}, \bibinfo {author}
		{\bibfnamefont {E.}~\bibnamefont {Lucero}}, \bibinfo {author} {\bibfnamefont
			{M.}~\bibnamefont {Neeley}}, \bibinfo {author} {\bibfnamefont
			{D.}~\bibnamefont {Sank}}, \bibinfo {author} {\bibfnamefont {H.}~\bibnamefont
			{Wang}}, \bibinfo {author} {\bibfnamefont {M.}~\bibnamefont {Weides}}, \emph
		{et~al.},\ }\bibfield  {title} {\bibinfo {title} {Quantum ground state and
			single-phonon control of a mechanical resonator},\ }\href
	{https://doi.org/10.1038/nature08967} {\bibfield  {journal} {\bibinfo
			{journal} {Nature}\ }\textbf {\bibinfo {volume} {464}},\ \bibinfo {pages}
		{697} (\bibinfo {year} {2010})}\BibitemShut {NoStop}%
	\bibitem [{\citenamefont {Chan}\ \emph {et~al.}(2011)\citenamefont {Chan},
		\citenamefont {Alegre}, \citenamefont {Safavi-Naeini}, \citenamefont {Hill},
		\citenamefont {Krause}, \citenamefont {Gr{\"o}blacher}, \citenamefont
		{Aspelmeyer},\ and\ \citenamefont {Painter}}]{chan2011laser}%
	\BibitemOpen
	\bibfield  {author} {\bibinfo {author} {\bibfnamefont {J.}~\bibnamefont
			{Chan}}, \bibinfo {author} {\bibfnamefont {T.~M.}\ \bibnamefont {Alegre}},
		\bibinfo {author} {\bibfnamefont {A.~H.}\ \bibnamefont {Safavi-Naeini}},
		\bibinfo {author} {\bibfnamefont {J.~T.}\ \bibnamefont {Hill}}, \bibinfo
		{author} {\bibfnamefont {A.}~\bibnamefont {Krause}}, \bibinfo {author}
		{\bibfnamefont {S.}~\bibnamefont {Gr{\"o}blacher}}, \bibinfo {author}
		{\bibfnamefont {M.}~\bibnamefont {Aspelmeyer}},\ and\ \bibinfo {author}
		{\bibfnamefont {O.}~\bibnamefont {Painter}},\ }\bibfield  {title} {\bibinfo
		{title} {Laser cooling of a nanomechanical oscillator into its quantum ground
			state},\ }\href {https://doi.org/10.1038/nature10461} {\bibfield  {journal}
		{\bibinfo  {journal} {Nature}\ }\textbf {\bibinfo {volume} {478}},\ \bibinfo
		{pages} {89} (\bibinfo {year} {2011})}\BibitemShut {NoStop}%
	\bibitem [{\citenamefont {Verhagen}\ \emph {et~al.}(2012)\citenamefont
		{Verhagen}, \citenamefont {Del{\'e}glise}, \citenamefont {Weis},
		\citenamefont {Schliesser},\ and\ \citenamefont
		{Kippenberg}}]{verhagen2012quantum}%
	\BibitemOpen
	\bibfield  {author} {\bibinfo {author} {\bibfnamefont {E.}~\bibnamefont
			{Verhagen}}, \bibinfo {author} {\bibfnamefont {S.}~\bibnamefont
			{Del{\'e}glise}}, \bibinfo {author} {\bibfnamefont {S.}~\bibnamefont {Weis}},
		\bibinfo {author} {\bibfnamefont {A.}~\bibnamefont {Schliesser}},\ and\
		\bibinfo {author} {\bibfnamefont {T.~J.}\ \bibnamefont {Kippenberg}},\
	}\bibfield  {title} {\bibinfo {title} {Quantum-coherent coupling of a
			mechanical oscillator to an optical cavity mode},\ }\href
	{https://doi.org/10.1038/nature10787} {\bibfield  {journal} {\bibinfo
			{journal} {Nature}\ }\textbf {\bibinfo {volume} {482}},\ \bibinfo {pages}
		{63} (\bibinfo {year} {2012})}\BibitemShut {NoStop}%
	\bibitem [{\citenamefont {Safavi-Naeini}\ \emph {et~al.}(2013)\citenamefont
		{Safavi-Naeini}, \citenamefont {Gr{\"o}blacher}, \citenamefont {Hill},
		\citenamefont {Chan}, \citenamefont {Aspelmeyer},\ and\ \citenamefont
		{Painter}}]{safavi2013squeezed}%
	\BibitemOpen
	\bibfield  {author} {\bibinfo {author} {\bibfnamefont {A.~H.}\ \bibnamefont
			{Safavi-Naeini}}, \bibinfo {author} {\bibfnamefont {S.}~\bibnamefont
			{Gr{\"o}blacher}}, \bibinfo {author} {\bibfnamefont {J.~T.}\ \bibnamefont
			{Hill}}, \bibinfo {author} {\bibfnamefont {J.}~\bibnamefont {Chan}}, \bibinfo
		{author} {\bibfnamefont {M.}~\bibnamefont {Aspelmeyer}},\ and\ \bibinfo
		{author} {\bibfnamefont {O.}~\bibnamefont {Painter}},\ }\bibfield  {title}
	{\bibinfo {title} {Squeezed light from a silicon micromechanical resonator},\
	}\href {https://doi.org/10.1038/nature12307} {\bibfield  {journal} {\bibinfo
			{journal} {Nature}\ }\textbf {\bibinfo {volume} {500}},\ \bibinfo {pages}
		{185} (\bibinfo {year} {2013})}\BibitemShut {NoStop}%
	\bibitem [{\citenamefont {Huang}\ \emph {et~al.}(2024)\citenamefont {Huang},
		\citenamefont {Beccari}, \citenamefont {Engelsen},\ and\ \citenamefont
		{Kippenberg}}]{huang2024room}%
	\BibitemOpen
	\bibfield  {author} {\bibinfo {author} {\bibfnamefont {G.}~\bibnamefont
			{Huang}}, \bibinfo {author} {\bibfnamefont {A.}~\bibnamefont {Beccari}},
		\bibinfo {author} {\bibfnamefont {N.~J.}\ \bibnamefont {Engelsen}},\ and\
		\bibinfo {author} {\bibfnamefont {T.~J.}\ \bibnamefont {Kippenberg}},\
	}\bibfield  {title} {\bibinfo {title} {Room-temperature quantum optomechanics
			using an ultralow noise cavity},\ }\href
	{https://doi.org/10.1038/s41586-023-06997-3} {\bibfield  {journal} {\bibinfo
			{journal} {Nature}\ }\textbf {\bibinfo {volume} {626}},\ \bibinfo {pages}
		{512} (\bibinfo {year} {2024})}\BibitemShut {NoStop}%
	\bibitem [{\citenamefont {Teufel}\ \emph
		{et~al.}(2011{\natexlab{a}})\citenamefont {Teufel}, \citenamefont {Donner},
		\citenamefont {Li}, \citenamefont {Harlow}, \citenamefont {Allman},
		\citenamefont {Cicak}, \citenamefont {Sirois}, \citenamefont {Whittaker},
		\citenamefont {Lehnert},\ and\ \citenamefont
		{Simmonds}}]{teufel2011sideband}%
	\BibitemOpen
	\bibfield  {author} {\bibinfo {author} {\bibfnamefont {J.~D.}\ \bibnamefont
			{Teufel}}, \bibinfo {author} {\bibfnamefont {T.}~\bibnamefont {Donner}},
		\bibinfo {author} {\bibfnamefont {D.}~\bibnamefont {Li}}, \bibinfo {author}
		{\bibfnamefont {J.~W.}\ \bibnamefont {Harlow}}, \bibinfo {author}
		{\bibfnamefont {M.}~\bibnamefont {Allman}}, \bibinfo {author} {\bibfnamefont
			{K.}~\bibnamefont {Cicak}}, \bibinfo {author} {\bibfnamefont {A.~J.}\
			\bibnamefont {Sirois}}, \bibinfo {author} {\bibfnamefont {J.~D.}\
			\bibnamefont {Whittaker}}, \bibinfo {author} {\bibfnamefont {K.~W.}\
			\bibnamefont {Lehnert}},\ and\ \bibinfo {author} {\bibfnamefont {R.~W.}\
			\bibnamefont {Simmonds}},\ }\bibfield  {title} {\bibinfo {title} {Sideband
			cooling of micromechanical motion to the quantum ground state},\ }\href
	{https://doi.org/10.1038/nature10261} {\bibfield  {journal} {\bibinfo
			{journal} {Nature}\ }\textbf {\bibinfo {volume} {475}},\ \bibinfo {pages}
		{359} (\bibinfo {year} {2011}{\natexlab{a}})}\BibitemShut {NoStop}%
	\bibitem [{\citenamefont {Ockeloen-Korppi}\ \emph {et~al.}(2018)\citenamefont
		{Ockeloen-Korppi}, \citenamefont {Damsk{\"a}gg}, \citenamefont
		{Pirkkalainen}, \citenamefont {Asjad}, \citenamefont {Clerk} \emph
		{et~al.}}]{ockeloen2018stabilized}%
	\BibitemOpen
	\bibfield  {author} {\bibinfo {author} {\bibfnamefont {C.}~\bibnamefont
			{Ockeloen-Korppi}}, \bibinfo {author} {\bibfnamefont {E.}~\bibnamefont
			{Damsk{\"a}gg}}, \bibinfo {author} {\bibfnamefont {J.-M.}\ \bibnamefont
			{Pirkkalainen}}, \bibinfo {author} {\bibfnamefont {M.}~\bibnamefont {Asjad}},
		\bibinfo {author} {\bibfnamefont {A.}~\bibnamefont {Clerk}}, \emph {et~al.},\
	}\bibfield  {title} {\bibinfo {title} {Stabilized entanglement of massive
			mechanical oscillators},\ }\href {https://doi.org/10.1038/s41586-018-0038-x}
	{\bibfield  {journal} {\bibinfo  {journal} {Nature}\ }\textbf {\bibinfo
			{volume} {556}},\ \bibinfo {pages} {478} (\bibinfo {year}
		{2018})}\BibitemShut {NoStop}%
	\bibitem [{\citenamefont {Kotler}\ \emph {et~al.}(2021)\citenamefont {Kotler},
		\citenamefont {Peterson}, \citenamefont {Shojaee}, \citenamefont {Lecocq},
		\citenamefont {Cicak}, \citenamefont {Kwiatkowski}, \citenamefont {Geller},
		\citenamefont {Glancy}, \citenamefont {Knill}, \citenamefont {Simmonds} \emph
		{et~al.}}]{kotler2021direct}%
	\BibitemOpen
	\bibfield  {author} {\bibinfo {author} {\bibfnamefont {S.}~\bibnamefont
			{Kotler}}, \bibinfo {author} {\bibfnamefont {G.~A.}\ \bibnamefont
			{Peterson}}, \bibinfo {author} {\bibfnamefont {E.}~\bibnamefont {Shojaee}},
		\bibinfo {author} {\bibfnamefont {F.}~\bibnamefont {Lecocq}}, \bibinfo
		{author} {\bibfnamefont {K.}~\bibnamefont {Cicak}}, \bibinfo {author}
		{\bibfnamefont {A.}~\bibnamefont {Kwiatkowski}}, \bibinfo {author}
		{\bibfnamefont {S.}~\bibnamefont {Geller}}, \bibinfo {author} {\bibfnamefont
			{S.}~\bibnamefont {Glancy}}, \bibinfo {author} {\bibfnamefont
			{E.}~\bibnamefont {Knill}}, \bibinfo {author} {\bibfnamefont {R.~W.}\
			\bibnamefont {Simmonds}}, \emph {et~al.},\ }\bibfield  {title} {\bibinfo
		{title} {Direct observation of deterministic macroscopic entanglement},\
	}\href {https://doi.org/10.1126/science.abf2998} {\bibfield  {journal}
		{\bibinfo  {journal} {Science}\ }\textbf {\bibinfo {volume} {372}},\ \bibinfo
		{pages} {622} (\bibinfo {year} {2021})}\BibitemShut {NoStop}%
	\bibitem [{\citenamefont {Wollman}\ \emph {et~al.}(2015)\citenamefont
		{Wollman}, \citenamefont {Lei}, \citenamefont {Weinstein}, \citenamefont
		{Suh}, \citenamefont {Kronwald}, \citenamefont {Marquardt}, \citenamefont
		{Clerk},\ and\ \citenamefont {Schwab}}]{wollman2015quantum}%
	\BibitemOpen
	\bibfield  {author} {\bibinfo {author} {\bibfnamefont {E.~E.}\ \bibnamefont
			{Wollman}}, \bibinfo {author} {\bibfnamefont {C.}~\bibnamefont {Lei}},
		\bibinfo {author} {\bibfnamefont {A.}~\bibnamefont {Weinstein}}, \bibinfo
		{author} {\bibfnamefont {J.}~\bibnamefont {Suh}}, \bibinfo {author}
		{\bibfnamefont {A.}~\bibnamefont {Kronwald}}, \bibinfo {author}
		{\bibfnamefont {F.}~\bibnamefont {Marquardt}}, \bibinfo {author}
		{\bibfnamefont {A.~A.}\ \bibnamefont {Clerk}},\ and\ \bibinfo {author}
		{\bibfnamefont {K.}~\bibnamefont {Schwab}},\ }\bibfield  {title} {\bibinfo
		{title} {Quantum squeezing of motion in a mechanical resonator},\ }\href
	{https://doi.org/10.1126/science.aac5138} {\bibfield  {journal} {\bibinfo
			{journal} {Science}\ }\textbf {\bibinfo {volume} {349}},\ \bibinfo {pages}
		{952} (\bibinfo {year} {2015})}\BibitemShut {NoStop}%
	\bibitem [{\citenamefont {Youssefi}\ \emph {et~al.}(2023)\citenamefont
		{Youssefi}, \citenamefont {Kono}, \citenamefont {Chegnizadeh},\ and\
		\citenamefont {Kippenberg}}]{youssefi2023squeezed}%
	\BibitemOpen
	\bibfield  {author} {\bibinfo {author} {\bibfnamefont {A.}~\bibnamefont
			{Youssefi}}, \bibinfo {author} {\bibfnamefont {S.}~\bibnamefont {Kono}},
		\bibinfo {author} {\bibfnamefont {M.}~\bibnamefont {Chegnizadeh}},\ and\
		\bibinfo {author} {\bibfnamefont {T.~J.}\ \bibnamefont {Kippenberg}},\
	}\bibfield  {title} {\bibinfo {title} {A squeezed mechanical oscillator with
			millisecond quantum decoherence},\ }\href
	{https://doi.org/10.1038/s41567-023-02135-y} {\bibfield  {journal} {\bibinfo
			{journal} {Nature Physics}\ }\textbf {\bibinfo {volume} {19}},\ \bibinfo
		{pages} {1697} (\bibinfo {year} {2023})}\BibitemShut {NoStop}%
	\bibitem [{\citenamefont {Youssefi}\ \emph {et~al.}(2022)\citenamefont
		{Youssefi}, \citenamefont {Kono}, \citenamefont {Bancora}, \citenamefont
		{Chegnizadeh}, \citenamefont {Pan}, \citenamefont {Vovk},\ and\ \citenamefont
		{Kippenberg}}]{youssefiTopological2022}%
	\BibitemOpen
	\bibfield  {author} {\bibinfo {author} {\bibfnamefont {A.}~\bibnamefont
			{Youssefi}}, \bibinfo {author} {\bibfnamefont {S.}~\bibnamefont {Kono}},
		\bibinfo {author} {\bibfnamefont {A.}~\bibnamefont {Bancora}}, \bibinfo
		{author} {\bibfnamefont {M.}~\bibnamefont {Chegnizadeh}}, \bibinfo {author}
		{\bibfnamefont {J.}~\bibnamefont {Pan}}, \bibinfo {author} {\bibfnamefont
			{T.}~\bibnamefont {Vovk}},\ and\ \bibinfo {author} {\bibfnamefont {T.~J.}\
			\bibnamefont {Kippenberg}},\ }\bibfield  {title} {\bibinfo {title}
		{Topological lattices realized in superconducting circuit optomechanics},\
	}\href {https://doi.org/10.1038/s41586-022-05367-9} {\bibfield  {journal}
		{\bibinfo  {journal} {Nature}\ }\textbf {\bibinfo {volume} {612}},\ \bibinfo
		{pages} {666} (\bibinfo {year} {2022})}\BibitemShut {NoStop}%
	\bibitem [{\citenamefont {Xuereb}\ \emph {et~al.}(2012)\citenamefont {Xuereb},
		\citenamefont {Genes},\ and\ \citenamefont {Dantan}}]{xuereb2012strong}%
	\BibitemOpen
	\bibfield  {author} {\bibinfo {author} {\bibfnamefont {A.}~\bibnamefont
			{Xuereb}}, \bibinfo {author} {\bibfnamefont {C.}~\bibnamefont {Genes}},\ and\
		\bibinfo {author} {\bibfnamefont {A.}~\bibnamefont {Dantan}},\ }\bibfield
	{title} {\bibinfo {title} {Strong coupling and long-range collective
			interactions in optomechanical arrays},\ }\href
	{https://doi.org/10.1103/PhysRevLett.109.223601} {\bibfield  {journal}
		{\bibinfo  {journal} {Physical review letters}\ }\textbf {\bibinfo {volume}
			{109}},\ \bibinfo {pages} {223601} (\bibinfo {year} {2012})}\BibitemShut
	{NoStop}%
	\bibitem [{\citenamefont {Heinrich}\ \emph {et~al.}(2011)\citenamefont
		{Heinrich}, \citenamefont {Ludwig}, \citenamefont {Qian}, \citenamefont
		{Kubala},\ and\ \citenamefont {Marquardt}}]{heinrichCollective2011}%
	\BibitemOpen
	\bibfield  {author} {\bibinfo {author} {\bibfnamefont {G.}~\bibnamefont
			{Heinrich}}, \bibinfo {author} {\bibfnamefont {M.}~\bibnamefont {Ludwig}},
		\bibinfo {author} {\bibfnamefont {J.}~\bibnamefont {Qian}}, \bibinfo {author}
		{\bibfnamefont {B.}~\bibnamefont {Kubala}},\ and\ \bibinfo {author}
		{\bibfnamefont {F.}~\bibnamefont {Marquardt}},\ }\bibfield  {title} {\bibinfo
		{title} {Collective dynamics in optomechanical arrays},\ }\href
	{https://doi.org/10.1103/PhysRevLett.107.043603} {\bibfield  {journal}
		{\bibinfo  {journal} {Physical Review Letters}\ }\textbf {\bibinfo {volume}
			{107}},\ \bibinfo {pages} {043603} (\bibinfo {year} {2011})}\BibitemShut
	{NoStop}%
	\bibitem [{\citenamefont {Peano}\ \emph {et~al.}(2015)\citenamefont {Peano},
		\citenamefont {Brendel}, \citenamefont {Schmidt},\ and\ \citenamefont
		{Marquardt}}]{peano2015topological}%
	\BibitemOpen
	\bibfield  {author} {\bibinfo {author} {\bibfnamefont {V.}~\bibnamefont
			{Peano}}, \bibinfo {author} {\bibfnamefont {C.}~\bibnamefont {Brendel}},
		\bibinfo {author} {\bibfnamefont {M.}~\bibnamefont {Schmidt}},\ and\ \bibinfo
		{author} {\bibfnamefont {F.}~\bibnamefont {Marquardt}},\ }\bibfield  {title}
	{\bibinfo {title} {Topological phases of sound and light},\ }\href
	{https://doi.org/10.1103/PhysRevX.5.031011} {\bibfield  {journal} {\bibinfo
			{journal} {Physical Review X}\ }\textbf {\bibinfo {volume} {5}},\ \bibinfo
		{pages} {031011} (\bibinfo {year} {2015})}\BibitemShut {NoStop}%
	\bibitem [{\citenamefont {Shkarin}\ \emph {et~al.}(2014)\citenamefont
		{Shkarin}, \citenamefont {Flowers-Jacobs}, \citenamefont {Hoch},
		\citenamefont {Kashkanova}, \citenamefont {Deutsch}, \citenamefont
		{Reichel},\ and\ \citenamefont {Harris}}]{shkarin2014optically}%
	\BibitemOpen
	\bibfield  {author} {\bibinfo {author} {\bibfnamefont {A.}~\bibnamefont
			{Shkarin}}, \bibinfo {author} {\bibfnamefont {N.}~\bibnamefont
			{Flowers-Jacobs}}, \bibinfo {author} {\bibfnamefont {S.}~\bibnamefont
			{Hoch}}, \bibinfo {author} {\bibfnamefont {A.}~\bibnamefont {Kashkanova}},
		\bibinfo {author} {\bibfnamefont {C.}~\bibnamefont {Deutsch}}, \bibinfo
		{author} {\bibfnamefont {J.}~\bibnamefont {Reichel}},\ and\ \bibinfo {author}
		{\bibfnamefont {J.}~\bibnamefont {Harris}},\ }\bibfield  {title} {\bibinfo
		{title} {Optically mediated hybridization between two mechanical modes},\
	}\href {https://doi.org/10.1103/physrevlett.112.013602} {\bibfield  {journal}
		{\bibinfo  {journal} {Physical review letters}\ }\textbf {\bibinfo {volume}
			{112}},\ \bibinfo {pages} {013602} (\bibinfo {year} {2014})}\BibitemShut
	{NoStop}%
	\bibitem [{\citenamefont {Massel}\ \emph {et~al.}(2012)\citenamefont {Massel},
		\citenamefont {Cho}, \citenamefont {Pirkkalainen}, \citenamefont {Hakonen},
		\citenamefont {Heikkil{\"a}},\ and\ \citenamefont
		{Sillanp{\"a}{\"a}}}]{massel2012multimode}%
	\BibitemOpen
	\bibfield  {author} {\bibinfo {author} {\bibfnamefont {F.}~\bibnamefont
			{Massel}}, \bibinfo {author} {\bibfnamefont {S.~U.}\ \bibnamefont {Cho}},
		\bibinfo {author} {\bibfnamefont {J.-M.}\ \bibnamefont {Pirkkalainen}},
		\bibinfo {author} {\bibfnamefont {P.~J.}\ \bibnamefont {Hakonen}}, \bibinfo
		{author} {\bibfnamefont {T.~T.}\ \bibnamefont {Heikkil{\"a}}},\ and\ \bibinfo
		{author} {\bibfnamefont {M.~A.}\ \bibnamefont {Sillanp{\"a}{\"a}}},\
	}\bibfield  {title} {\bibinfo {title} {Multimode circuit optomechanics near
			the quantum limit},\ }\href {https://doi.org/10.1038/ncomms1993} {\bibfield
		{journal} {\bibinfo  {journal} {Nature communications}\ }\textbf {\bibinfo
			{volume} {3}},\ \bibinfo {pages} {987} (\bibinfo {year} {2012})}\BibitemShut
	{NoStop}%
	\bibitem [{\citenamefont {Ockeloen-Korppi}\ \emph {et~al.}(2019)\citenamefont
		{Ockeloen-Korppi}, \citenamefont {Gely}, \citenamefont {Damsk{\"a}gg},
		\citenamefont {Jenkins}, \citenamefont {Steele},\ and\ \citenamefont
		{Sillanp{\"a}{\"a}}}]{ockeloen2019sideband}%
	\BibitemOpen
	\bibfield  {author} {\bibinfo {author} {\bibfnamefont {C.}~\bibnamefont
			{Ockeloen-Korppi}}, \bibinfo {author} {\bibfnamefont {M.}~\bibnamefont
			{Gely}}, \bibinfo {author} {\bibfnamefont {E.}~\bibnamefont {Damsk{\"a}gg}},
		\bibinfo {author} {\bibfnamefont {M.}~\bibnamefont {Jenkins}}, \bibinfo
		{author} {\bibfnamefont {G.}~\bibnamefont {Steele}},\ and\ \bibinfo {author}
		{\bibfnamefont {M.}~\bibnamefont {Sillanp{\"a}{\"a}}},\ }\bibfield  {title}
	{\bibinfo {title} {Sideband cooling of nearly degenerate micromechanical
			oscillators in a multimode optomechanical system},\ }\href
	{https://doi.org/10.1103/PhysRevA.99.023826} {\bibfield  {journal} {\bibinfo
			{journal} {Physical Review A}\ }\textbf {\bibinfo {volume} {99}},\ \bibinfo
		{pages} {023826} (\bibinfo {year} {2019})}\BibitemShut {NoStop}%
	\bibitem [{\citenamefont {Kharel}\ \emph {et~al.}(2022)\citenamefont {Kharel},
		\citenamefont {Chu}, \citenamefont {Mason}, \citenamefont {Kittlaus},
		\citenamefont {Otterstrom}, \citenamefont {Gertler},\ and\ \citenamefont
		{Rakich}}]{kharel2022multimode}%
	\BibitemOpen
	\bibfield  {author} {\bibinfo {author} {\bibfnamefont {P.}~\bibnamefont
			{Kharel}}, \bibinfo {author} {\bibfnamefont {Y.}~\bibnamefont {Chu}},
		\bibinfo {author} {\bibfnamefont {D.}~\bibnamefont {Mason}}, \bibinfo
		{author} {\bibfnamefont {E.~A.}\ \bibnamefont {Kittlaus}}, \bibinfo {author}
		{\bibfnamefont {N.~T.}\ \bibnamefont {Otterstrom}}, \bibinfo {author}
		{\bibfnamefont {S.}~\bibnamefont {Gertler}},\ and\ \bibinfo {author}
		{\bibfnamefont {P.~T.}\ \bibnamefont {Rakich}},\ }\bibfield  {title}
	{\bibinfo {title} {Multimode strong coupling in cavity optomechanics},\
	}\href {https://doi.org/10.1103/physrevapplied.18.024054} {\bibfield
		{journal} {\bibinfo  {journal} {Physical Review Applied}\ }\textbf {\bibinfo
			{volume} {18}},\ \bibinfo {pages} {024054} (\bibinfo {year}
		{2022})}\BibitemShut {NoStop}%
	\bibitem [{\citenamefont {Xu}\ \emph {et~al.}(2016)\citenamefont {Xu},
		\citenamefont {Mason}, \citenamefont {Jiang},\ and\ \citenamefont
		{Harris}}]{xu2016topological}%
	\BibitemOpen
	\bibfield  {author} {\bibinfo {author} {\bibfnamefont {H.}~\bibnamefont
			{Xu}}, \bibinfo {author} {\bibfnamefont {D.}~\bibnamefont {Mason}}, \bibinfo
		{author} {\bibfnamefont {L.}~\bibnamefont {Jiang}},\ and\ \bibinfo {author}
		{\bibfnamefont {J.}~\bibnamefont {Harris}},\ }\bibfield  {title} {\bibinfo
		{title} {Topological energy transfer in an optomechanical system with
			exceptional points},\ }\href {https://doi.org/10.1038/nature18604} {\bibfield
		{journal} {\bibinfo  {journal} {Nature}\ }\textbf {\bibinfo {volume}
			{537}},\ \bibinfo {pages} {80} (\bibinfo {year} {2016})}\BibitemShut
	{NoStop}%
	\bibitem [{\citenamefont {Patil}\ \emph {et~al.}(2022)\citenamefont {Patil},
		\citenamefont {H{\"o}ller}, \citenamefont {Henry}, \citenamefont {Guria},
		\citenamefont {Zhang}, \citenamefont {Jiang}, \citenamefont {Kralj},
		\citenamefont {Read},\ and\ \citenamefont {Harris}}]{patil2022measuring}%
	\BibitemOpen
	\bibfield  {author} {\bibinfo {author} {\bibfnamefont {Y.~S.}\ \bibnamefont
			{Patil}}, \bibinfo {author} {\bibfnamefont {J.}~\bibnamefont {H{\"o}ller}},
		\bibinfo {author} {\bibfnamefont {P.~A.}\ \bibnamefont {Henry}}, \bibinfo
		{author} {\bibfnamefont {C.}~\bibnamefont {Guria}}, \bibinfo {author}
		{\bibfnamefont {Y.}~\bibnamefont {Zhang}}, \bibinfo {author} {\bibfnamefont
			{L.}~\bibnamefont {Jiang}}, \bibinfo {author} {\bibfnamefont
			{N.}~\bibnamefont {Kralj}}, \bibinfo {author} {\bibfnamefont
			{N.}~\bibnamefont {Read}},\ and\ \bibinfo {author} {\bibfnamefont {J.~G.}\
			\bibnamefont {Harris}},\ }\bibfield  {title} {\bibinfo {title} {Measuring the
			knot of non-hermitian degeneracies and non-commuting braids},\ }\href
	{https://doi.org/10.1038/s41586-022-04796-w} {\bibfield  {journal} {\bibinfo
			{journal} {Nature}\ }\textbf {\bibinfo {volume} {607}},\ \bibinfo {pages}
		{271} (\bibinfo {year} {2022})}\BibitemShut {NoStop}%
	\bibitem [{\citenamefont {Rieser}\ \emph {et~al.}(2022)\citenamefont {Rieser},
		\citenamefont {Ciampini}, \citenamefont {Rudolph}, \citenamefont {Kiesel},
		\citenamefont {Hornberger}, \citenamefont {Stickler}, \citenamefont
		{Aspelmeyer},\ and\ \citenamefont {Deli{\'c}}}]{rieser2022tunable}%
	\BibitemOpen
	\bibfield  {author} {\bibinfo {author} {\bibfnamefont {J.}~\bibnamefont
			{Rieser}}, \bibinfo {author} {\bibfnamefont {M.~A.}\ \bibnamefont
			{Ciampini}}, \bibinfo {author} {\bibfnamefont {H.}~\bibnamefont {Rudolph}},
		\bibinfo {author} {\bibfnamefont {N.}~\bibnamefont {Kiesel}}, \bibinfo
		{author} {\bibfnamefont {K.}~\bibnamefont {Hornberger}}, \bibinfo {author}
		{\bibfnamefont {B.~A.}\ \bibnamefont {Stickler}}, \bibinfo {author}
		{\bibfnamefont {M.}~\bibnamefont {Aspelmeyer}},\ and\ \bibinfo {author}
		{\bibfnamefont {U.}~\bibnamefont {Deli{\'c}}},\ }\bibfield  {title} {\bibinfo
		{title} {Tunable light-induced dipole-dipole interaction between optically
			levitated nanoparticles},\ }\href {https://doi.org/10.1126/science.abp9941}
	{\bibfield  {journal} {\bibinfo  {journal} {Science}\ }\textbf {\bibinfo
			{volume} {377}},\ \bibinfo {pages} {987} (\bibinfo {year}
		{2022})}\BibitemShut {NoStop}%
	\bibitem [{\citenamefont {Vijayan}\ \emph {et~al.}(2024)\citenamefont
		{Vijayan}, \citenamefont {Piotrowski}, \citenamefont {Gonzalez-Ballestero},
		\citenamefont {Weber}, \citenamefont {Romero-Isart},\ and\ \citenamefont
		{Novotny}}]{vijayan2024cavity}%
	\BibitemOpen
	\bibfield  {author} {\bibinfo {author} {\bibfnamefont {J.}~\bibnamefont
			{Vijayan}}, \bibinfo {author} {\bibfnamefont {J.}~\bibnamefont {Piotrowski}},
		\bibinfo {author} {\bibfnamefont {C.}~\bibnamefont {Gonzalez-Ballestero}},
		\bibinfo {author} {\bibfnamefont {K.}~\bibnamefont {Weber}}, \bibinfo
		{author} {\bibfnamefont {O.}~\bibnamefont {Romero-Isart}},\ and\ \bibinfo
		{author} {\bibfnamefont {L.}~\bibnamefont {Novotny}},\ }\bibfield  {title}
	{\bibinfo {title} {Cavity-mediated long-range interactions in levitated
			optomechanics},\ }\href {https://doi.org/10.1038/s41567-024-02405-3}
	{\bibfield  {journal} {\bibinfo  {journal} {Nature Physics}\ }\textbf
		{\bibinfo {volume} {20}},\ \bibinfo {pages} {1} (\bibinfo {year}
		{2024})}\BibitemShut {NoStop}%
	\bibitem [{SI()}]{SI}%
	\BibitemOpen
	\href@noop {} {\bibinfo {title} {see supplementary materials.}}\BibitemShut
	{Stop}%
	\bibitem [{\citenamefont {Weinstein}\ \emph {et~al.}(2014)\citenamefont
		{Weinstein}, \citenamefont {Lei}, \citenamefont {Wollman}, \citenamefont
		{Suh}, \citenamefont {Metelmann}, \citenamefont {Clerk},\ and\ \citenamefont
		{Schwab}}]{weinstein2014observation}%
	\BibitemOpen
	\bibfield  {author} {\bibinfo {author} {\bibfnamefont {A.}~\bibnamefont
			{Weinstein}}, \bibinfo {author} {\bibfnamefont {C.}~\bibnamefont {Lei}},
		\bibinfo {author} {\bibfnamefont {E.}~\bibnamefont {Wollman}}, \bibinfo
		{author} {\bibfnamefont {J.}~\bibnamefont {Suh}}, \bibinfo {author}
		{\bibfnamefont {A.}~\bibnamefont {Metelmann}}, \bibinfo {author}
		{\bibfnamefont {A.}~\bibnamefont {Clerk}},\ and\ \bibinfo {author}
		{\bibfnamefont {K.}~\bibnamefont {Schwab}},\ }\bibfield  {title} {\bibinfo
		{title} {Observation and interpretation of motional sideband asymmetry in a
			quantum electromechanical device},\ }\href
	{https://doi.org/10.1103/PhysRevX.4.041003} {\bibfield  {journal} {\bibinfo
			{journal} {Physical Review X}\ }\textbf {\bibinfo {volume} {4}},\ \bibinfo
		{pages} {041003} (\bibinfo {year} {2014})}\BibitemShut {NoStop}%
	\bibitem [{\citenamefont {Weis}\ \emph {et~al.}(2010)\citenamefont {Weis},
		\citenamefont {Rivi{\`e}re}, \citenamefont {Del{\'e}glise}, \citenamefont
		{Gavartin}, \citenamefont {Arcizet}, \citenamefont {Schliesser},\ and\
		\citenamefont {Kippenberg}}]{weis2010optomechanically}%
	\BibitemOpen
	\bibfield  {author} {\bibinfo {author} {\bibfnamefont {S.}~\bibnamefont
			{Weis}}, \bibinfo {author} {\bibfnamefont {R.}~\bibnamefont {Rivi{\`e}re}},
		\bibinfo {author} {\bibfnamefont {S.}~\bibnamefont {Del{\'e}glise}}, \bibinfo
		{author} {\bibfnamefont {E.}~\bibnamefont {Gavartin}}, \bibinfo {author}
		{\bibfnamefont {O.}~\bibnamefont {Arcizet}}, \bibinfo {author} {\bibfnamefont
			{A.}~\bibnamefont {Schliesser}},\ and\ \bibinfo {author} {\bibfnamefont
			{T.~J.}\ \bibnamefont {Kippenberg}},\ }\bibfield  {title} {\bibinfo {title}
		{Optomechanically induced transparency},\ }\href
	{https://doi.org/10.1126/science.1195596} {\bibfield  {journal} {\bibinfo
			{journal} {Science}\ }\textbf {\bibinfo {volume} {330}},\ \bibinfo {pages}
		{1520} (\bibinfo {year} {2010})}\BibitemShut {NoStop}%
	\bibitem [{\citenamefont {Bernier}\ \emph {et~al.}(2017)\citenamefont
		{Bernier}, \citenamefont {Toth}, \citenamefont {Koottandavida}, \citenamefont
		{Ioannou}, \citenamefont {Malz}, \citenamefont {Nunnenkamp}, \citenamefont
		{Feofanov},\ and\ \citenamefont {Kippenberg}}]{bernier2017nonreciprocal}%
	\BibitemOpen
	\bibfield  {author} {\bibinfo {author} {\bibfnamefont {N.~R.}\ \bibnamefont
			{Bernier}}, \bibinfo {author} {\bibfnamefont {L.~D.}\ \bibnamefont {Toth}},
		\bibinfo {author} {\bibfnamefont {A.}~\bibnamefont {Koottandavida}}, \bibinfo
		{author} {\bibfnamefont {M.~A.}\ \bibnamefont {Ioannou}}, \bibinfo {author}
		{\bibfnamefont {D.}~\bibnamefont {Malz}}, \bibinfo {author} {\bibfnamefont
			{A.}~\bibnamefont {Nunnenkamp}}, \bibinfo {author} {\bibfnamefont
			{A.}~\bibnamefont {Feofanov}},\ and\ \bibinfo {author} {\bibfnamefont
			{T.}~\bibnamefont {Kippenberg}},\ }\bibfield  {title} {\bibinfo {title}
		{Nonreciprocal reconfigurable microwave optomechanical circuit},\ }\href
	{https://doi.org/10.1038/s41467-017-00447-1} {\bibfield  {journal} {\bibinfo
			{journal} {Nature communications}\ }\textbf {\bibinfo {volume} {8}},\
		\bibinfo {pages} {604} (\bibinfo {year} {2017})}\BibitemShut {NoStop}%
	\bibitem [{\citenamefont {Teufel}\ \emph
		{et~al.}(2011{\natexlab{b}})\citenamefont {Teufel}, \citenamefont {Li},
		\citenamefont {Allman}, \citenamefont {Cicak}, \citenamefont {Sirois},
		\citenamefont {Whittaker},\ and\ \citenamefont
		{Simmonds}}]{teufel2011circuit}%
	\BibitemOpen
	\bibfield  {author} {\bibinfo {author} {\bibfnamefont {J.~D.}\ \bibnamefont
			{Teufel}}, \bibinfo {author} {\bibfnamefont {D.}~\bibnamefont {Li}}, \bibinfo
		{author} {\bibfnamefont {M.~S.}\ \bibnamefont {Allman}}, \bibinfo {author}
		{\bibfnamefont {K.}~\bibnamefont {Cicak}}, \bibinfo {author} {\bibfnamefont
			{A.}~\bibnamefont {Sirois}}, \bibinfo {author} {\bibfnamefont {J.~D.}\
			\bibnamefont {Whittaker}},\ and\ \bibinfo {author} {\bibfnamefont
			{R.}~\bibnamefont {Simmonds}},\ }\bibfield  {title} {\bibinfo {title}
		{Circuit cavity electromechanics in the strong-coupling regime},\ }\href
	{https://doi.org/10.1038/nature09898} {\bibfield  {journal} {\bibinfo
			{journal} {Nature}\ }\textbf {\bibinfo {volume} {471}},\ \bibinfo {pages}
		{204} (\bibinfo {year} {2011}{\natexlab{b}})}\BibitemShut {NoStop}%
	\bibitem [{\citenamefont {Safavi-Naeini}\ \emph {et~al.}(2012)\citenamefont
		{Safavi-Naeini}, \citenamefont {Chan}, \citenamefont {Hill}, \citenamefont
		{Alegre}, \citenamefont {Krause},\ and\ \citenamefont
		{Painter}}]{safavi2012observation}%
	\BibitemOpen
	\bibfield  {author} {\bibinfo {author} {\bibfnamefont {A.~H.}\ \bibnamefont
			{Safavi-Naeini}}, \bibinfo {author} {\bibfnamefont {J.}~\bibnamefont {Chan}},
		\bibinfo {author} {\bibfnamefont {J.~T.}\ \bibnamefont {Hill}}, \bibinfo
		{author} {\bibfnamefont {T.~P.~M.}\ \bibnamefont {Alegre}}, \bibinfo {author}
		{\bibfnamefont {A.}~\bibnamefont {Krause}},\ and\ \bibinfo {author}
		{\bibfnamefont {O.}~\bibnamefont {Painter}},\ }\bibfield  {title} {\bibinfo
		{title} {Observation of quantum motion of a nanomechanical resonator},\
	}\href {https://doi.org/10.1103/PhysRevLett.108.033602} {\bibfield  {journal}
		{\bibinfo  {journal} {Physical Review Letters}\ }\textbf {\bibinfo {volume}
			{108}},\ \bibinfo {pages} {033602} (\bibinfo {year} {2012})}\BibitemShut
	{NoStop}%
	\bibitem [{\citenamefont {Qiu}\ \emph {et~al.}(2020)\citenamefont {Qiu},
		\citenamefont {Shomroni}, \citenamefont {Seidler},\ and\ \citenamefont
		{Kippenberg}}]{qiu2020laser}%
	\BibitemOpen
	\bibfield  {author} {\bibinfo {author} {\bibfnamefont {L.}~\bibnamefont
			{Qiu}}, \bibinfo {author} {\bibfnamefont {I.}~\bibnamefont {Shomroni}},
		\bibinfo {author} {\bibfnamefont {P.}~\bibnamefont {Seidler}},\ and\ \bibinfo
		{author} {\bibfnamefont {T.~J.}\ \bibnamefont {Kippenberg}},\ }\bibfield
	{title} {\bibinfo {title} {Laser cooling of a nanomechanical oscillator to
			its zero-point energy},\ }\href
	{https://doi.org/10.1103/PhysRevLett.124.173601} {\bibfield  {journal}
		{\bibinfo  {journal} {Physical review letters}\ }\textbf {\bibinfo {volume}
			{124}},\ \bibinfo {pages} {173601} (\bibinfo {year} {2020})}\BibitemShut
	{NoStop}%
	\bibitem [{\citenamefont {Vitali}\ \emph {et~al.}(2007)\citenamefont {Vitali},
		\citenamefont {Mancini},\ and\ \citenamefont
		{Tombesi}}]{vitali2007stationary}%
	\BibitemOpen
	\bibfield  {author} {\bibinfo {author} {\bibfnamefont {D.}~\bibnamefont
			{Vitali}}, \bibinfo {author} {\bibfnamefont {S.}~\bibnamefont {Mancini}},\
		and\ \bibinfo {author} {\bibfnamefont {P.}~\bibnamefont {Tombesi}},\
	}\bibfield  {title} {\bibinfo {title} {Stationary entanglement between two
			movable mirrors in a classically driven fabry--perot cavity},\ }\href
	{https://doi.org/10.1088/1751-8113/40/28/S14} {\bibfield  {journal} {\bibinfo
			{journal} {Journal of Physics A: Mathematical and Theoretical}\ }\textbf
		{\bibinfo {volume} {40}},\ \bibinfo {pages} {8055} (\bibinfo {year}
		{2007})}\BibitemShut {NoStop}%
	\bibitem [{\citenamefont {Hartmann}\ and\ \citenamefont
		{Plenio}(2008)}]{hartmann2008steady}%
	\BibitemOpen
	\bibfield  {author} {\bibinfo {author} {\bibfnamefont {M.~J.}\ \bibnamefont
			{Hartmann}}\ and\ \bibinfo {author} {\bibfnamefont {M.~B.}\ \bibnamefont
			{Plenio}},\ }\bibfield  {title} {\bibinfo {title} {Steady state entanglement
			in the mechanical vibrations of two dielectric membranes},\ }\href
	{https://doi.org/10.1103/PhysRevLett.101.200503} {\bibfield  {journal}
		{\bibinfo  {journal} {Physical Review Letters}\ }\textbf {\bibinfo {volume}
			{101}},\ \bibinfo {pages} {200503} (\bibinfo {year} {2008})}\BibitemShut
	{NoStop}%
	\bibitem [{\citenamefont {Lai}\ \emph {et~al.}(2022)\citenamefont {Lai},
		\citenamefont {Liao}, \citenamefont {Miranowicz},\ and\ \citenamefont
		{Nori}}]{lai2022noise}%
	\BibitemOpen
	\bibfield  {author} {\bibinfo {author} {\bibfnamefont {D.-G.}\ \bibnamefont
			{Lai}}, \bibinfo {author} {\bibfnamefont {J.-Q.}\ \bibnamefont {Liao}},
		\bibinfo {author} {\bibfnamefont {A.}~\bibnamefont {Miranowicz}},\ and\
		\bibinfo {author} {\bibfnamefont {F.}~\bibnamefont {Nori}},\ }\bibfield
	{title} {\bibinfo {title} {Noise-tolerant optomechanical entanglement via
			synthetic magnetism},\ }\href
	{https://doi.org/10.1103/PhysRevLett.129.063602} {\bibfield  {journal}
		{\bibinfo  {journal} {Physical Review Letters}\ }\textbf {\bibinfo {volume}
			{129}},\ \bibinfo {pages} {063602} (\bibinfo {year} {2022})}\BibitemShut
	{NoStop}%
	\bibitem [{\citenamefont {Leggett}\ \emph {et~al.}(1987)\citenamefont
		{Leggett}, \citenamefont {Chakravarty}, \citenamefont {Dorsey}, \citenamefont
		{Fisher}, \citenamefont {Garg},\ and\ \citenamefont
		{Zwerger}}]{leggett1987dynamics}%
	\BibitemOpen
	\bibfield  {author} {\bibinfo {author} {\bibfnamefont {A.~J.}\ \bibnamefont
			{Leggett}}, \bibinfo {author} {\bibfnamefont {S.}~\bibnamefont
			{Chakravarty}}, \bibinfo {author} {\bibfnamefont {A.~T.}\ \bibnamefont
			{Dorsey}}, \bibinfo {author} {\bibfnamefont {M.~P.}\ \bibnamefont {Fisher}},
		\bibinfo {author} {\bibfnamefont {A.}~\bibnamefont {Garg}},\ and\ \bibinfo
		{author} {\bibfnamefont {W.}~\bibnamefont {Zwerger}},\ }\bibfield  {title}
	{\bibinfo {title} {Dynamics of the dissipative two-state system},\ }\href
	{https://doi.org/10.1103/RevModPhys.59.1} {\bibfield  {journal} {\bibinfo
			{journal} {Reviews of Modern Physics}\ }\textbf {\bibinfo {volume} {59}},\
		\bibinfo {pages} {1} (\bibinfo {year} {1987})}\BibitemShut {NoStop}%
	\bibitem [{\citenamefont {Chegnizadeh}\ \emph {et~al.}(2024)\citenamefont
		{Chegnizadeh}, \citenamefont {Scigliuzzo}, \citenamefont {Youssefi},
		\citenamefont {Kono}, \citenamefont {Guzovskii},\ and\ \citenamefont
		{Kippenberg}}]{dataset}%
	\BibitemOpen
	\bibfield  {author} {\bibinfo {author} {\bibfnamefont {M.}~\bibnamefont
			{Chegnizadeh}}, \bibinfo {author} {\bibfnamefont {M.}~\bibnamefont
			{Scigliuzzo}}, \bibinfo {author} {\bibfnamefont {A.}~\bibnamefont
			{Youssefi}}, \bibinfo {author} {\bibfnamefont {S.}~\bibnamefont {Kono}},
		\bibinfo {author} {\bibfnamefont {E.}~\bibnamefont {Guzovskii}},\ and\
		\bibinfo {author} {\bibfnamefont {T.}~\bibnamefont {Kippenberg}},\ }\href
	{https://doi.org/10.5281/zenodo.13941432} {\bibinfo {title} {{Data and codes
				for the article "Quantum collective motion of macroscopic mechanical
				oscillators"}}},\ \bibinfo {howpublished} {Zenodo} (\bibinfo {year}
	{2024})\BibitemShut {NoStop}%
	\bibitem [{\citenamefont {Devoret}(1995)}]{devoret1995quantum}%
	\BibitemOpen
	\bibfield  {author} {\bibinfo {author} {\bibfnamefont {M.~H. e.~a.}\
			\bibnamefont {Devoret}},\ }\bibfield  {title} {\bibinfo {title} {Quantum
			fluctuations in electrical circuits},\ }\href@noop {} {\bibfield  {journal}
		{\bibinfo  {journal} {Les Houches, Session LXIII}\ } (\bibinfo {year}
		{1995})}\BibitemShut {NoStop}%
	\bibitem [{\citenamefont {Blais}\ \emph {et~al.}(2021)\citenamefont {Blais},
		\citenamefont {Grimsmo},\ and\ \citenamefont {Wallraff}}]{blais2021circuit}%
	\BibitemOpen
	\bibfield  {author} {\bibinfo {author} {\bibfnamefont {A.}~\bibnamefont
			{Blais}}, \bibinfo {author} {\bibfnamefont {A.~L.}\ \bibnamefont {Grimsmo}},\
		and\ \bibinfo {author} {\bibfnamefont {A.}~\bibnamefont {Wallraff}},\
	}\bibfield  {title} {\bibinfo {title} {Circuit quantum electrodynamics},\
	}\href {https://doi.org/10.1103/RevModPhys.93.025005} {\bibfield  {journal}
		{\bibinfo  {journal} {Reviews of Modern Physics}\ }\textbf {\bibinfo {volume}
			{93}},\ \bibinfo {pages} {025005} (\bibinfo {year} {2021})}\BibitemShut
	{NoStop}%
	\bibitem [{\citenamefont {Toth}\ \emph {et~al.}(2017)\citenamefont {Toth},
		\citenamefont {Bernier}, \citenamefont {Nunnenkamp}, \citenamefont
		{Feofanov},\ and\ \citenamefont {Kippenberg}}]{toth2017dissipative}%
	\BibitemOpen
	\bibfield  {author} {\bibinfo {author} {\bibfnamefont {L.~D.}\ \bibnamefont
			{Toth}}, \bibinfo {author} {\bibfnamefont {N.~R.}\ \bibnamefont {Bernier}},
		\bibinfo {author} {\bibfnamefont {A.}~\bibnamefont {Nunnenkamp}}, \bibinfo
		{author} {\bibfnamefont {A.}~\bibnamefont {Feofanov}},\ and\ \bibinfo
		{author} {\bibfnamefont {T.}~\bibnamefont {Kippenberg}},\ }\bibfield  {title}
	{\bibinfo {title} {A dissipative quantum reservoir for microwave light using
			a mechanical oscillator},\ }\href {https://doi.org/10.1038/nphys4121}
	{\bibfield  {journal} {\bibinfo  {journal} {Nature Physics}\ }\textbf
		{\bibinfo {volume} {13}},\ \bibinfo {pages} {787} (\bibinfo {year}
		{2017})}\BibitemShut {NoStop}%
	\bibitem [{\citenamefont {Bernier}(2019)}]{bernier2019multimode}%
	\BibitemOpen
	\bibfield  {author} {\bibinfo {author} {\bibfnamefont {N.~R.}\ \bibnamefont
			{Bernier}},\ }\emph {\bibinfo {title} {Multimode microwave circuit
			optomechanics as a platform to study coupled quantum harmonic oscillators}},\
	\href@noop {} {Ph.D. thesis} (\bibinfo {year} {2019})\BibitemShut {NoStop}%
	\bibitem [{\citenamefont {Bowen}\ and\ \citenamefont
		{Milburn}(2015)}]{bowen2015quantum}%
	\BibitemOpen
	\bibfield  {author} {\bibinfo {author} {\bibfnamefont {W.~P.}\ \bibnamefont
			{Bowen}}\ and\ \bibinfo {author} {\bibfnamefont {G.~J.}\ \bibnamefont
			{Milburn}},\ }\href@noop {} {\emph {\bibinfo {title} {Quantum
				optomechanics}}}\ (\bibinfo  {publisher} {CRC press},\ \bibinfo {year}
	{2015})\BibitemShut {NoStop}%
	\bibitem [{\citenamefont {Joshi}\ \emph {et~al.}(2021)\citenamefont {Joshi},
		\citenamefont {Sauerwein}, \citenamefont {Youssefi}, \citenamefont {Uhrich},\
		and\ \citenamefont {Kippenberg}}]{joshi2021automated}%
	\BibitemOpen
	\bibfield  {author} {\bibinfo {author} {\bibfnamefont {Y.~J.}\ \bibnamefont
			{Joshi}}, \bibinfo {author} {\bibfnamefont {N.}~\bibnamefont {Sauerwein}},
		\bibinfo {author} {\bibfnamefont {A.}~\bibnamefont {Youssefi}}, \bibinfo
		{author} {\bibfnamefont {P.}~\bibnamefont {Uhrich}},\ and\ \bibinfo {author}
		{\bibfnamefont {T.~J.}\ \bibnamefont {Kippenberg}},\ }\bibfield  {title}
	{\bibinfo {title} {Automated wide-ranged finely tunable microwave cavity for
			narrowband phase noise filtering},\ }\bibfield  {journal} {\bibinfo
		{journal} {Review of Scientific Instruments}\ }\textbf {\bibinfo {volume}
		{92}},\ \href {https://doi.org/10.1063/5.0034696} {10.1063/5.0034696}
	(\bibinfo {year} {2021})\BibitemShut {NoStop}%
	\bibitem [{\citenamefont {Rabl}\ \emph {et~al.}(2009)\citenamefont {Rabl},
		\citenamefont {Genes}, \citenamefont {Hammerer},\ and\ \citenamefont
		{Aspelmeyer}}]{rabl2009phase}%
	\BibitemOpen
	\bibfield  {author} {\bibinfo {author} {\bibfnamefont {P.}~\bibnamefont
			{Rabl}}, \bibinfo {author} {\bibfnamefont {C.}~\bibnamefont {Genes}},
		\bibinfo {author} {\bibfnamefont {K.}~\bibnamefont {Hammerer}},\ and\
		\bibinfo {author} {\bibfnamefont {M.}~\bibnamefont {Aspelmeyer}},\ }\bibfield
	{title} {\bibinfo {title} {Phase-noise induced limitations on cooling and
			coherent evolution in optomechanical systems},\ }\href
	{https://doi.org/10.1103/PhysRevA.80.063819} {\bibfield  {journal} {\bibinfo
			{journal} {Physical Review A}\ }\textbf {\bibinfo {volume} {80}},\ \bibinfo
		{pages} {063819} (\bibinfo {year} {2009})}\BibitemShut {NoStop}%
	\bibitem [{\citenamefont {Kono}\ \emph {et~al.}(2024)\citenamefont {Kono},
		\citenamefont {Pan}, \citenamefont {Chegnizadeh}, \citenamefont {Wang},
		\citenamefont {Youssefi}, \citenamefont {Scigliuzzo},\ and\ \citenamefont
		{Kippenberg}}]{kono2023mechanically}%
	\BibitemOpen
	\bibfield  {author} {\bibinfo {author} {\bibfnamefont {S.}~\bibnamefont
			{Kono}}, \bibinfo {author} {\bibfnamefont {J.}~\bibnamefont {Pan}}, \bibinfo
		{author} {\bibfnamefont {M.}~\bibnamefont {Chegnizadeh}}, \bibinfo {author}
		{\bibfnamefont {X.}~\bibnamefont {Wang}}, \bibinfo {author} {\bibfnamefont
			{A.}~\bibnamefont {Youssefi}}, \bibinfo {author} {\bibfnamefont
			{M.}~\bibnamefont {Scigliuzzo}},\ and\ \bibinfo {author} {\bibfnamefont
			{T.~J.}\ \bibnamefont {Kippenberg}},\ }\bibfield  {title} {\bibinfo {title}
		{Mechanically induced correlated errors on superconducting qubits with
			relaxation times exceeding 0.4 ms},\ }\href
	{https://doi.org/10.1038/s41467-024-48230-3} {\bibfield  {journal} {\bibinfo
			{journal} {Nature Communications}\ }\textbf {\bibinfo {volume} {15}},\
		\bibinfo {pages} {3950} (\bibinfo {year} {2024})}\BibitemShut {NoStop}%
	\bibitem [{\citenamefont {Yeh}(2006)}]{yeh2006optical}%
	\BibitemOpen
	\bibfield  {author} {\bibinfo {author} {\bibfnamefont {P.}~\bibnamefont
			{Yeh}},\ }\href@noop {} {\emph {\bibinfo {title} {Optical Waves in Layered
				Media}}}\ (\bibinfo  {publisher} {Wiley Online Library},\ \bibinfo {year}
	{2006})\BibitemShut {NoStop}%
	\bibitem [{\citenamefont {Rieger}\ \emph {et~al.}(2023)\citenamefont {Rieger},
		\citenamefont {G{\"u}nzler}, \citenamefont {Spiecker}, \citenamefont
		{Nambisan}, \citenamefont {Wernsdorfer},\ and\ \citenamefont
		{Pop}}]{rieger2023fano}%
	\BibitemOpen
	\bibfield  {author} {\bibinfo {author} {\bibfnamefont {D.}~\bibnamefont
			{Rieger}}, \bibinfo {author} {\bibfnamefont {S.}~\bibnamefont {G{\"u}nzler}},
		\bibinfo {author} {\bibfnamefont {M.}~\bibnamefont {Spiecker}}, \bibinfo
		{author} {\bibfnamefont {A.}~\bibnamefont {Nambisan}}, \bibinfo {author}
		{\bibfnamefont {W.}~\bibnamefont {Wernsdorfer}},\ and\ \bibinfo {author}
		{\bibfnamefont {I.}~\bibnamefont {Pop}},\ }\bibfield  {title} {\bibinfo
		{title} {Fano interference in microwave resonator measurements},\ }\href
	{https://doi.org/10.1103/PhysRevApplied.20.014059} {\bibfield  {journal}
		{\bibinfo  {journal} {Physical Review Applied}\ }\textbf {\bibinfo {volume}
			{20}},\ \bibinfo {pages} {014059} (\bibinfo {year} {2023})}\BibitemShut
	{NoStop}%
	\bibitem [{\citenamefont {Eichler}\ and\ \citenamefont
		{Wallraff}(2014)}]{eichler2014controlling}%
	\BibitemOpen
	\bibfield  {author} {\bibinfo {author} {\bibfnamefont {C.}~\bibnamefont
			{Eichler}}\ and\ \bibinfo {author} {\bibfnamefont {A.}~\bibnamefont
			{Wallraff}},\ }\bibfield  {title} {\bibinfo {title} {Controlling the dynamic
			range of a josephson parametric amplifier},\ }\href
	{https://doi.org/10.1140/epjqt2} {\bibfield  {journal} {\bibinfo  {journal}
			{EPJ Quantum Technology}\ }\textbf {\bibinfo {volume} {1}},\ \bibinfo {pages}
		{1} (\bibinfo {year} {2014})}\BibitemShut {NoStop}%
	\bibitem [{\citenamefont {Zemlicka}\ \emph {et~al.}(2023)\citenamefont
		{Zemlicka}, \citenamefont {Redchenko}, \citenamefont {Peruzzo}, \citenamefont
		{Hassani}, \citenamefont {Trioni}, \citenamefont {Barzanjeh},\ and\
		\citenamefont {Fink}}]{zemlicka2023compact}%
	\BibitemOpen
	\bibfield  {author} {\bibinfo {author} {\bibfnamefont {M.}~\bibnamefont
			{Zemlicka}}, \bibinfo {author} {\bibfnamefont {E.}~\bibnamefont {Redchenko}},
		\bibinfo {author} {\bibfnamefont {M.}~\bibnamefont {Peruzzo}}, \bibinfo
		{author} {\bibfnamefont {F.}~\bibnamefont {Hassani}}, \bibinfo {author}
		{\bibfnamefont {A.}~\bibnamefont {Trioni}}, \bibinfo {author} {\bibfnamefont
			{S.}~\bibnamefont {Barzanjeh}},\ and\ \bibinfo {author} {\bibfnamefont
			{J.~M.}\ \bibnamefont {Fink}},\ }\bibfield  {title} {\bibinfo {title}
		{Compact vacuum-gap transmon qubits: Selective and sensitive probes for
			superconductor surface losses},\ }\href
	{https://doi.org/10.1103/PhysRevApplied.20.044054} {\bibfield  {journal}
		{\bibinfo  {journal} {Physical Review Applied}\ }\textbf {\bibinfo {volume}
			{20}},\ \bibinfo {pages} {044054} (\bibinfo {year} {2023})}\BibitemShut
	{NoStop}%
	\bibitem [{\citenamefont {Cicak}\ \emph {et~al.}(2009)\citenamefont {Cicak},
		\citenamefont {Allman}, \citenamefont {Strong}, \citenamefont {Osborn},\ and\
		\citenamefont {Simmonds}}]{cicak2009vacuum}%
	\BibitemOpen
	\bibfield  {author} {\bibinfo {author} {\bibfnamefont {K.}~\bibnamefont
			{Cicak}}, \bibinfo {author} {\bibfnamefont {M.~S.}\ \bibnamefont {Allman}},
		\bibinfo {author} {\bibfnamefont {J.~A.}\ \bibnamefont {Strong}}, \bibinfo
		{author} {\bibfnamefont {K.~D.}\ \bibnamefont {Osborn}},\ and\ \bibinfo
		{author} {\bibfnamefont {R.~W.}\ \bibnamefont {Simmonds}},\ }\bibfield
	{title} {\bibinfo {title} {Vacuum-gap capacitors for low-loss superconducting
			resonant circuits},\ }\href {https://doi.org/10.1109/TASC.2009.2019665}
	{\bibfield  {journal} {\bibinfo  {journal} {IEEE transactions on applied
				superconductivity}\ }\textbf {\bibinfo {volume} {19}},\ \bibinfo {pages}
		{948} (\bibinfo {year} {2009})}\BibitemShut {NoStop}%
	\bibitem [{\citenamefont {Scigliuzzo}\ \emph {et~al.}(2020)\citenamefont
		{Scigliuzzo}, \citenamefont {Bruhat}, \citenamefont {Bengtsson},
		\citenamefont {Burnett}, \citenamefont {Roudsari},\ and\ \citenamefont
		{Delsing}}]{scigliuzzo2020phononic}%
	\BibitemOpen
	\bibfield  {author} {\bibinfo {author} {\bibfnamefont {M.}~\bibnamefont
			{Scigliuzzo}}, \bibinfo {author} {\bibfnamefont {L.~E.}\ \bibnamefont
			{Bruhat}}, \bibinfo {author} {\bibfnamefont {A.}~\bibnamefont {Bengtsson}},
		\bibinfo {author} {\bibfnamefont {J.~J.}\ \bibnamefont {Burnett}}, \bibinfo
		{author} {\bibfnamefont {A.~F.}\ \bibnamefont {Roudsari}},\ and\ \bibinfo
		{author} {\bibfnamefont {P.}~\bibnamefont {Delsing}},\ }\bibfield  {title}
	{\bibinfo {title} {Phononic loss in superconducting resonators on
			piezoelectric substrates},\ }\href {https://doi.org/10.1088/1367-2630/ab8044}
	{\bibfield  {journal} {\bibinfo  {journal} {New Journal of Physics}\ }\textbf
		{\bibinfo {volume} {22}},\ \bibinfo {pages} {053027} (\bibinfo {year}
		{2020})}\BibitemShut {NoStop}%
	\bibitem [{\citenamefont {Clark}\ \emph {et~al.}(2017)\citenamefont {Clark},
		\citenamefont {Lecocq}, \citenamefont {Simmonds}, \citenamefont {Aumentado},\
		and\ \citenamefont {Teufel}}]{clark2017sideband}%
	\BibitemOpen
	\bibfield  {author} {\bibinfo {author} {\bibfnamefont {J.~B.}\ \bibnamefont
			{Clark}}, \bibinfo {author} {\bibfnamefont {F.}~\bibnamefont {Lecocq}},
		\bibinfo {author} {\bibfnamefont {R.~W.}\ \bibnamefont {Simmonds}}, \bibinfo
		{author} {\bibfnamefont {J.}~\bibnamefont {Aumentado}},\ and\ \bibinfo
		{author} {\bibfnamefont {J.~D.}\ \bibnamefont {Teufel}},\ }\bibfield  {title}
	{\bibinfo {title} {Sideband cooling beyond the quantum backaction limit with
			squeezed light},\ }\href {https://doi.org/10.1038/nature20604} {\bibfield
		{journal} {\bibinfo  {journal} {Nature}\ }\textbf {\bibinfo {volume} {541}},\
		\bibinfo {pages} {191} (\bibinfo {year} {2017})}\BibitemShut {NoStop}%
\end{thebibliography}
\end{document}